\documentclass[12pt,a4paper]{article}

\usepackage[left=2.5cm,right=2.5cm,top=3cm,bottom=3cm,bindingoffset=0cm]{geometry}

\usepackage{amsmath}
\usepackage{stmaryrd} 
\usepackage{amsfonts}
\usepackage{amssymb}
\usepackage{graphicx}
\usepackage{url}
\usepackage{natbib}
\bibpunct{(}{)}{;}{a}{,}{,}
\usepackage[small,bf]{caption}

\usepackage{pdflscape}

\usepackage{todonotes}

\usepackage{ulem}
\usepackage{color}

\usepackage{amsthm}
\theoremstyle{definition}
\newtheorem{remark}{Remark}
\theoremstyle{definition}

\newcommand{\vect}[1]{\textbf{#1}}
\newcommand{\obs}{\hat}

\newcommand{\bx}{\textbf{x}}

\newcommand{\bxh}{\hat{\bx}}
\newcommand{\bxs}{\textbf{x}^{\star}}
\newcommand{\bz}{\textbf{z}}

\newcommand{\bK}{\textbf{K}}
\newcommand{\bL}{\textbf{L}}
\newcommand{\bT}{\textbf{T}}

\newcommand{\bm}{\textbf{m}}
\newcommand{\bc}{\hat{\textbf{c}}}
\newcommand{\bcs}{\textbf{c}^{\star}}

\renewcommand{\bf}{\textbf{f}}
\newcommand{\bfs}{\textbf{f}^{\star}}
\newcommand{\ind}{\mathbb{I}}
\newcommand{\nsigma}{\sigma_{\epsilon}}

\newcommand{\bnu}{\boldsymbol{\nu}}

\newcommand{\Cov}{\mathrm{Cov}}

\newcommand{\R}{\mathbb{R}}
\newcommand{\E}{\mathbb{E}}

\DeclareMathOperator*{\argmax}{arg\,max}

\title{\textbf{ A Probabilistic Approach to Nonparametric Local Volatility}}
\author{Martin Tegn\'{e}r\thanks{Correspondence to: \texttt{martin.tegner@eng.ox.ac.uk}. We thank 
Chris Oates, Mike Osborne and Christoph Reisinger for helpful comments and discussions.}\, and Stephen Roberts \\ 
Department of Engineering Science \\
Oxford-Man Institute \\
University of Oxford, UK}

\date{}

\begin{document}

\maketitle

\begin{abstract}
\noindent The local volatility model is a  widely used for pricing and hedging financial derivatives. While its main appeal is its capability of reproducing any given surface of observed option prices---it provides a perfect fit---the essential component  is a latent function which can be uniquely determined only in the limit of infinite data. To (re)construct this function, numerous calibration methods have been suggested involving steps of interpolation and extrapolation, most often of parametric form and with point-estimate representations. We look at the calibration problem in a probabilistic framework with a  nonparametric approach based on a Gaussian process prior. This immediately gives a way of encoding prior beliefs about the local volatility function and a hypothesis model which is highly flexible yet not prone to overfitting. Besides providing a method for calibrating a (range of) point-estimate(s), we  draw posterior inference from the distribution over local volatility. This leads to a better understanding of  uncertainty associated with the calibration in particular, and with the model in general. Further, we infer dynamical properties of local volatility by augmenting the hypothesis space with a time dimension. Ideally, this provides predictive distributions not only locally, but also for entire surfaces forward in time. We  apply our approach to S\&P 500 market data. 
\end{abstract}

\section{Introduction}

In the context of option pricing, the local volatility model introduced by \cite{derman1994} and  \cite{dupire1994pricing} is well celebrated, and a versatile generalisation of the foundational Black-Scholes model (\cite{black1973pricing} and  \cite{merton1973theory}). 
From the latter, the constant diffusion coefficient  of the underlying stock price---the \textit{volatility}---is replaced by a function of current time and stock price---the \textit{local volatility function}. As volatility effectively determines how a diffusion model  prices options on the underlying stock, going from a single- to an infinite-dimensional parameter is a tactical move: While Black-Scholes fits only one observed option price perfectly, the local volatility function provides flexibility to  fit a whole \textit{surface} of prices. Dupire's formula even provides an explicit recipe for how this can be done. However, the latent local volatility function can only be uniquely retrieved  if one has access to a \textit{continuum} of option prices at all strikes and maturities. Here begins the challenge of local volatility modelling.

\begin{figure}
\makebox[\textwidth][c]{  
\includegraphics[scale=0.55,trim=50 50 50 50,clip]{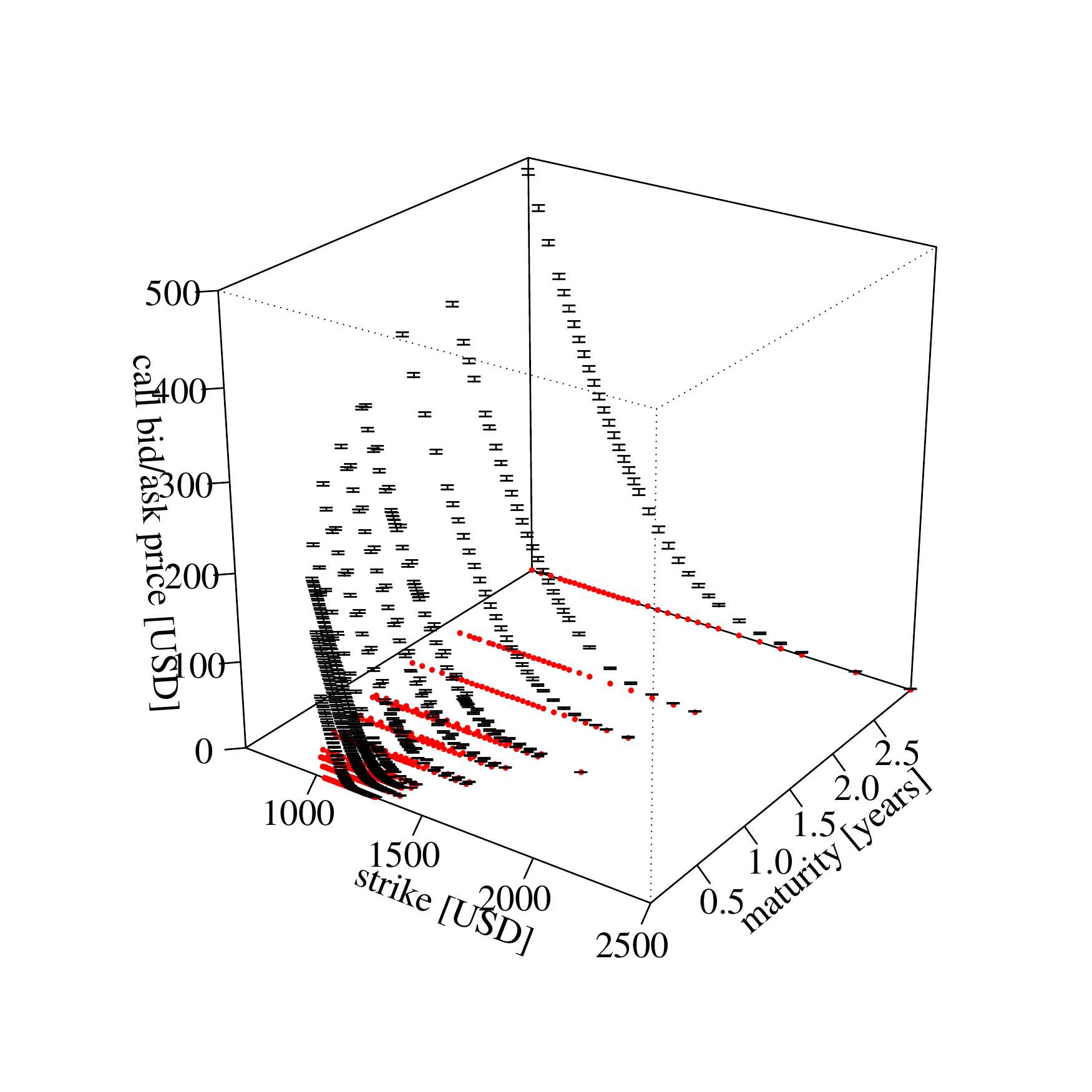}
\includegraphics[scale=0.55,trim=50 50 50 50,clip]{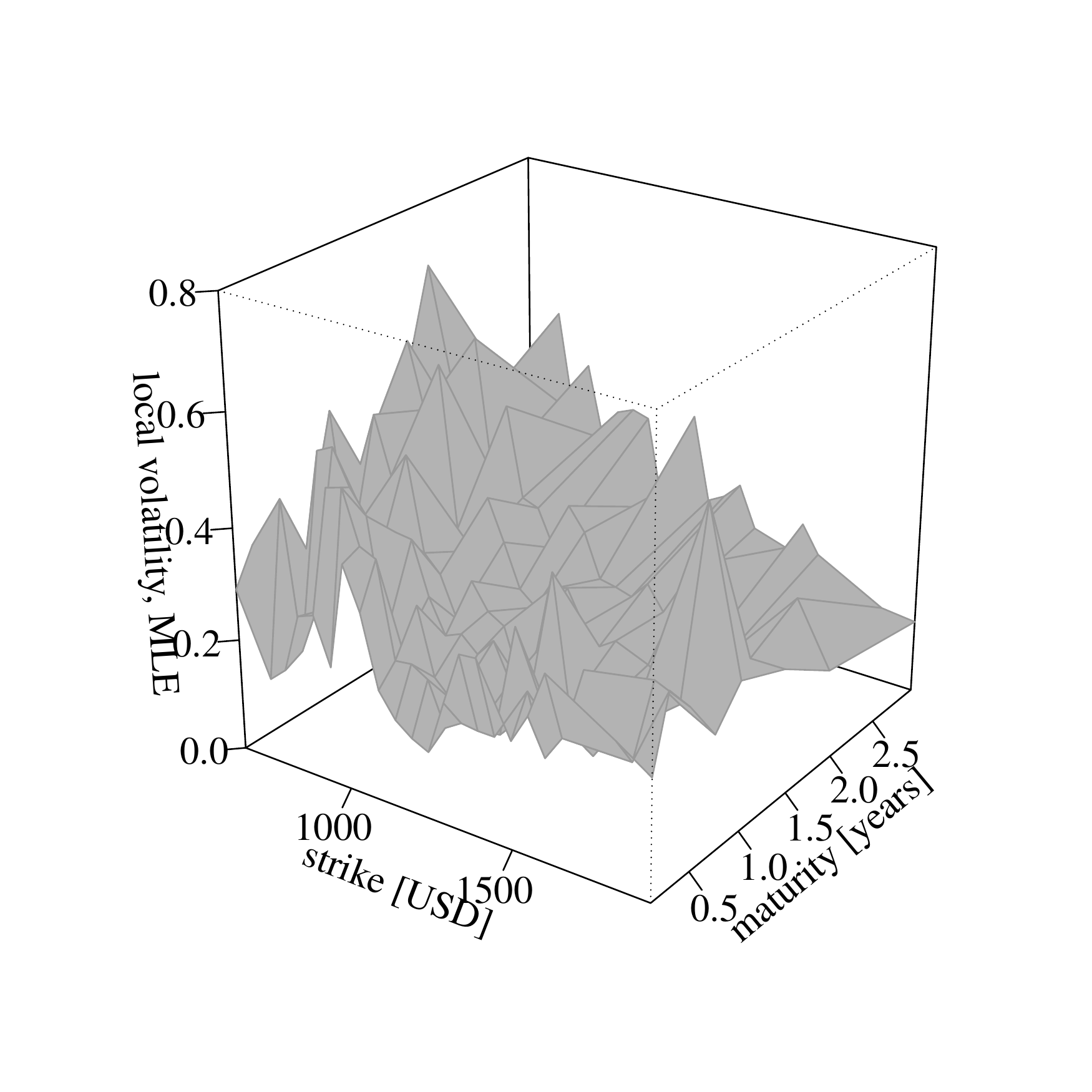}
}
\caption{\textbf{Left:} Market prices of S\&P 500 call option:  best bid--ask quotes as of closing 4 January 2010. \textbf{Right:} Calibrated local volatility surface from optimising squared price errors \eqref{eqnG} over all  local volatility points. The surface achieves the same error as the \textit{maximum a posteriori} estimate shown in Figure \ref{fig2}.}
\label{fig9}
\end{figure}

With a discrete and finite 
set of market prices that can be utilised for calibration (Figure \ref{fig9}, (left) illustrates one set of data we use in this paper) local volatility modelling becomes tantamount to proposing a finite-dimensional approximation to the local volatility function along with a routine that details how to calibrate the suggested model to market prices. Two general approaches appear to form common practice: either, one works with observable option prices (or equivalently, implied volatilities) with an interpolation/extrapolation scheme such that Dupire's formula can be employed for a direct calculation of local volatility (\cite{kahale2004arbitrage}, \cite{benko2007extracting}, \cite{fengler2009arbitrage}, \cite{fengler2011semi}, \cite{glaser2012arbitrage}, to mention just but a few). Alternatively, a parametrised functional form is assumed for the non-observable local volatility function and calibrated to market prices by minimisation of a suitable objective function based on  model-to-market discrepancy (see for example \cite{jackson1998computation}, \cite{luo2010local}, \cite{andreasen2011volatility}, \cite{lipton2011filling}, \cite{reghai2012local}).

One challenge with the first approach is that it requires the call price surface to be interpolated in an arbitrage-free manner.  Indeed, constructing an interpolation method that produces option prices consistent with observed data is essentially equivalent to constructing (and estimating) an option price model; and so we are back to square one. Furthermore, the resulting local volatility model is generally non-robust and non-smooth, 
both of which are problems that make further employment (hedging, pricing exotics, quantifying risks, etc.) under this approach questionable (\cite{modelCalibration}, \cite{hirsa2003effect}).

Proposing a functional form for local volatility and  formulating the calibration problem as one of  targetting a  model-to-market distance is perhaps a more appealing approach, even if it also has the challenge of designing an appropriate model. 
As for the number of  parameters of the functional form, choosing too few might lead to  under-parametrised models unable to fit market data. At the other extreme, choosing as many parameters as observed option prices makes the ill-posedness of the problem severe: With finite market data, a solution to the minimal model-to-market distance under some objective is  non-unique, and generally non-smooth and non-stable. For a vivid illustration of this point, Figure \ref{fig9}, (right) shows a local volatility surface obtained from optimisation\footnote{We use a stochastic gradient method for this purpose, \cite{spall1998implementation}.} over all points of the surface that correspond to market data.   On top of this is the fact that the optimisation problem is non-trivial. Common objectives, such as weighted least-squares, yield non-convex maps with respect to local volatility and  are also unsuitable for gradient-based optimisers due to dimensionality. To this end, regularisation has been proposed to both address the ill-posedness of the problem and to impose structure on the local volatility parameter (\cite{lagnado1997technique}, \cite{chiarella2000calibration}, \cite{crepey2003calibration}, \cite{cezaro2008convex}).

This paper adapts the latter approach and attempts to address its shortcomings. Firstly, the non-uniqueness of an optimal local volatility is tackled by acknowledging that \textit{parameter uncertainty} pertains to the model, whereby the  focus of  calibration is shifted from an optimal point-estimate to a probability distribution over local volatility functions.  This is reasonable  from a statistical perspective, by realising that common optimisation objectives correspond to likelihoods of  (generalised) Gaussian noise models, and further by the fact that market prices are only observed up to a bid-ask spread.

A probabilistic formulation  for local volatility appears in \cite{cont2004recovering}, who suggest an algorithm for generating  samples of a parametrised model that yields option prices within bid-ask spreads of market data. While this sample is intended for a coherent measure of model risk (\cite{cont2006model}), we  adopt a  Bayesian formalism and concentrate on drawing  inference from 
the joint posterior.
In this regard, our approach is similar in spirit to what has been proposed in a recent paper by \cite{gupta2014robust} for the parametric model of \cite{jackson1998computation}. Most notably, we extend their Bayesian idea  by proposing a nonparametric functional prior for local volatility, and leverage the rich framework of Gaussian processes for this purpose. We also draw inference about the observation model for a full Bayesian treatment. Fixing parameters of the latter, as in  \cite{gupta2014robust},  influences the posterior over local volatility and critically leads to biased uncertainty estimates.

Gaussian processes are nonparametric models which offer a  generic tool for estimation that is particularly suitable for Bayesian setups.
 They are widely used by the machine learning community due to their expressiveness, tractability and robustness to overfitting (\cite{rasmussen2006gaussian}, \cite{roberts2013gaussian}). In the light of local volatility modelling and calibration, we find a Bayesian Gaussian-process approach beneficial for several reasons: 

\begin{itemize}
\item
The ability to explicitly encode prior beliefs about local volatility in a straightforward way. For instance, it is widely understood that a smooth surface is desirable, especially for further hedging and pricing purposes. Such functional characteristics are easily encoded by specifying a suitable mean and covariance kernel. Comparing with the prior formulation outlined in \cite{gupta2014robust}, we  find a Gaussian process specification  more intuitive, flexible and direct. Indeed, even more so in contrast to calibration approaches which use regularisation.

\item
A Gaussian process model provides full posterior predictions, both interpolations within and extrapolations outside input strike-maturity data. Moreover, predictions come with an associated notion of uncertainty which is appealing from a model-risk perspective. We utilise this prediction ability to estimate volatility points both internally on the surface, and for prediction of the entire local volatility surface forward in time. 

\item
The  inference process for Gaussian processes  is well studied, i.e. the  practical steps of model calibration to data. We  utilise methods widely used in the machine learning literature for efficient sampling.  

\item
Finally, we use  the fact that  inferring a latent local volatility surface, by minimising an objective of  option prices, effectively corresponds to a likelihood model; most notably with additive Gaussian noise for the  squared error objective. In fact, our setup is flexible enough to accommodate any choice of calibration instruments and noise model, where the latter takes into account that one observse bid-ask quotes in place of idealised non-arbitrage prices. The noise variance (and more generally  correlation structure) is then associated with the size of the bid-ask spread and it is naturally included in the Bayesian inference process.

\end{itemize}

The  rest of the paper is organised as follows. Section \ref{secLV} reviews  the local volatility model and discusses practical aspects that motivate our approach. The latter is detailed in Section \ref{secPF} along with predictive equations and numerical methods for inference. Section \ref{secEX} contains experiments, Section \ref{secLVdynamics} an extension on local volatility  over time. Finally, Section \ref{secC} concludes.

\section{Local volatility modelling}\label{secLV}
To set the scene, we consider a financial market model consisting of a risk-free money account and a risky asset whose price process is modelled by the local volatility model. Since we are  interested in option pricing in this paper, we  focus on the risk-neutral perspective and describe some theoretical and practical aspects thereof in the following sections.

\subsection{{European option pricing} }
The local volatility model is formulated as the class of diffusion processes under which the asset price $S$ has risk-neutral dynamics under $Q$
\begin{equation} \label{eqnSDE}
dS_t = (r-q) S_t dt + \sigma(t,S_t)S_tdW_t,\quad t\in[0,\bar{T}]
\end{equation}
where the latent \textit{local volatility function}, $\sigma:\mathcal{X}\rightarrow \R$, on the domain $\mathcal{X}=[0,\bar{T}]\times\R^+$, essentially represents the model. 

For a current time $t$, a European call option with strike $K$ and maturity $T\geq t$ is defined by its payoff $(S_T-K)^+$ at maturity where 
$S_T$ is unknown for $T>t$. If $S$ follows  the local volatility model,  the  price of the option at time $t$ is given by a pricing functional
\begin{equation*}
(t,S_t,T,K,\sigma) \mapsto C(t,S_t,T,K,\sigma)
\end{equation*}
which satisfies Black-Scholes' equation in the variables $(t,s)$
\begin{equation}\label{eqnBS}
\frac{\partial C}{\partial t} + \frac{1}{2}\sigma^2(t,s)s^2\frac{\partial^2C}{\partial s^2} + (r-q)s \frac{\partial C}{\partial s}	-rC = 0
\end{equation}
with terminal condition $C(T,s)=(s-K)^+$.  This is due to no-arbitrage pricing theory (see e.g. \cite{bjork2009arbitrage}) and equation (\ref{eqnBS}) can equivalently be represented by the  risk-neutral  pricing formula 
\begin{equation} \label{eqnRNPF}
C(t,s) = \E^Q\left[\left. e^{-r(T-t)} (S_T-K)^+ \right| S_t = s \right].
\end{equation}
More generally, the Markovian pricing formula \eqref{eqnRNPF} yields the price of any contingent claim by replacing the  payoff of the call option  
with the corresponding payoff of the claim. 
For  European-style exercises, (\ref{eqnRNPF}) also admits a representation (\ref{eqnBS}) with the terminal condition replaced by appropriate boundary conditions associated with  the  claim. 

The main appeal of the local volatility model, at least from a theoretical viewpoint, is that it is capable of reproducing any set $\{c_i\}$ of time-$t$ observable  call  prices.\footnote{Generally speaking, one can reproduce $\{c_i \}$ only if it lies within the range of prices attainable by the model. For local volatility, all sets of prices are attainable as long as they are \textit{consistent} in the sense of no static arbitrage opportunities (see \cite{carr2005note}). This is  encoded in Dupires formula (\ref{eqnDupireLV}) since $C$ must be non-decreasing in maturity (otherwise conversion arbitrage) and convex in strike (otherwise butterfly arbitrage) for the local volatility to be admissible.   }   
To each price $c_i$, there is an associated  maturity and strike price which we collect with an input variable $x=(T,K)\in\mathcal{X}$. Thus, there exists a local volatility function $\sigma$ 
such that
\begin{equation}\label{eqn1}
c_i = C(x_i,\sigma),\quad \forall i 
\end{equation}
for any set $\{c_i \}$ with corresponding strike-maturity inputs $\{x_i\}$.  In \eqref{eqn1} and in what follows, we suppress the current state variable $(t,S_t)=(0,S_0)$ from the call price functional and simply write $C(x,\sigma)\equiv C(T,K,\sigma)$. Hence, we consider the market as seen from a single date only. We will have more to say about time evolution of volatility in Section \ref{secLVdynamics}.

In view of (\ref{eqn1}) one would expect that the knowledge about $\sigma$ grows with  number of observations. Indeed, given a consistent continuum of call prices $\{C(T,K)$\,:\,$(T,K)\in\mathcal{X}\}$, or equivalently, a price function $C:\mathcal{X}\rightarrow\R$,
 Dupire's formula (\cite{dupire1994pricing}) gives the unique local volatility function
\begin{equation}\label{eqnDupireLV}
\sigma(T,K) = \sqrt{ \frac{\frac{\partial C}{\partial T}-(r-q)\left(C-K \frac{\partial C}{\partial K} \right)}{\frac{1}{2} K^2 \frac{\partial^2C}{\partial K^2}} }
\end{equation}
 compatible with this infinite set of prices. The converse is also true: given a local volatility function, Dupire's \textit{forward} equation gives the {unique} call price function
\begin{equation}\label{eqnDupire}
\frac{\partial C}{\partial T} + (r-q)K\frac{\partial C}{\partial K} - \frac{K^2\sigma^2(T,K)}{2}\frac{\partial^2C}{\partial K^2}+qC = 0
\end{equation}
with initial condition $C(0,K) = (s-K)^+$. This is a restatement of the \textit{backward} equation (\ref{eqnBS}) in the variables $(T,K)$ and there is thus a one-to-one correspondence between a local volatility function and a call price function.

\begin{remark}
The local volatility model in (\ref{eqnSDE}) is formulated with constant  risk-free rate of return $r$ and dividend yield $q$. This is for convenience only, and the methodology suggested in this paper applies equally to settings with time-dependent rates and dividends, both continuous and discrete, as long as the set-up allows for a pricing equation that can be solved at least numerically.
\end{remark}

\subsection{Local volatility in practice}\label{seqLVIP}

As for most option pricing models, there is no known closed-form solution to the call price equation \eqref{eqnBS} nor \eqref{eqnDupire}, and one has to revert to numerical methods. To this end, Dupire's equation advances a second major appeal of the local volatility model: for a given input set of strike-maturities, 
a numerical solution of (\ref{eqnDupire}) gives the whole \textit{surface} of corresponding model prices 
simultaneously. This is an attractive feature from a practical viewpoint since one is  interested in computing a large set of prices corresponding to market quoted options  when performing \textit{calibration} of the model.

Mode-to-market calibration is an inverse problem commonly  formalised with least squares. Given a finite set of market data of call prices $\{\obs{c}_i,\hat{x}_i\}_{i=1}^{n}$, 
find a local volatility function such that the square error functional 
\begin{equation}\label{eqnG}
G(\sigma) = 
 \sum_{i=1}^{n} \left(C(\obs{x}_i,\sigma) - \obs{c}_i	\right)^2
\end{equation}
is minimised. Weighted least squares is also common, where the weights may represent liquidity through inverse bid-ask spreads, or the inverse Black-Scholes vega for a first order approximation of the square error in terms of implied volatility.

 In view of (\ref{eqn1}), it should be possible to find a $\sigma$ such that the error \eqref{eqnG} is zero.\footnote{At least, this could be expected when the data is a set of consistent  prices observed without noise. In practice, however, prices are observed up to their bid-ask quotes such that neither consistent nor noise-free prices are readily available.} 
However, due to what is effectively expressed in (\ref{eqnDupireLV}), one can not expect that there is a \textit{unique} optimal $\sigma$ to the calibration problem of minimising (\ref{eqnG}). This could only be achieved in the limit of infinite data.\footnote{Indeed, as mentioned in the introduction, the calibration problem is \textit{ill-posed} in the sense that there exist several solutions to the minimisation of (\ref{eqnG}). What further complicates is that $\sigma\mapsto G(\sigma)$ is neither convex nor suitable for gradient-based optimisers, see  the discussion in \cite{cont2004recovering}.  } In other words,  when calibrating to finite data, one can hope for  a local volatility function to be pinned down 
 only at a finite set of strike-maturities 
(not necessarily the same as market quoted strike-maturities) subject to uncertainty.

Dupire's formula also suggests a converse estimation approach in at least two flavours: either non-parametrically by approximating the derivatives of  (\ref{eqnDupireLV}) with finite differences, or by assuming  a parametric form $\tilde{C}$  in place of $C$, fitting its parameters to data and then plugging derivatives of $\tilde{C}$ into (\ref{eqnDupireLV}). However, while the former  is known to output ``spiky'' local volatility surfaces along with model-to-market errors that are far from minimal (a finite difference approximation is not even guaranteed to preserve the {positiveness} of local volatility), the latter brings questions about overfitting, interpolation/extrapolation behaviour as well as choice of parametric form---design choices which arguably should be made with  minimisation of a model-to-market error  in mind. Further, with a fixed choice of $\tilde{C}$, note that an increasing dataset  affiliates the estimation of the parameters of $\tilde{C}$, not necessarily the 
minimisation of (\ref{eqnG}). In all, both  approaches are  subject to the ambiguity present under finite data. The latter in that as long as the parametric form fits well with observations from the call price surface, one has liberty to chose its functional form. For the former, one is free to choose and tune the approximation methods for first- and second order derivatives. Finally, even if these methods appear more direct as they provide means of \textit{computing} the local volatility function  in place of \textit{imputing}, they should, at least in principle, support an equivalent formulation as an optimisation problem of the error functional (\ref{eqnG}) since this explicates an objective of minimising a distance to observed call prices.

In what follows we  build on the calibration approach based on the model-to-market measure. We propose to cast local volatility modelling in a probabilistic framework, such that the  calibration problem effectively corresponds to optimising a regularised version of (\ref{eqnG}). Conceptually, this  gives a clear probabilistic representation for a \textit{prior} local volatility model (corresponding to the regularising term) and results in a \textit{posterior distribution} over local volatility, as opposed to a point-estimate obtained from the  optimum under regularisation.

\section{A probabilistic framework}\label{secPF}

Following the  approach of inverse calibration through optimisation with  a square error, the next section outlines how a probabilistic framework can be adopted for local volatility with a nonparametric Gaussian process model. We  detail posterior predictive equations for local volatility and call prices, and end the section by proposing a numerical method.

\subsection{The calibration problem recast}

In practice,  prices of options are quoted only up to bid-ask spreads from which it is common the use the mid-market price for calibration. To this end, we assume a set of observable mid-market call prices ${\bc}=\{\obs{c}_i\}_{i=1}^n$  at a market set of strikes and maturities 
$\obs{\bx}=\{\obs{x}_i\}_{i=1}^n$ to be generated from the noise model
\begin{equation}\label{eqnNM}
c_i = C(\obs{x}_i,\sigma) + \epsilon_i 
\end{equation}
for a common local volatility function. Hence $C(\cdot,\sigma)$ represents the \textit{fair} price surface from which the mid-market prices are noisy observations. For simplicity, we assume $\{\epsilon_i\}$ is independent  Gaussian noise with variance $\nsigma^2$, and note  that our approach generalises easily to heteroscedastic noise structured over $\mathcal{X}$. This leads to a log-likelihood function
\begin{equation}\label{eqlikelihood1}
\log p(\bc|\sigma,\nsigma) = -\frac{1}{2\nsigma^2}\sum_{i=1}^n \left(  C(\obs{x}_i,\sigma) - \obs{c}_i \right)^2 - \frac{n}{2}\log\left(2\pi \nsigma^2\right).
\end{equation}
In terms of maximising the  likelihood with respect to the local volatility function, this is equivalent with minimising the sum of squared errors (\ref{eqnG}). 

As mentioned in the introduction, a standard approach to modelling is to model the local volatility function as a parametric function 
$\sigma(T,K)\equiv\sigma_{\theta}(T,K)$ with a finite-dimensional parameter $\theta$,\footnote{Examples include \cite{gupta2014robust}, \cite{cont2004recovering}, \cite{luo2010local} and \cite{jackson1998computation}. } and then to calibrate by minimising $\theta\mapsto G(\sigma_{\theta})$.
\footnote{It is also common to calibrate parts of $\theta$ sequentially to subsets of $\bc$ to obtain a more direct correspondence between calibrated parameters and particular sets of strikes-maturities; for instance slices of common maturity  (see the discussion in \cite{luo2010local}). However, since a particular model price $C(T_i,K_i,\sigma)$ depends on $\sigma(T,K)$ for all $K\geq0$, $T\leq T_i$, a \textit{partial} calibration to local parts of the price surface always yields a worse fit than a global optimum.}
With this as an example, taking the calibration problem to a probabilistic setting
 means postulating a \textit{prior} distribution $p(\theta)$ over the parameters 
based on prior knowledge about $\theta$. Note that it is  possible to include the noise variance $\nsigma$ in $\theta$. Conditioning on observed data, Bayes' rule then gives the \textit{posterior} distribution over parameters
\begin{equation*}
p(\theta|\bc) = \frac{p(\bc|\theta)p(\theta)}{p(\bc)}
\end{equation*}
where $p(\bc|\theta)$ is s the likelihood \eqref{eqlikelihood1} written as a function of $\theta$ and the \textit{marginal likelihood} is given by $p(\bc)=\int p(\bc|\theta)p(\theta) d\theta$.

In a Bayesian framework, the inference process considers the  posterior distribution over $\theta$ and not only  point-estimates. For  summary measures, one may use the \textit{maximum a posteriori} (MAP) estimator
\begin{equation*}
\theta_{\text{MAP}} =\argmax _{\theta} p(\theta|\bc).
\end{equation*}
Another common choice is the mean of $\theta$ with respect to the posterior distribution,  which is referred to as Bayes' estimator. For a notion of uncertainty attached to the parameter estimate, quantiles of the posterior distribution 
give \textit{credible intervals}.  We will make common use of the posterior mean plus/minus two standard deviations as a  more robust  alternative to quantile estimates.

\subsection{Probabilistic nonparametric local volatility}
At the base of our approach is a random  nonparametric local volatility function. 
To this end, we assume a Gaussian process \textit{functional prior}  with input space $\mathcal{X}$ (\cite{rasmussen2006gaussian})
\begin{eqnarray*}
&f\sim\mathcal{GP}(m,k(x,x')), \\
& \sigma =  \phi(f)
\end{eqnarray*}
where  $\phi$ is a positive function applied pointwise for imposing positivity of the local volatility function. 
The Gaussian process defines a distribution over real valued functions on $\mathcal{X}$  where 
any finite set of function evaluations of $f$ follow the multivariate normal distribution---see  \eqref{eqnfprior}.
In particular, prior beliefs are encoded in a  mean $m$ and covariance function
\begin{equation*}
k(x,x') = \Cov(f(x),f(x'))
\end{equation*}
defined for any  pair of inputs  in the strike-maturity space $\mathcal{X}$. We denote the parameters of the covariance function with $\kappa$, which together with $m$ and 
$\nsigma$ are referred to as the model's {hyperparameters}. Note that the mean and covariance function uniquely specify a Gaussian process. We use a constant $m$ for convenience although it is possible to use a parametrised function for more informed  beliefs about the mean.

The observed prices $\bc$ are conditioned on the function evaluated at   an input set which we denote with $\bx=\{x_i\}_{i=1}^N$, that is, on the surface\footnote{With $f({\bx})$ we mean the set of $f$-values at every point of the finite set $\bx$, i.e.  $f({\bx})\equiv\{f(x)\}_{x\in{\bx}}$. } $\bf = f({\bx})$, such that we have the log likelihood
\begin{equation}\label{eqnll2}
\log p(\bc|\bf,\nsigma) = -\frac{1}{2\nsigma^2}\sum_{i=1}^n  \left(  C(\obs{x}_i,\bf) - \obs{c}_i \right)^2  - \frac{n}{2}\log\left(2\pi \nsigma^2\right)
\end{equation}
In \eqref{eqnll2}, the call price  is written directly as a function of $\bf$, i.e. implicitly as a composition with the link function 
$\bf\mapsto\phi(\bf)=\boldsymbol{\sigma}$ where $\boldsymbol{\sigma}=\sigma(\bx)$.

There are a few  aspects to point out here. First, considering the likelihood as a function of $\bf=f(\bx)$ instead of $f(\cdot)$---i.e. a finite set of function evaluations in place of the  function object itself---is {without loss of generality} (see Remark \ref{remPred}) although a  necessity since we can only work with finite sets of function values in practice. When working with local volatility models, the call price function is given only up to a numerical solver of (\ref{eqnDupire}) 
which in turn relies on local volatility values taken  at a finite input set. Schematically, for a finite difference solver, we have 
\begin{equation}\label{eqnFDmap}
\bf\mapsto C(\bx,\bf).
\end{equation}
Second, while the finite difference mapping  (\ref{eqnFDmap}) outputs call prices over a strike-maturity set  corresponding to the set of input local volatility values, $\bx$ does not necessarily coincide with the observed market set $\obs{\bx}$. Typically, $\obs{\bx}$  is scattered over the strike-maturity space while a finite difference grid  is a regular Cartesian product
\begin{equation}\label{eqnCP}
\bx=\bT\times\bK=\{(T_i,K_j):i=1,\dots,I,j=1,\dots,J\}.
\end{equation} 
As long as $\bx$ is constructed such that $\obs{\bx}\subseteq\bx$, ($n\leq N$) we may evaluate the likelihood (\ref{eqnll2}) which depends on call prices over the market set only---see Figure \ref{fig1} for an example. Note, however, that the likelihood does indeed depend on every point in $\bx$ (on every value of $\bf$) since the model price at every strike-maturity  point $x$ is  dependent on the entire local volatility function up to the maturity of $x$, albeit with a stronger dependency \textit{locally} around $x$ due to the smoothness properties of \eqref{eqnDupire}. Ultimately, we want to infer posterior knowledge about the local volatility function not only at the points of the market set, but also for strike-maturities lying outside, and quantify how uncertain such predictions might be. Equations for this purpose will be given in Section \ref{secPred} (see also Remark \ref{remPred}). 

Third, the likelihood (\ref{eqnll2}) is neither factorisable nor a Gaussian with respect to $\bf$ (two properties commonly assumed in  Gaussian process regression). This  hinders analytical tractability and efficient computations.  In effect, what we face here is a nonlinear regression problem in a latent space where  observations are noisy outputs from a  transformation of an unobservable function. 


At the next level of inference, hyperparameters 
are considered unknown up to some \textit{hyperprior}. We then obtain the {joint posterior} for hyperparameters and function values
\begin{equation}\label{eqnJointPost}
p(\bf,\kappa,m,\nsigma|\bc) = \frac{1}{p(\bc)}\underbrace{ p(\bc|\bf,\nsigma) }_\text{likelihood} \underbrace{ p(\bf|\kappa,m) }_{\bf-\text{prior}} \underbrace{ p(\kappa,m,\nsigma) }_\text{hyperprior}
\end{equation}
where $p(\bf|\kappa,m)$ is short for the prior over $\bf$ induced by the Gaussian process. By definition
\begin{equation}\label{eqnfprior}
\bf\sim\mathcal{N}(m,\bK_{\bf\bf})
\end{equation}
 with $\bK_{\bf\bf}$  the covariance matrix given by evaluating the covariance function at all pairs of  input points $\bx$, that is, with elements
\begin{equation*}
[\bK_{\bf\bf}]_{i,j}=k(x_i,x_j). 
\end{equation*}

Eventually we write $\bK_{\kappa}\equiv\bK_{\bf\bf}$ to stress that it is a matrix valued function of $\kappa$. We use a product covariance function of squared exponentials
\begin{gather}\label{eqnSE}
k(x_i,x_j;\kappa) = \sigma_f^2 \, k_\text{SE}(T_i,T_j;l_T)\, k_\text{SE}(K_i,K_j;l_K), \\ 
k_\text{SE}(x,x';l) = \exp\left(-\frac{(x-x')^2}{2l^2}   \right),  \nonumber 
\end{gather}
with parameters collected in $\kappa = (l_T,l_K,\sigma_f)$. Since \eqref{eqnSE} is smooth with finite derivatives at zero of all orders, $f$ will have continuous sample-function derivatives of all orders (\cite{lindgren2012stationary}) and we thus impose a strong smoothness property on the local volatility function. 
This is a desire commonly expressed in the literature, especially when the model is considered for derivative hedging and pricing purposes (e.g. \cite{jackson1998computation}).  We also consider a product Mat\'{e}rn 3/2 covariance function with factors

\begin{equation}\label{eqnMat32}
k_{\text{Mat32}}(x,x';l) =  \left(1+\frac{\sqrt{3}|x-x'|}{l}  \right) \exp\left(-\frac{\sqrt{3}|x-x'|}{l}   \right)
\end{equation}


which yields sample paths of $f$ that are continuously differentiable. This covariance will potentially achieve  higher likelihood values (i.e. a closer fit in terms of model-to-market errors), but at the cost of losing smoothness of the local volatility surface.


For $\kappa$, and for the remaining hyperparameters, we assume  scaled sigmoid Gaussian priors; each of $l_T,l_K,\sigma_f,m,\nsigma$ is generated independently as (with $\theta$    a generic parameter)
\begin{equation}\label{eqnSSG}
\theta =\theta_{\min} + \frac{\theta_{\max}-\theta_{\min}}{1+\exp(-\xi)},\quad \xi\sim\mathcal{N}(\mu_{\xi},\sigma_{\xi}).
\end{equation}
The motivation behind this choice of hyperprior is that it gives a convenient parametrisation. Each parameter is constrained to take values in $(\theta_{\min},\theta_{\max})$ while  $\mu_{\xi}$ and $\sigma_{\xi}$ provide flexibility over the distributional shape on this interval. This makes the specification of desirable hyperprior assumptions straightforward. A limited support also makes the model more prone to identifiability issues and therefore improves sampling efficiency, see \cite{aki}. For convenience, we henceforth use $\mu_{\xi}=0$, $\sigma_{\xi}=1$, and write $\theta = \text{ssg}(\xi)$ for the transformation in \eqref{eqnSSG}.

In the Bayesian framework, the model calibration problem effectively amounts to generating a sample of  local-volatility surfaces, 
$\{\bf^{(1)},\bf^{(2)},\bf^{(3)},\dots\}$, from the joint posterior (\ref{eqnJointPost}). We  detail an algorithm for this  in Section \ref{secNum}. From this sample, one can extract the sample counterpart of a MAP-surface (the ``point-estimate'' surface which achieves highest posterior likelihood), calculate a posterior mean-surface as well as quantifying uncertainty in the calibrated local volatility by means of  credible intervals.
Furthermore it gives a principled way for 
predicting  volatility  at unseen inputs  consistent with observed data. The posterior predictive distribution  is outlined in Section \ref{secPred}.  

\begin{remark}

To gain some insight into the inference process, we consider the likelihood written as
\begin{equation}\label{eqSSE}
\log p(\bc|\nsigma,SSE) = -\frac{1}{2\nsigma^2}SSE -\frac{n}{2}\log(2\pi\nsigma^2)
\end{equation}
with the sum of squared errors, $SSE = \sum_{i=1}^n(C(\obs{x}_i,\bf) - \obs{c}_i)^2$. The first term of \eqref{eqSSE}, the ``data-fit",  assigns probability  mass to small values of $SSE$ (the likelihood is an exponential distribution over  $SSE$ with its mode at zero) and large variances with $\nsigma$. The second term, the ``model-complexity", on the other hand prefers small $\nsigma$. As a result, if we  ignore the prior over $(\nsigma,\bf)$, the likelihood strictly works to  minimise the model-to-data error whilst balancing  with an appropriate noise variance to accommodate for variability in the data. 
Similarly, the Gaussian process prior
\begin{equation*}
\log p(\bf|\kappa,m) = -(\bf-m)^\top \bK_\kappa^{-1}(\bf-m) - \frac{1}{2}\log(|2\pi\bK_\kappa|)
\end{equation*}
 punishes complex models through the second term. The first term assigns probability to  surfaces referenced to the mean, i.e. it balances the likelihood over $\bf$, while it also tunes an appropriate covariance matrix.
In all, the inference process automatically performs a trade-off between model-fit (bias) and complexity (variance), in a principled manner. This will also be the theme of the next section.
\end{remark}

\subsection{Bayesian model selection}

We have proposed the use of different covariance functions which effectively define separate  members of a local volatility model-family. To discriminate between different such members in a principled way   we consider  model selection from a Bayesian perspective.

Consider yet another level of inference with a discrete set of possible model structures $\{\mathcal{M}_i\}$. Before seeing data, these are distributed according to a prior, $p(\mathcal{M}_i)$, which  may  be uniform in lack of  preferences. Given data, Bayes' rule yields the posterior over models
\begin{equation*}
p(\mathcal{M}_i|\bc) = \frac{p(\bc|\mathcal{M}_i)p(\mathcal{M}_i)}{p(\bc)}
\end{equation*}
where $p(\bc) = \sum_i p(\bc|\mathcal{M}_i)p(\mathcal{M}_i)$. The key factor in this expression is the model evidence, the probability of the data under each model 
\begin{equation}\label{eqME}
p(\bc|\mathcal{M}_i) = \int \underbrace{ p(\bc|\bf,\nsigma) }_\text{likelihood} \underbrace{ p(\bf|\kappa,m,\mathcal{M}_i) }_{\bf-\text{prior }\mathcal{M}_i} \underbrace{ p(\kappa,m,\nsigma) }_\text{hyperprior} d\bf\,d\kappa\,dm\,d\nsigma
\end{equation}
where we keep the hyperprior (and likelihood) invariant across models---cf. \eqref{eqnJointPost}. 

A selection process based on the model evidence   incorporates an automatic trade-off between model fit and model complexity, such that the least complex model which is able to explain the data is favoured. This is the  \textit{Occam's razor} effect (\cite{mackay2003information},  \cite{rasmussen2001occam}): A simple model can account for a smaller {range} of possible data sets. Since its evidence is a distribution over data normalising to unity, is has {large} probabilities attached to sets within its range. Conversely, a complex model may account for a wide range of data but with relatively smaller probabilities,  as it has to spread its probability mass across a larger range. The latter is the reason for the  evidence not necessarily favouring (complex) models with the best data fit---their complexity is penalised as well---but one which yields a balance between the two.

\subsection{Predicting local volatility and call prices}\label{secPred}

To make predictions  at a strike-maturity point $x^{\star}$ (at a set $\bxs$) which is not 
included in the input grid $\bx$, we have the posterior predictive distribution of $\bfs=f(\bxs)$
\begin{equation}\label{eqnPredf}
p(\bfs|\bc) 
 = \int p(\bfs|\bf,\kappa,m,\nsigma,\bc)\underbrace{ p(\bf,\kappa,m,\nsigma|\bc)}_{\text{joint posterior}} d\kappa\,dm\,d\nsigma\,d\bf.
\end{equation}
The conditional distribution $p(\bfs|\bf,\kappa,m,\nsigma,\bc)$ is usually not dependent on data. Here 
\begin{equation}\label{berra}
p(\bfs|\bf,\kappa,m,\nsigma,\bc) = \underbrace{ p(\bfs|\bf,\kappa,m) }_{\text{cond. prior}}  \underbrace{ \frac{p(\bc|\bfs,\bf,\nsigma)}{p(\bc|\bf,\nsigma)} }_{\text{likelihood ratio}}.
\end{equation}
Except for the case when predicting at maturities $T^{\star}$ greater that the latest maturity of $\obs{\bx}$,  the likelihood ratio in \eqref{berra} does not cancel out. This is because \eqref{eqnll2} is non-factorisable over the local volatility surface: a model price $C(\obs{x},\bf,\bfs)$ depends on all values in $\{\bf,\bfs\}$ taken at maturities less than $\obs{T},$ even if the dependency is strongest on function values taken locally around $\obs{x}$. 

The conditional prior in \eqref{berra} is the predictive distribution for a Gaussian process (\cite{rasmussen2006gaussian})
\begin{align}\label{eqnGPpred}
 p(\bfs|\bf,\kappa,m) & = \mathcal{N}(\bfs|\bm_{\bfs|\bf},\bK_{\bfs|\bf})
\end{align}
with
\begin{align}\label{eqnGPpred2}
\bm_{\bfs|\bf} = m + \bK_{\bfs\bf} {\bK_{\bf\bf}}^{-1}(\bf-m),  \quad
\bK_{\bfs|\bf} = \bK_{\bfs\bfs} - \bK_{\bfs\bf} {\bK_{\bf\bf}}^{-1}  \bK_{\bf\bfs} .
\end{align}

From the posterior predictive distribution we  obtain the distribution for predicting  unseen call prices $\bcs$ at corresponding strike-maturities $\bxs$
\begin{equation}\label{eqBT1}
p(\bcs|\bc) = \int \underbrace{ p(\bcs|\bf,\bfs,\nsigma) }_{\text{data distribution}}p(\bfs|\bf,\kappa,m,\nsigma,\bc)p(\bf,\kappa,m,\nsigma|\bc)d\kappa\,dm\,d\nsigma\,d\bf\,d\bfs.
\end{equation}
The data distribution is a spherical Gaussian as follows from the noise model (\ref{eqnNM})
\begin{equation}\label{bertil2}
p(\bcs|\bf,\bfs,\nsigma) = \mathcal{N}(\bcs|C(\{\bfs,\bf\},\bxs),\nsigma^2\ind)
\end{equation}
with $\ind$ denoting the identity matrix of appropriate dimension.

\begin{remark}\label{remPred}
The joint distribution that is marginalised in \eqref{eqnPredf} is equal to
\begin{equation}\label{eqNewView}
\frac{1}{p(\bc)}p(\bc|\bfs,\bf,\nsigma)p(\bfs,\bf|\kappa,m)p(\kappa,m,\nsigma)
\end{equation}
which is the  posterior over $(\bfs,\bf,\kappa,m,\nsigma)$ given $\bc$. Consider $\bxs$ to be the inputs of $\bx$ which are \textit{not} included in $\bxh$, such that $\{\bfs,f(\bxh)\}=\bf$, and replace $\bf$ with $f(\bxh)$ in \eqref{eqNewView}. We thus have that  local volatility at  \textit{unquoted} (unobserved) strike-maturities can either be inferred from the posterior \eqref{eqnJointPost} or, equivalently, by predicting these points with respect to the joint posterior over $(f(\bxh),\kappa,m,\nsigma)$ from \eqref{eqnPredf}, i.e. with respect to local volatility at market-quoted (observed) inputs only. The convenient consequence of this fact is that we can choose to either include our predictions $\bxs$ in the input set $\bx$ and generate jointly their functional values by posterior sampling, or we can first  sample from the posterior over values at (the grid which spans) the market set $\bxh$ and  then infer the predictive distribution over $\bxs$ in a second step.
\end{remark}

\subsection{Numerical method}\label{secNum}

A posterior distribution of the form \eqref{eqnJointPost} is  intractable in general, with the case of Gaussian likelihoods standing as a rare exception (under known hyperparameters). 
To this end, we will represent the posterior with samples generated by Markov chain Monte Carlo (MCMC), which is an asymptotically exact method (see e.g. \cite{robert2004monte}). Several methods also exists for deterministic approximations where a tractable distribution is chosen within a suitable family, to be close to the posterior. 
These approximations do, however, depend on factorising likelihoods to be effective, and are {known to be \textit{exclusive}  in fitting  mass to a (local) mode of the posterior, hence potentially underestimating uncertainty.} As well, when employing a calibrated model in practice, we still require posterior samples.

To sample $(\bf,\kappa,m,\nsigma)$ from the joint posterior (\ref{eqnJointPost}), we use 
 blocked Gibbs sampling (\cite{geman1984stochastic}) and alternate between updating the variable of function values  conditional on hyperparameters, and hyperparameters conditional on function values, respectively. This gives a Markov chain with the joint target as the limiting distribution (\cite{tierney1994markov}).

For the update of functional values we use elliptical slice sampling  (\cite{murray2010elliptical}). The model is first re-parametrised to have a zero-mean prior with parameter $\kappa$, while $m$ is absorbed into the likelihood
\begin{gather}
\bf|\kappa\sim p(\bf|\kappa)\equiv\mathcal{N}(0,\bK_{\kappa}), \quad \bc|\bf,m,\nsigma \sim  p(\bc|\bf,\nsigma,m)\equiv p(\bc|\bf+m,\nsigma), \nonumber \\ 
p(\bf,\kappa,m,\nsigma|\bc) = \frac{1}{p(\bc)} p(\bc|\bf,{\nsigma,m}) p(\bf|\kappa)p(\kappa)p(\nsigma)p(m). \label{eqESS}
\end{gather}
For hyperparameters fixed at some values, i.e. given $\bK_{\kappa}$ and $(\nsigma,m)$, the elliptical slice sampler  targets the  conditional\footnote{With $\propto p(\theta)$ we mean an unormalised density with normalising constant that does not depend on the target variable of interest. The normalising constant cancels out in a Metropolis-Hastings acceptance ratio for all the samplings steps we consider here.} $\propto p(\bc|\bf,\nsigma,m) p(\bf|\kappa)$ with a proposed transition $\bf\rightarrow\bf'$ defined by an ellipse
\begin{gather*}
\bf'(\theta) = \bf \cos(\theta) + \bar{\bf} \sin(\theta),\quad \bar{\bf}\sim\mathcal{N}(0,\bK_{\kappa}), \quad \theta\sim\text{Uniform}[0,2\pi].
\end{gather*}
The proceeding angle $\theta$ is drawn uniformly over a bracket 
$[\theta_{\min},\theta_{\max}] = [\theta-2\pi,\theta]$, 
which is shrunk (``sliced'' from \cite{neal2003slice}) towards $\theta=0$ until the proposal is accepted---see \cite{murray2010elliptical} for further details.

Next we consider updating the covariance hyperparameter $\kappa$. 
Its conditional given functional values is $\propto p(\bf|\kappa)p(\kappa)$.  Targeting this density will, however, lead to a slow exploration of the support for $\kappa$.  This since the functional prior is very  informative of its hyperparameters in the sense that the two are  strongly correlated. Further, there is no direct guidance from  data through the likelihood in this update.  
To improve mixing, we therefore decouple the prior dependency between $\bf$ and $\kappa$ by the parametrisation
\begin{equation}\label{eqnWhite}
\bf(\kappa,\bnu) = \bL_{\kappa}\bnu,\quad \bL_{\kappa}{\bL_{\kappa}}^{\top} = \bK_{\kappa},\quad \bnu\sim\mathcal{N}(0,\ind),
\end{equation}
with a  Cholesky decomposition for the  covariance matrix square-root. This results in an equivalent  posterior
\begin{equation}\label{eqnJointPostII}
p(\bnu,\kappa,m,\nsigma|\bc) = \frac{1}{p(\bc)} \underbrace{ p(\bc|\bf(\bnu,\kappa),m,\nsigma) }_\text{likelihood} \underbrace{ p(\bnu) }_{\bnu-\text{prior}} { p(\kappa)p(m)p(\nsigma) }
\end{equation}
where we stress that the likelihood now depends on {covariance} hyperparameters through the deterministic transformation $\bf(\kappa,\bnu) = \bL_{\kappa}\bnu$, and that the prior for $\bnu$ is independent of the hyperprior. For a current state $(\bf,\kappa)$, the corresponding $\bnu$ is implied by $\bnu = \bL_{\kappa}^{-1}\bf$. Given $(\bnu,m,\nsigma)$, a proposed transition $\kappa\rightarrow\kappa'$ then targets the (unnormalised) conditional 
\begin{equation}\label{eqnWhite2}
p(\bc|\bf(\bnu,\kappa),m,\nsigma)p(\kappa)
\end{equation}
where functional values are indirectly updated through $\bf\rightarrow\bf'=\bL_{\kappa'}\bnu$. Further, due to the hyperprior assumption \eqref{eqnSSG} with $\kappa = \kappa(\bz) \equiv \text{ssg}(\bz)$ for a Gaussian $\bz\sim\mathcal{N}(0,\ind)$, we  employ the elliptical slice sampler on $p(\bz|\bnu,m,\nsigma,\bc)\propto p(\bc|\bf(\bnu,\kappa(\bz)),m,\nsigma)p(\bz)$.

Updating covariance hyperparameters based on \eqref{eqnWhite} works best when the likelihood dependency on $\bf$ is relatively weak, such that $\bf$ is  distributed almost according to its prior. This is discussed by \cite{murray2010slice}, who propose surrogate data slice sampling (SDSS) for models with strong likelihoods. In this case, proposals targeting \eqref{eqnWhite2} are often rejected such that mixing of the Markov chain for $\kappa$ is poor. As a remit, they augment the model with a noise variable centred around  function values, $\textbf{g}|\bf \sim\mathcal{N}(\bf,\textbf{S})$, where $\textbf{S}$ is a user specified diagonal covariance. They then make hyperparameter proposals  that updates $\bf$ conditional on its noisy version $\textbf{g}$, through the parametrisation
\begin{equation}\label{eqnWhiteNoise}
\bf(\kappa,\boldsymbol{\eta},\textbf{g}) = \bL_{\textbf{R}_{\kappa}} \boldsymbol{\eta} + \textbf{m}_{\kappa,\textbf{g}},\quad \bL_{\textbf{R}_{\kappa}}{ \bL_{\textbf{R}_{\kappa}} }^{\top} = \textbf{R}_{\kappa},\quad \boldsymbol{\eta}\sim\mathcal{N}(0,\ind),
\end{equation}
where $\textbf{R}_{\kappa}$ and $\textbf{m}_{\kappa,\textbf{g}}$ from the conditional $\bf|\textbf{g},\kappa \sim\mathcal{N}(\textbf{m}_{\kappa,\textbf{g}},\textbf{R}_{\kappa})$ are given by
\begin{gather*}
\textbf{R}_{\kappa} = \textbf{S} - \textbf{S}(\textbf{S}+\bK_{\kappa})^{-1}\textbf{S}, \quad 
\textbf{m}_{\kappa,\textbf{g}} = \textbf{R}_{\kappa} \textbf{S}^{-1}\textbf{g}.
\end{gather*}
In practice, a current state $(\bf,\kappa,m,\nsigma)$ is first augmented by drawing $\textbf{g}$ from $\mathcal{N}(\bf,\textbf{S})$. The implied variable is calculated, $\boldsymbol{\eta}=\bL_{\textbf{R}_{\kappa}}^{-1}(\bf - \textbf{m}_{\kappa,\textbf{g}} )$, whereby a proposal $\kappa'\rightarrow\kappa$ targets its conditional posterior given $(\boldsymbol{\eta},\textbf{g},m,\nsigma)$, 
\begin{equation*}
p(\kappa|\boldsymbol{\eta},\textbf{g},m,\nsigma,\bc) = \frac{1}{p(\bc)}p(\bc|\bf(\kappa,\boldsymbol{\eta},\textbf{g}),m,\nsigma)p(\textbf{g}|\kappa,\textbf{S})p(\boldsymbol{\eta})p(\kappa)p(m)p(\nsigma),
\end{equation*}
where the relevant term is $ p(\bc|\bf(\kappa,\boldsymbol{\eta},\textbf{g}),m,\nsigma)p(\textbf{g}|\kappa,\textbf{S})p(\kappa)$---see \cite{murray2010slice} for details. Again, due to our hyperparameter assumption, we perform this update in $\bz$-space with elliptical slice sampling.

The final step in our sampling scheme updates the hyperparameters $(m,\nsigma)$  of the likelihood. We target its  conditional $\propto p(\bc|\bf,\nsigma(\bz),m(\bz))p(\bz)$ with elliptical slice sampling on the Gaussian $\bz$. As the mean-level $m$ and $\nsigma$ are  one-dimensional parameters to which the likelihood is quite sensitive, this sampling step is relatively effective; it is also cheap since the target conditional does not involve the functional prior.

The three updates---functional values, covariance and likelihood parameters---summarises each sequential step 
of the Markov chain. We combine them for a hybrid strategy where we 
randomise between 
sampling based on \eqref{eqnWhite} and \eqref{eqnWhiteNoise} when updating $\kappa$ (see \cite{robert2010introducing}). Similarly, we follow the strategy of \cite{murray2010slice} and apply three computationally ``cheap" updates of $\bf$  between every expensive update of the covariance. 

\begin{figure}[!t]
\makebox[\textwidth][c]{  
\includegraphics[scale=0.44,trim=0 0 0 0,clip]{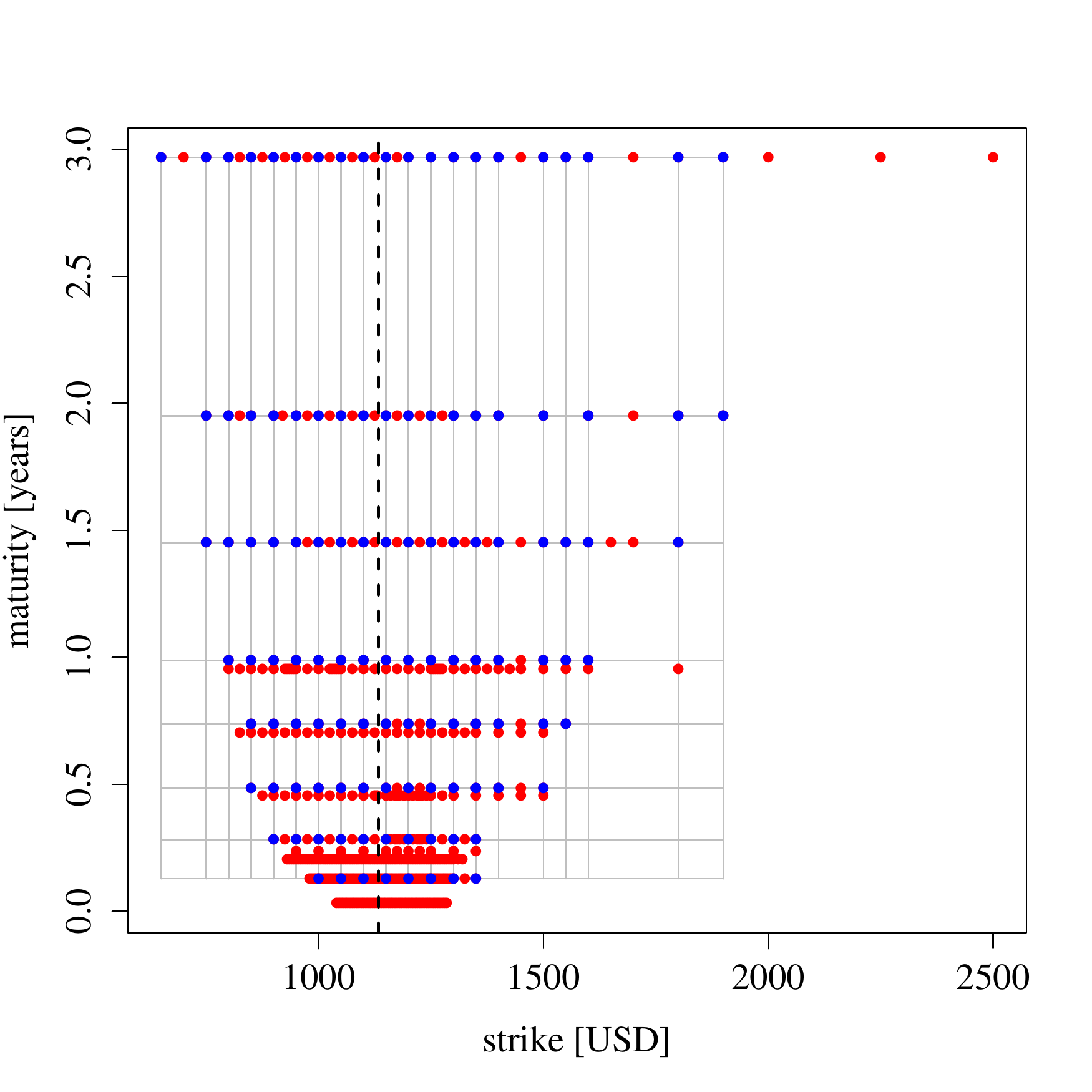}
\includegraphics[scale=0.55,trim=50 50 50 50,clip]{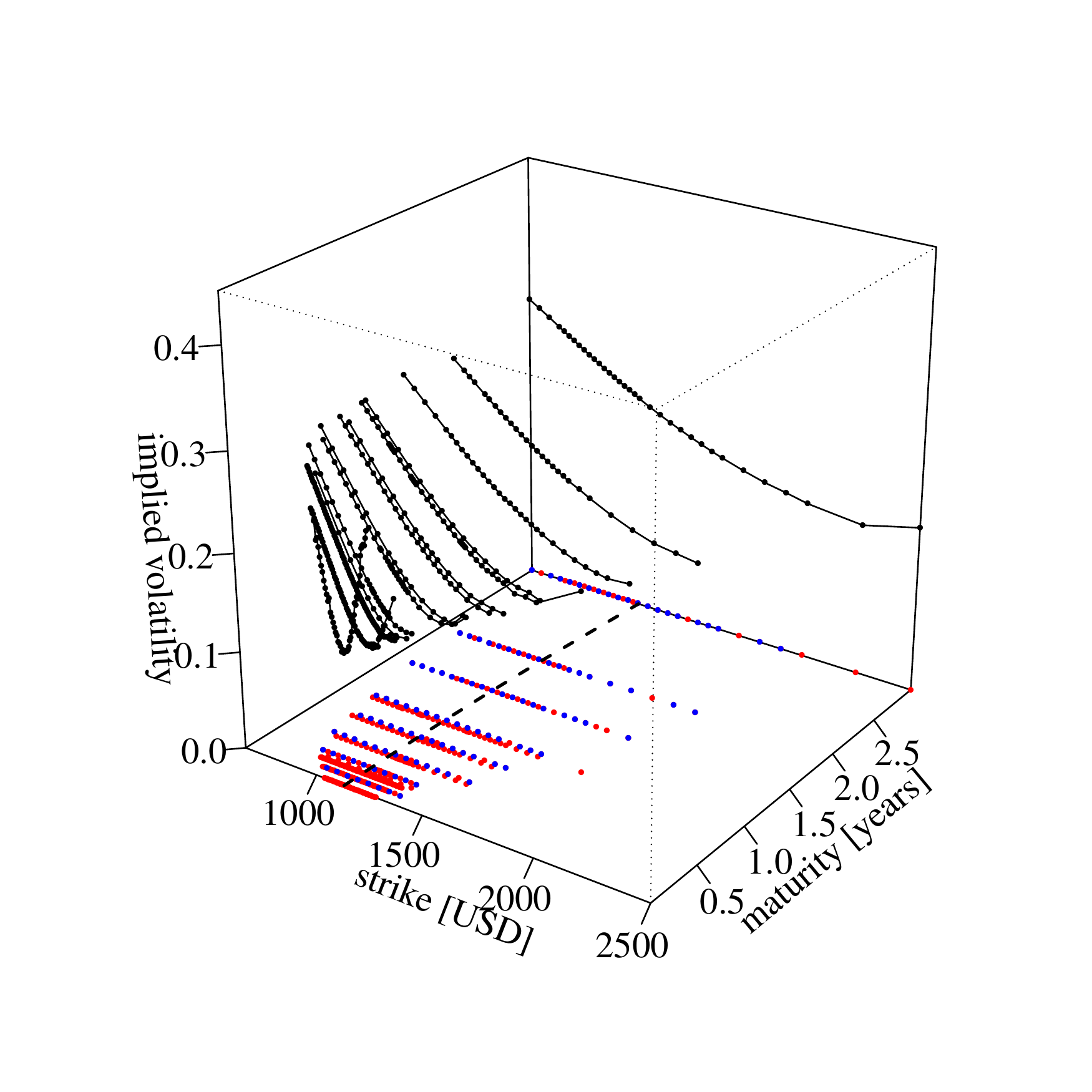}
}
\caption{\textbf{Left:} Strikes and maturities of observed S\&P 500 call options  as of 4 January 2010. The red and blue dots show all 473 market observations; blue dots show the subset $\obs{\bx}$ of  175  points used for the experiments (the market set). The strikes and maturities of this set span the grid shown in grey, from which all nodes are collected as the input set $\bx$ (i.e. the Cartesian product of all unique strikes and maturities available in $\obs{\bx}$). \textbf{Right:} Black-Scholes implied volatilities by mid-market prices (corresponding bid-ask prices shown in Figure \ref{fig9}). 
}
\label{fig1}
\end{figure}

To evaluate the likelihood, we need to calculate the mapping (\ref{eqnFDmap}) of a local volatility surface to a call price surface. To this end, we use a Crank-Nicolson finite difference scheme for a  numerical solution of Dupire's equation (\ref{eqnDupire}), see \cite{hirsa2012computational}. 
For this purpose, we construct the input set $\bx$ by taking the Cartesian product (\ref{eqnCP}) of all  observed maturities and strikes in the market set $\obs{\bx}$. This results in a grid with non-equidistant spacing in both dimensions as illustrated for market data by the  grey mesh in Figure \ref{fig1}. For boundary conditions across maturity, we use that the call price approaches $S-Ke^{-rT}$ for small $K$ and zero for large $K$. To improve the accuracy of these approximations, we extend the grid in the strike dimension with a flat extrapolation of the input local volatility surface  with moneyness $S/K$ down to 0.1 and up to 4, respectively.

In terms of computational complexity, the second update of covariance parameters is the most expensive with a cost $\mathcal{O}(N^3)$ for (standard) inversion of the covariance matrix each time a new $\kappa$ is proposed. Proposing a new $\bf$ also costs $\mathcal{O}(N^3)$ for the Cholesky decomposition of $\bK_{\kappa}$, although the same decomposition can be recycled for consecutive proposals. The third update is the cheapest since it only requires computing the likelihood. One such evaluation involves solving $I$ systems of $J\times J$ tridiagonal matrices for the Crank-Nicolson scheme of the call price, where $I$ ($J$) is the number of maturities (strikes) of the input set $\bx$,  i.e. a cost  $\mathcal{O}(IJ)=\mathcal{O}(N)$. In fact, we can readily exploit that $\bx$ is a Cartesian product and use Kronecker methods  when computing matrix decompositions and inversions, see \cite{saatcci2012scalable}. This reduces the cost to $\mathcal{O}(N^{3/2})$ for both the second (for each $\kappa$ proposal) and the first update (recycled throughout a $\bf$-update with elliptical slice sampling).

Finally, we consider sampling from the predictive distribution \eqref{eqnPredf}. For a prediction input $x^{\star}$ with  maturity $T^{\star}$ {smaller} than the latest maturity of $\bxh$ we 
approximate the likelihood ratio in \eqref{berra} to be constant (note that this is exact if $T^{\star}$ is \textit{larger} than the latest maturity of $\bxh$). 
We then generate predictions  by direct simulation ({ancestral pass}), i.e. from an approximating  Gaussian mixture 
\begin{equation}\label{eqID}
p(\bfs|\bc) \approx \sum_{i=1}^M \frac{1}{M}  p(\bfs|\bf^{(i)},\kappa^{(i)},m^{(i)})
\end{equation}
where $\{\bf^{(i)},\kappa^{(i)},m^{(i)}\}_{i=1}^M$ is a sample from the posterior. This approach is used in   \cite{wilson2010copula} and \cite{osbornebayesian}, and yields sampling from the exact predictive distribution (under constant likelihood ratio) in the limit $M\rightarrow\infty$. 

To sample from the complete  distribution \eqref{eqnPredf} we may add an importance step to the above procedure. First we generate a sample $\{{\bfs}^{(i)}\}$ from the Gaussian mixture \eqref{eqID}, i.e. we employ
\begin{equation}\label{eqIMP}
p(\bfs|\bf,\kappa,m)p(\bf,\kappa,m,\nsigma|\bc)
\end{equation}
as importance distribution. The unnormalised weight function 
\begin{equation}\label{eqISw}
w(\bfs,\bf,\nsigma) = \frac{p(\bc|\bfs,\bf,\nsigma)}{p(\bc|\bf,\nsigma)}
\end{equation}
evaluated at each sample point $({\bfs}^{(i)},\bf^{(i)},\nsigma^{(i)})$ is then used for resampling over $\{{\bfs}^{(i)}\}$. Importance sampling generally works well  when the variance of the importance weights is low, in our case when the likelihood $p(\bc|\bfs,\bf,\nsigma)$ is weak with respect to $\bfs$. Again this depends on how many (and how strongly) model prices corresponding to $\bc$ vary with $\bfs$, which is due to the locations of the prediction inputs $\bxs$. In the case of a sensitive likelihood, one may improve the resampling procedure by using annealed importance sampling 
(see \cite{neal2001annealed} for details).

With a  sample of local volatility predictions in hand, we  approach the predictive distribution over call prices \eqref{bertil2}. To this end, we first map $\{ {\bfs}^{(i)},{\bf}^{(i)} \}_{i=1}^M$ through $C(\cdot,\bxs)$ to obtain a sample $\{ {\textbf{C}^{\star}}^{(i)}\}_{i=1}^M$ from the predictive distribution over fair call prices. {We   use this in turn for representing the distribution over predicted data  \eqref{eqBT1} as a Gaussian centred around the mean over predicted fair prices and covariance matrix from the law of total variance, i.e. $\textbf{mean}( \{ {\nsigma^2}^{(i)}\ind\}_i)+\textbf{cov}(\{ {\textbf{C}^{\star}}^{(i)} \}_i)$.}

\section{Experiments}\label{secEX}

In the following sections, we put our approach into practice with an example based on market data from the S\&P 500 index. We draw posterior inference about  local volatility, and demonstrate the benefit of our probabilistic formulation with focus on {parameter uncertainty}, i.e. to what extent we can be confident about the local volatility estimate we learn from data. In our view, the Gaussian process framework provides a major benefit as the inference process is straightforward. Further, we demonstrate posterior prediction of both local volatility and call prices. We focus on uncertainty in the predictions, while still the setup  allows for straightforward and robust inference without additional extrapolation assumptions.

\subsection{Data and model set-up}\label{secBerra2}

Figure \ref{fig1} shows  quoted strikes and maturities, and implied volatilities from S\&P 500 call options  as observed on 4 January 2010. The data is pre-filtered   
following  \cite{constantinides2013puzzle}:  quotes are removed having any of (i) zero bid-prices, 
(ii) less than 7 days to maturity, or (iii) implied volatility below 0.05 or above 1. 
Corresponding bid and ask prices  are shown in  Figure \ref{fig9} (left).

As seen in  Figure \ref{fig1}, 
the quotation density is higher for options with short maturities (less than three months) and strikes close to the current spot price \$1133 shown with a vertical dashed line---options being close to  ``{at-the-money}". Across maturity, quotes are sparser at later maturities and the same holds for options away from the at-the-money strike level:  for calls deep ``{out-of-money}'' (strikes $>\$1750$) there is only a handful of quotes while no options are quoted deep ``{in-the-money}'' (strikes $<\$750$).

To reduce the computational cost, we  select a subsample from the full (pre-filtered) market data as the set  $\{\bc,\bxh \}$ we use for inference. This is illustrated in Figure \ref{fig1} where $\bxh$ is plotted with blue dots while red dots show excluded  data. The selection is based on a simple algorithm that sequentially  includes the strike level with the highest number of available quotes across maturity, until a fixed number of strikes are included in $\bxh$. For each selection, only strikes lying outside the  range of strikes available in the current $\bxh$ are considered as candidates. The set is then pruned, by sequentially excluding the maturity with the smallest number of quotes, until a fixed number of maturities remain in the set. The purpose of this procedure is to end up with a grid $\bx=\hat{\textbf{T}}\times\hat{\textbf{K}}$ (where $\hat{\textbf{T}}$ and $\hat{\textbf{K}}$ are strikes and maturities included in $\bxh$) of a reduced size = \#strikes $\times$ \#maturities, while the size of $\bxh$ (\#observations) is as large as possible. For our example, the original data has a grid size $103\times14=1442$ and 473 observations (33\%) while the selected subset $\bx$ is  of size $20\times8=160$ and $\bxh$ has 117 observations (73\%). The computational saving is in having to handle the covariance matrix associated with $\bf = f(\bx)$ of $N=160$  values while the full data would require  $N=1442$.

To set-up an operational Gaussian process model, we scale $\bx$ to lie in the unit square and set $\kappa_{i,\max}=1$, $\forall i$, for the covariance hyperprior, and $\sigma_{\epsilon,\max}=0.75$ and  $ m_{\max} = \log 0.5$ for the likelihood. These values stem from  heuristics and some initial runs of the MCMC algorithm. Length-scales around one (maximum) give  slowly varying volatility surfaces much smoother than the implied volatility of Figure \ref{fig1}; hence, prior $\kappa$-ranges of 0--1 should provide more than enough expressiveness. For the likelihood, the maximum noise standard-deviation  0.75 is of the same magnitude as the average bid-ask spread of 3.6. Note, however, that the noise models the deviance of the mid-market price from the unobserved \textit{fair} price, not the (half) bid-ask spread.

\subsection{Results}\label{seqBerra}

\begin{figure}[!t]
\makebox[\textwidth][c]{  
\includegraphics[scale=0.55,trim=50 50 50 50,clip]{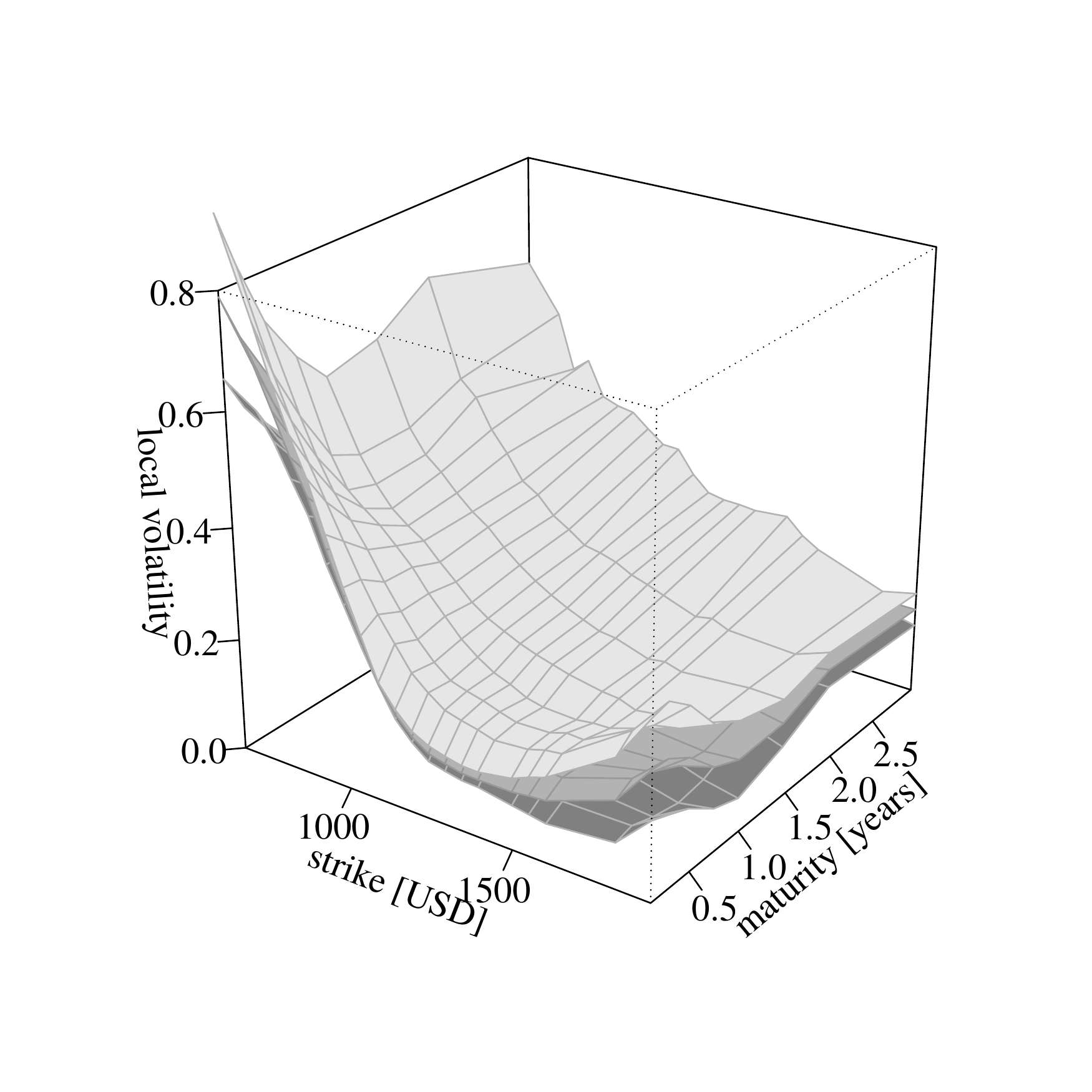}
\includegraphics[scale=0.55,trim=50 50 50 50,clip]{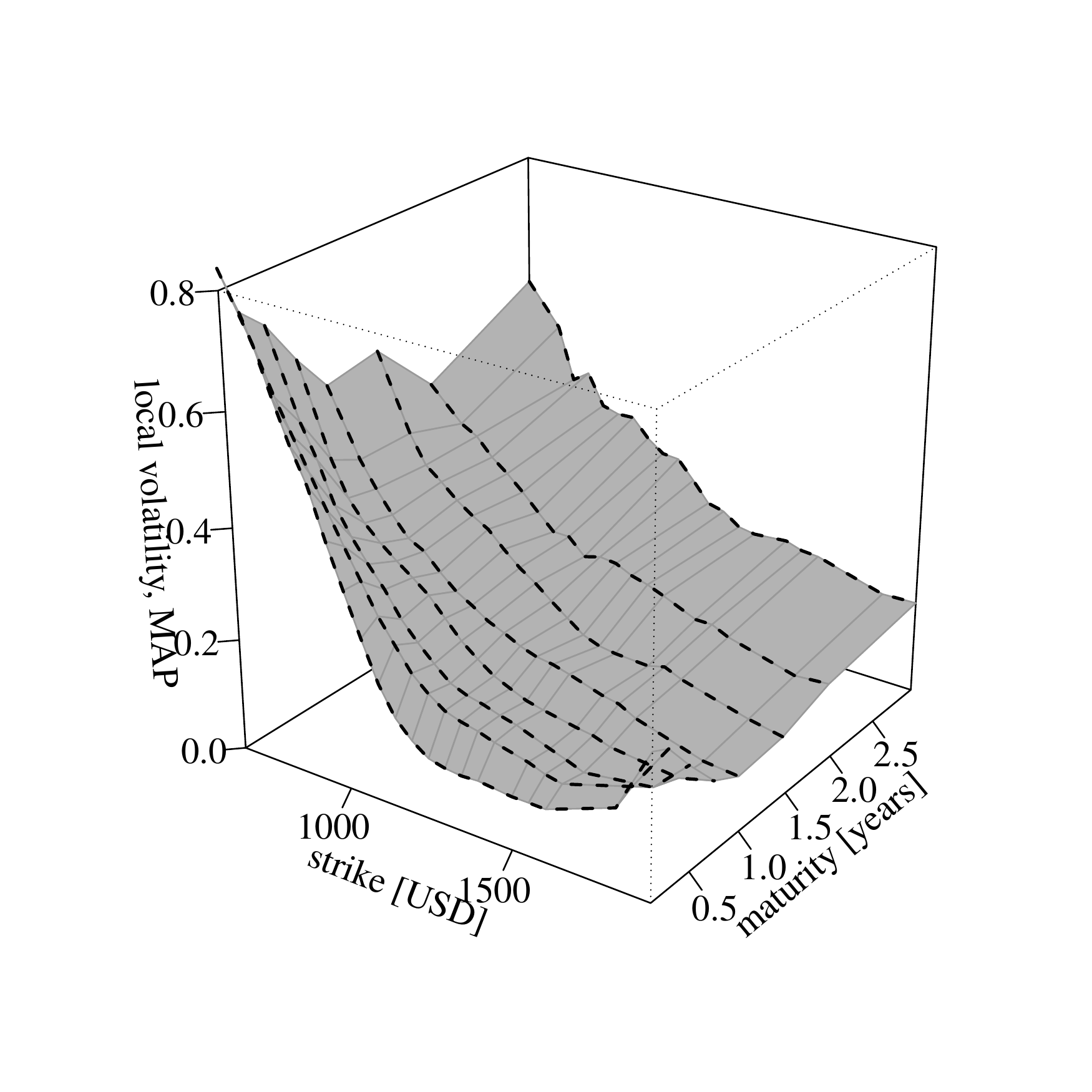}
}
\caption{Local volatility calibrated to S\&P 500 call prices. \textbf{Left:} credible envelope of $\pm2$ standard deviations around the sample mean. 
\textbf{Right:} maximum a posterior surface. Dashed black lines indicate cross sections taken at six maturities and shown in Figure \ref{fig2b}.}
\label{fig2}
\end{figure}

\paragraph{Calibration}
 In the first part of our experiment, we {calibrate} the local volatility model  to the  S\&P 500 mid-market call prices as described in Section \ref{secNum} 	by  sampling from the joint posterior (\ref{eqnJointPost}) under a squared exponential covariance. We generate 50,000 states from the Markov chain, 
discard the first 10,000 as burn-in and subset the reminder to a thinned sample of size $1000$.  Figure \ref{fig2} (left) represents the posterior with a credible region of $\pm$2 standard deviations around the mean as estimated pointwise from the sample.
 Figure \ref{fig2} (right) shows the empirical maximum a priori estimate (i.e. the local volatility surface that achieves the highest posterior likelihood). The {smoothness} of the calibrated local volatility surface(s) is apparent, as encoded by the prior squared exponential covariance function. For a vivid comparison,  Figure \ref{fig9} shows a local volatility surface from pointwise minimisation of the least squares objective  
which achieves the same error as the MAP estimate in Figure \ref{fig2}. 

\begin{figure}
\centering
\includegraphics[scale=0.5,trim=  0 65 25 50,clip]{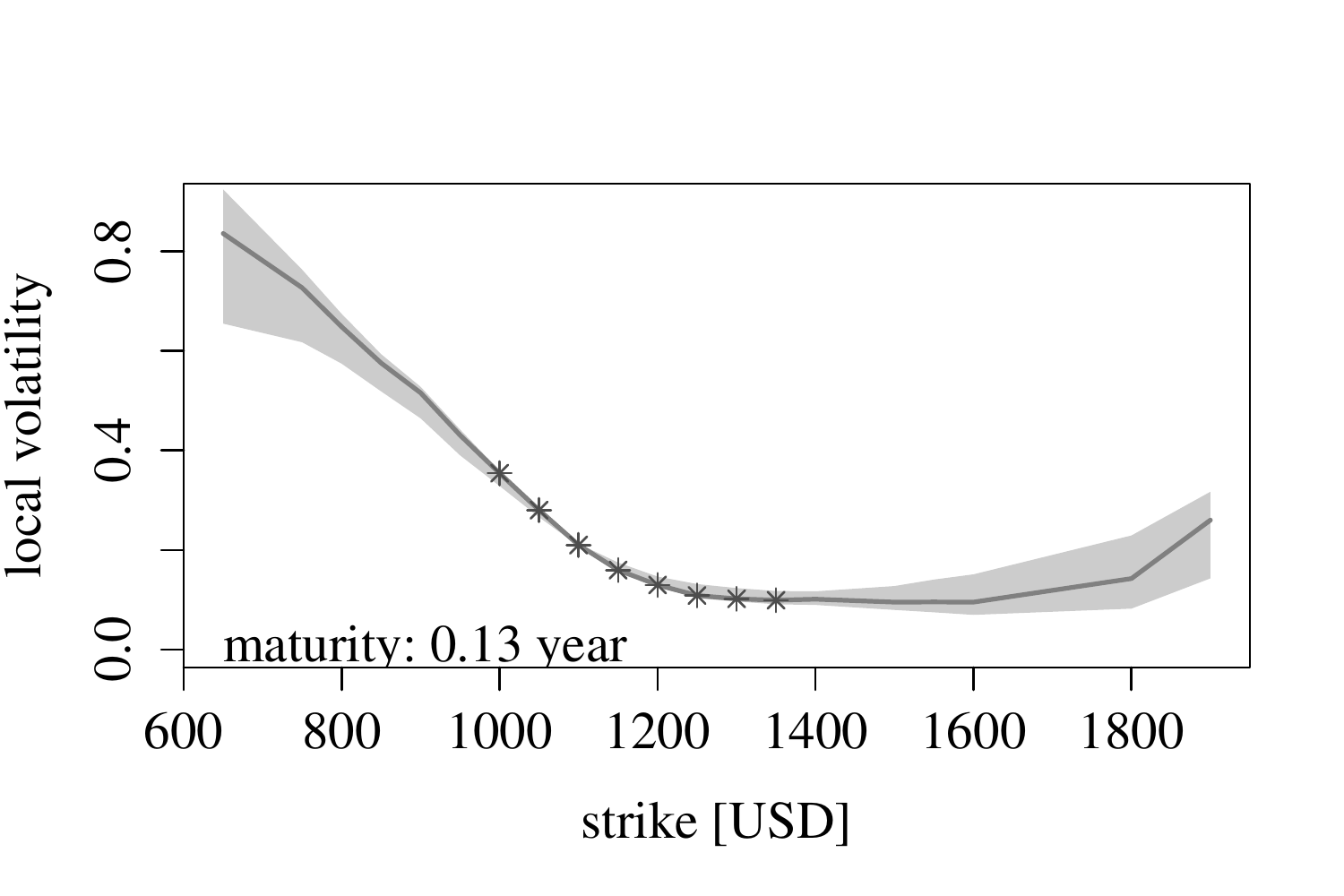} 
\includegraphics[scale=0.5,trim=46 65 25 50,clip]{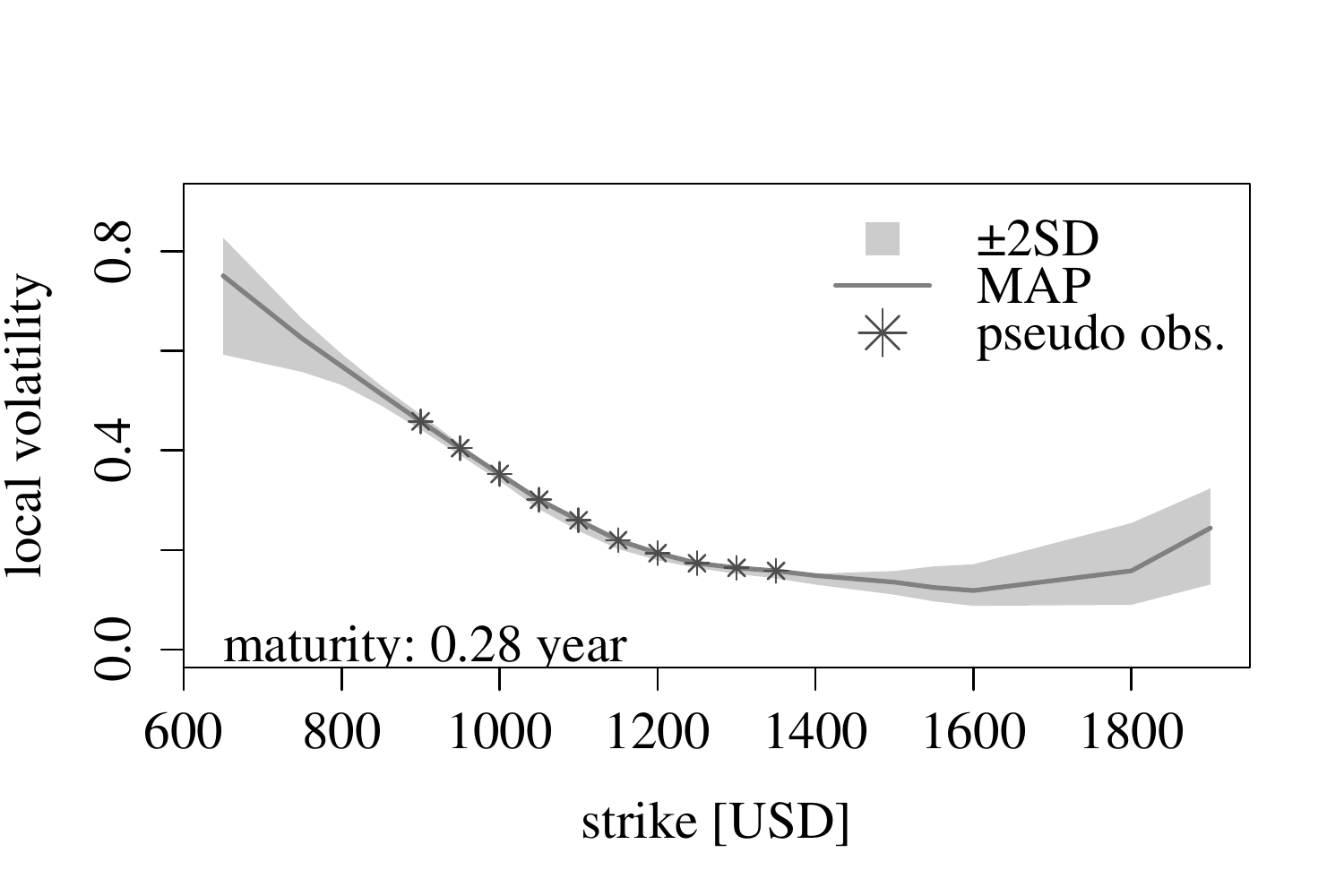} \\
\includegraphics[scale=0.5,trim=  0 65 25 50,clip]{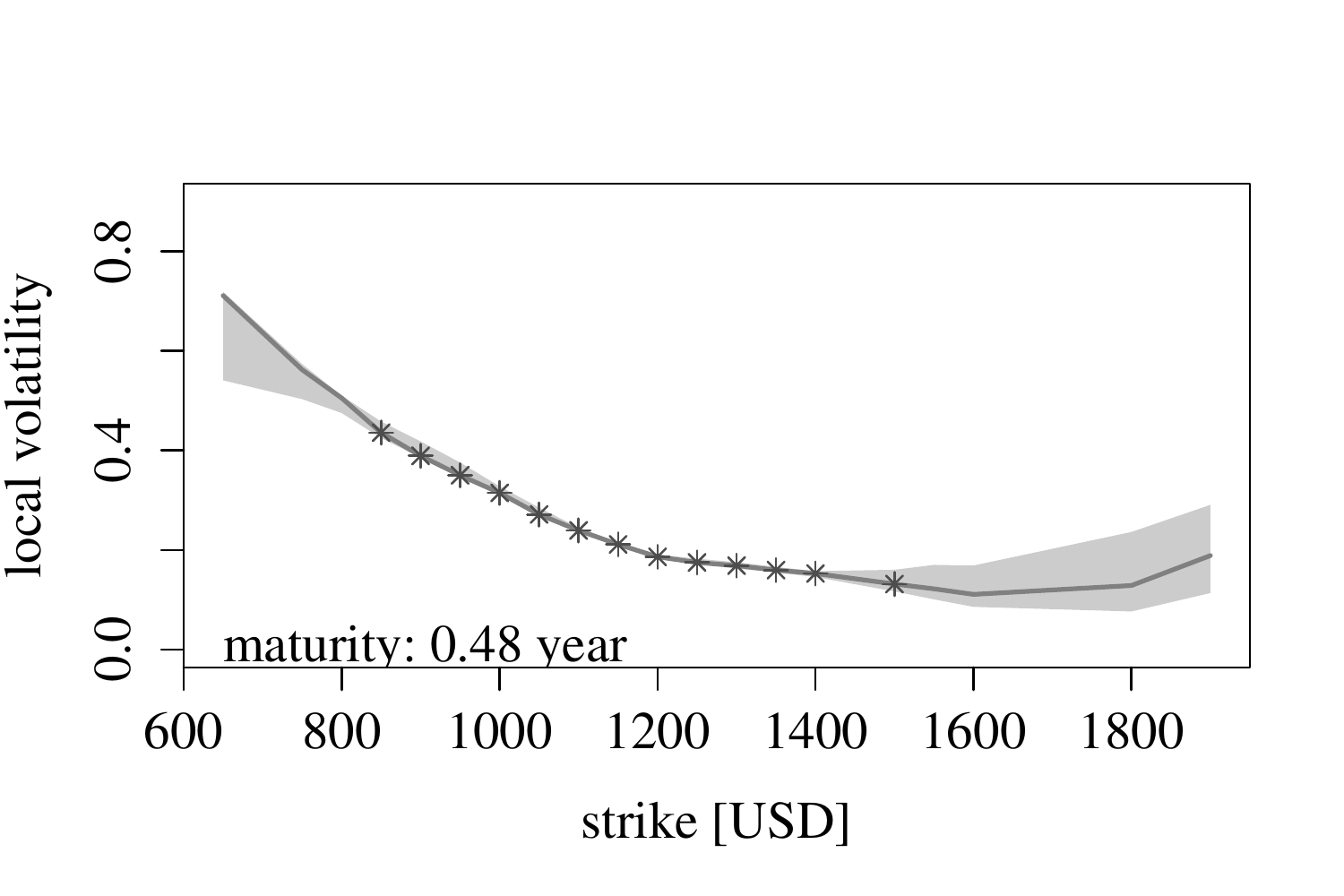} 
\includegraphics[scale=0.5,trim=46 65 25 50,clip]{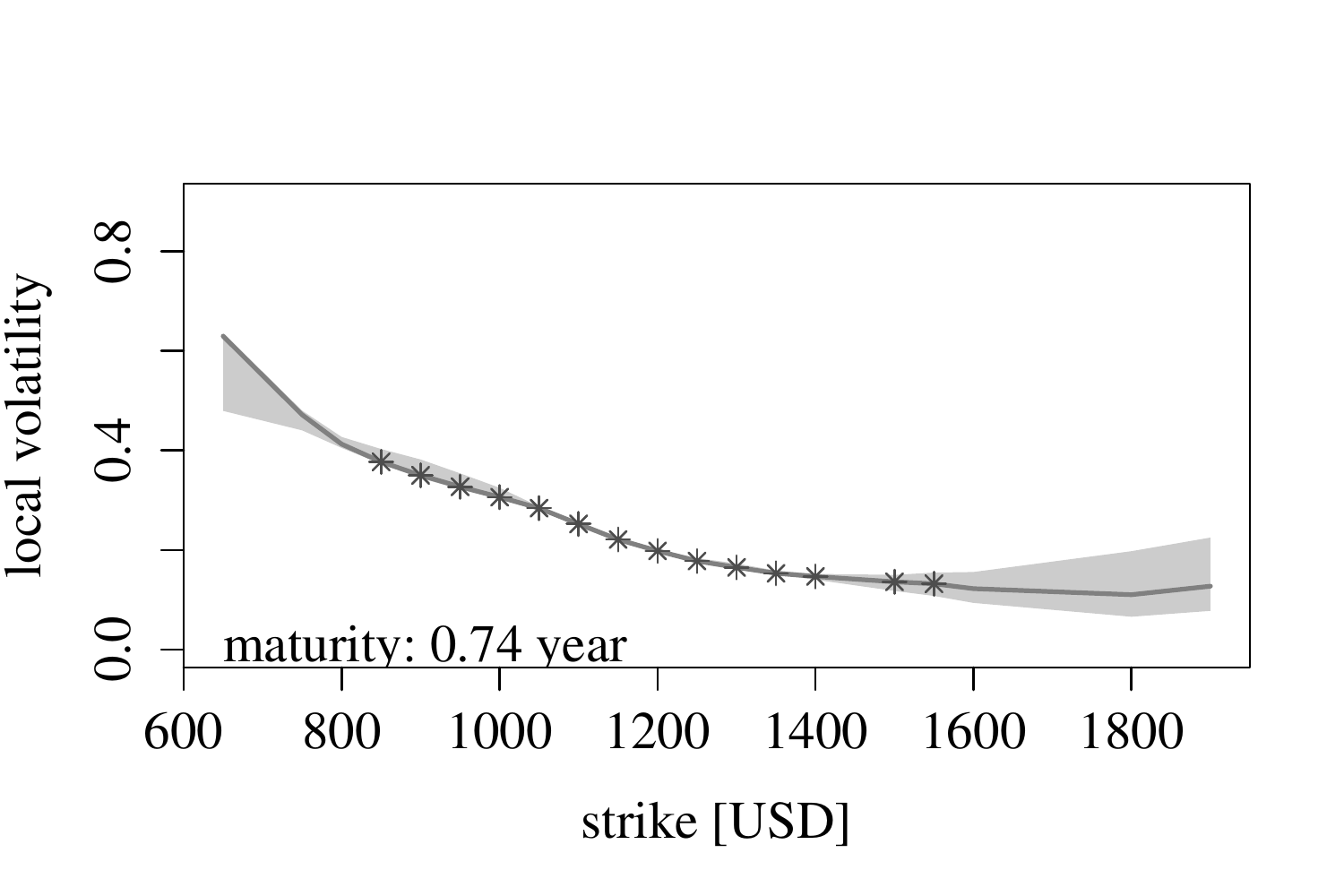} \\
\includegraphics[scale=0.5,trim=  0 65 25 50,clip]{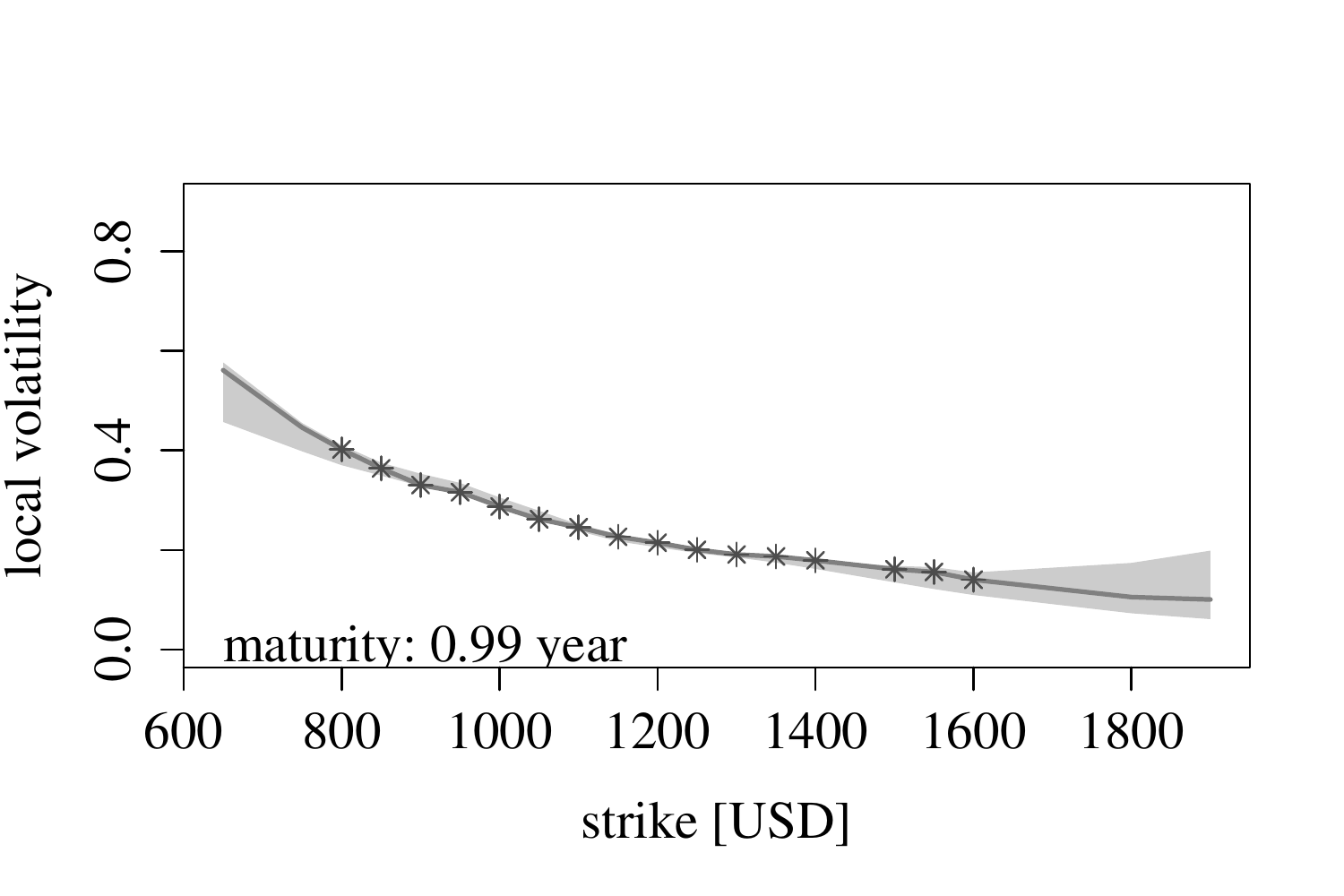} 
\includegraphics[scale=0.5,trim=46 65 25 50,clip]{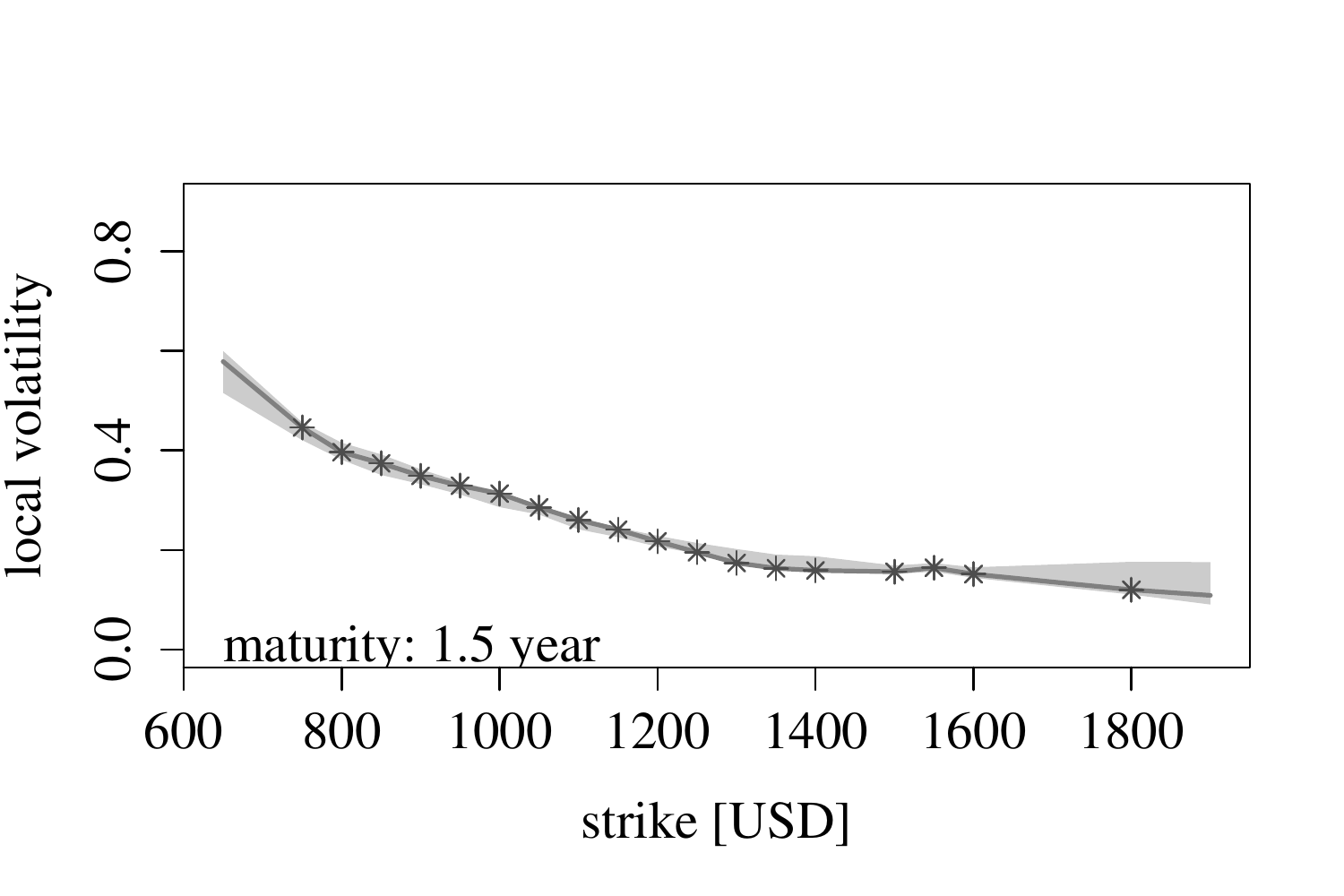} \\
\includegraphics[scale=0.5,trim=  0 10 25 50,clip]{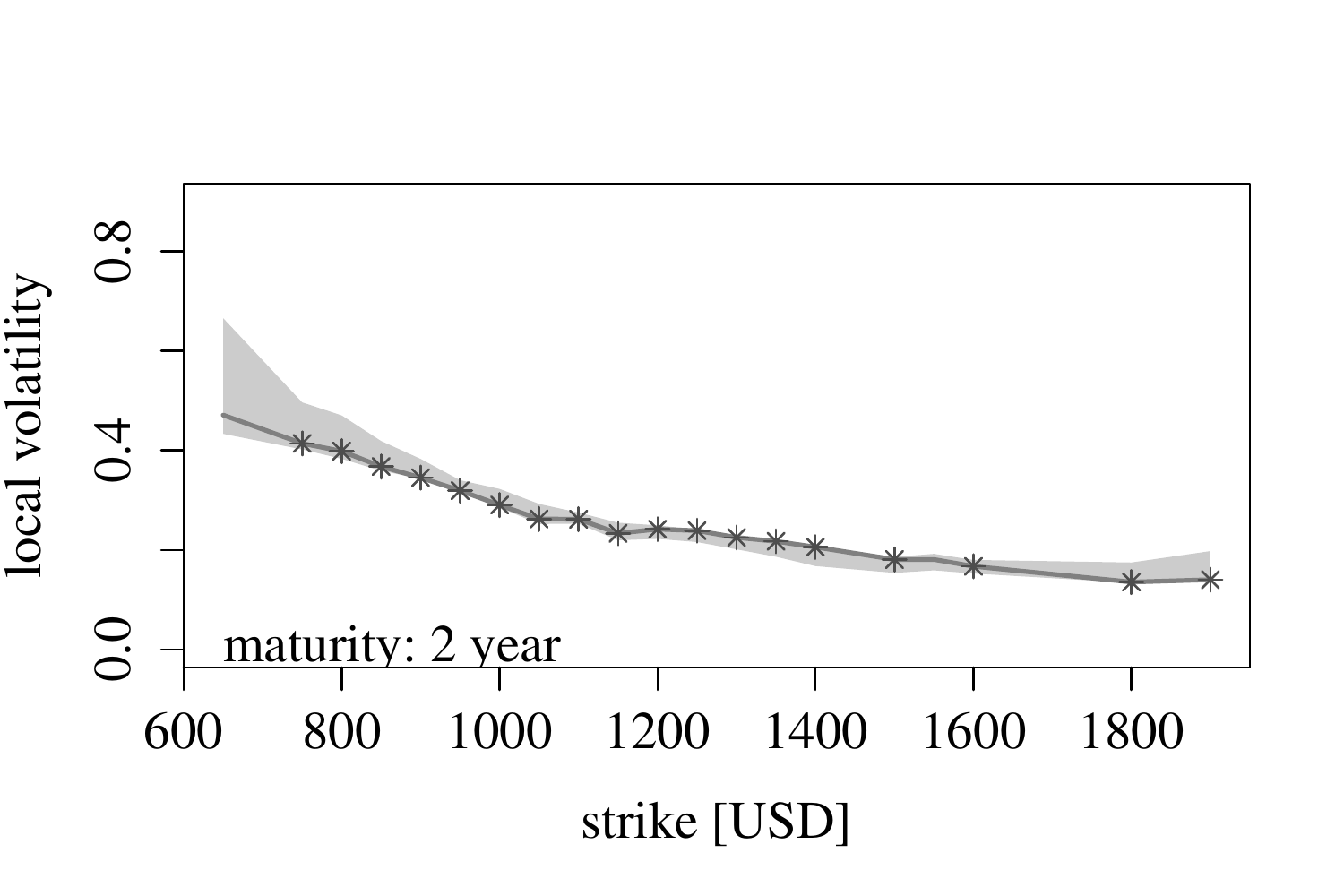} 
\includegraphics[scale=0.5,trim=46 10 25 50,clip]{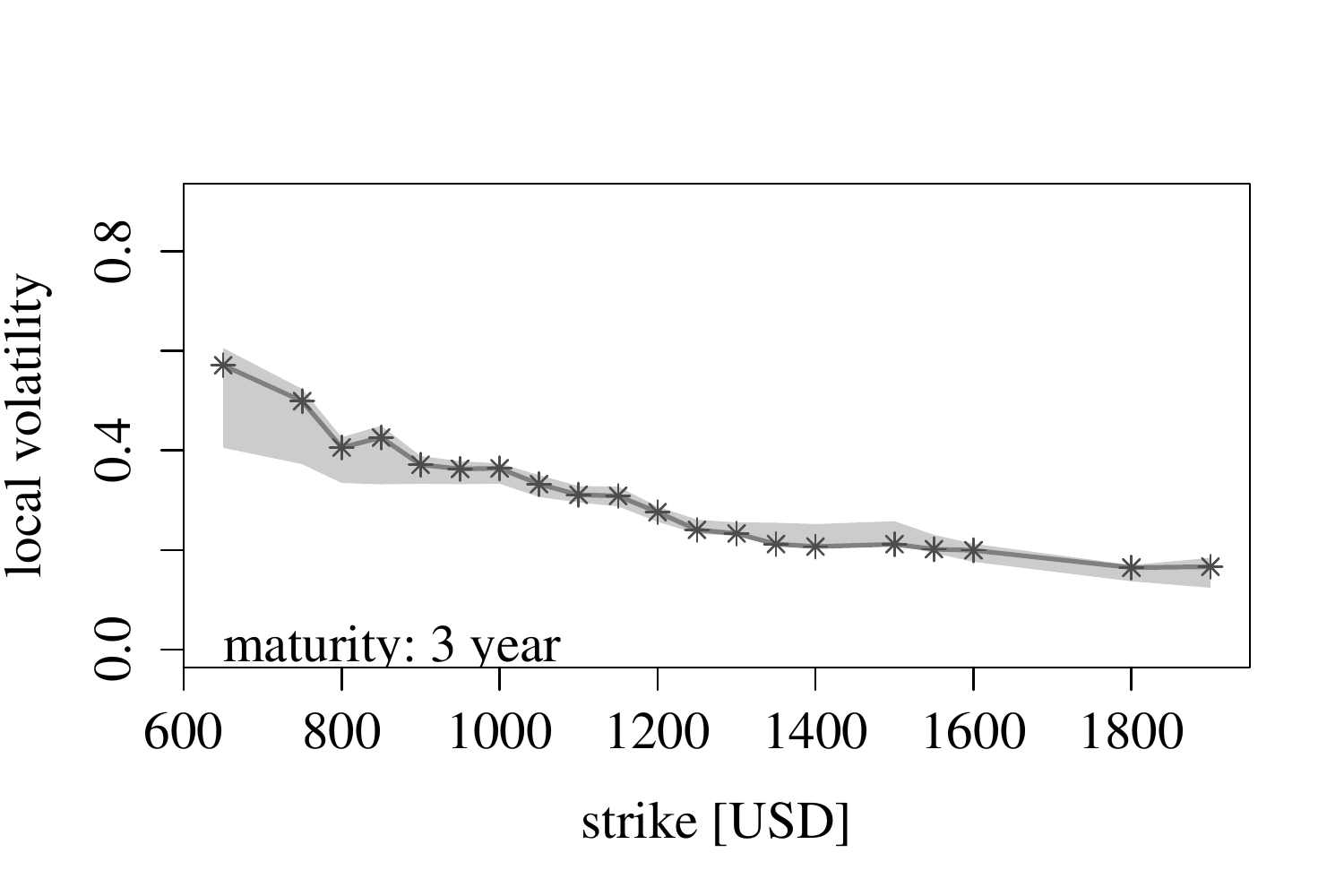} 
\caption{Posterior distribution over  local volatility as represented by the  maximum a posteriori surface (solid grey line) and a  credible interval of $\pm2$ standard deviations around the sample mean. 
 Asterisks show ``pseudo-volatilities''  to indicate where available values would have been located if they were observable.   
 }
\label{fig2b}
\end{figure}

The dashed lines in Figure \ref{fig2} (right) indicate  cross-sections of the local volatility surfaces  at 
maturities of 47 days (earliest maturity of the set), up to 3 years (latest maturity). The posterior  at these maturities is represented in Figure \ref{fig2b}. 

The shapes of the credible intervals in Figure \ref{fig2b} are symptomatic of two related features: non-linearity of the pricing/likelihood function and data availability across strikes and maturities. For short maturities ($T<6$ months) the option price is relatively insensitive to model parameters for strikes away from at-the-money. For $K \gg S_0$ there is not much chance of the option reaching in-the-money before its (short) time to maturity and hence its price is close to zero; the option is likely to expire worthless. Likewise, for  $K\ll S_0$ there is little chance of going out-of-money and the option's price is close to its intrinsic value $S_0-K$ (the reminding time-value is small, in both case). In effect, the price insensitivity makes the likelihood  uninformative about local volatility over these strike-maturity points, i.e. the probabilistic inference process attributes more uncertainty 
to the local volatility values. For the same reasons there is little trading in these options, 
which also contribute to an increased posterior uncertainty 
as there is lack of data. Market quotes are concentrated around at-the-money, $K=S_0$, and these options are also the most sensitive to changes in local volatility, since the volatility affects the prospects of moving out/in the money. The overall effects can be seen in the top panes of Figure \ref{fig2b}:  a high degree of estimation uncertainty is attached to the local volatility at large and small strikes, while close to at-the-money 
there is almost no posterior uncertainty in the estimate.

For later maturities the price sensitivity increases, also for strikes away from at-the-money. With time, the volatility has an increased effect on the probability that the option leaves its in- or out-of-money region---in view of \eqref{eqnSDE} it is a parameter of diffuseness of the underlying. For long maturities, local volatility therefore has higher impact on option prices across all strikes. In effect, credible intervals of  the bottom panes in Figure \ref{fig2b} show rather constant uncertainty across strikes. Compared to earlier maturities, the band is wider around at-the-money. Data availability is also a reason since there is a large gap in  market quotes between maturities of 2 and 3 years---see Figure \ref{fig1}.

The posterior sample of local volatility surfaces gives a probabilistic representation of the local volatility function. This is interesting in its own right as it provides information on what can be learned (and not learned) from call price data about the latent function, i.e. to what degree we may pin down $\sigma$-values over different strike-maturity points. The result---point-estimate(s) and associated uncertainty---can then be used when employing the model for secondary purposes. For example, if pricing a derivative which is sensitive to volatility over points with high calibration uncertainty, one should take this into account for a robust price-estimate. Next we use the posterior sample to re-price the  calibration instruments. 
 \begin{figure}
\centering
\includegraphics[scale=0.5,trim=  0 65 25 50,clip]{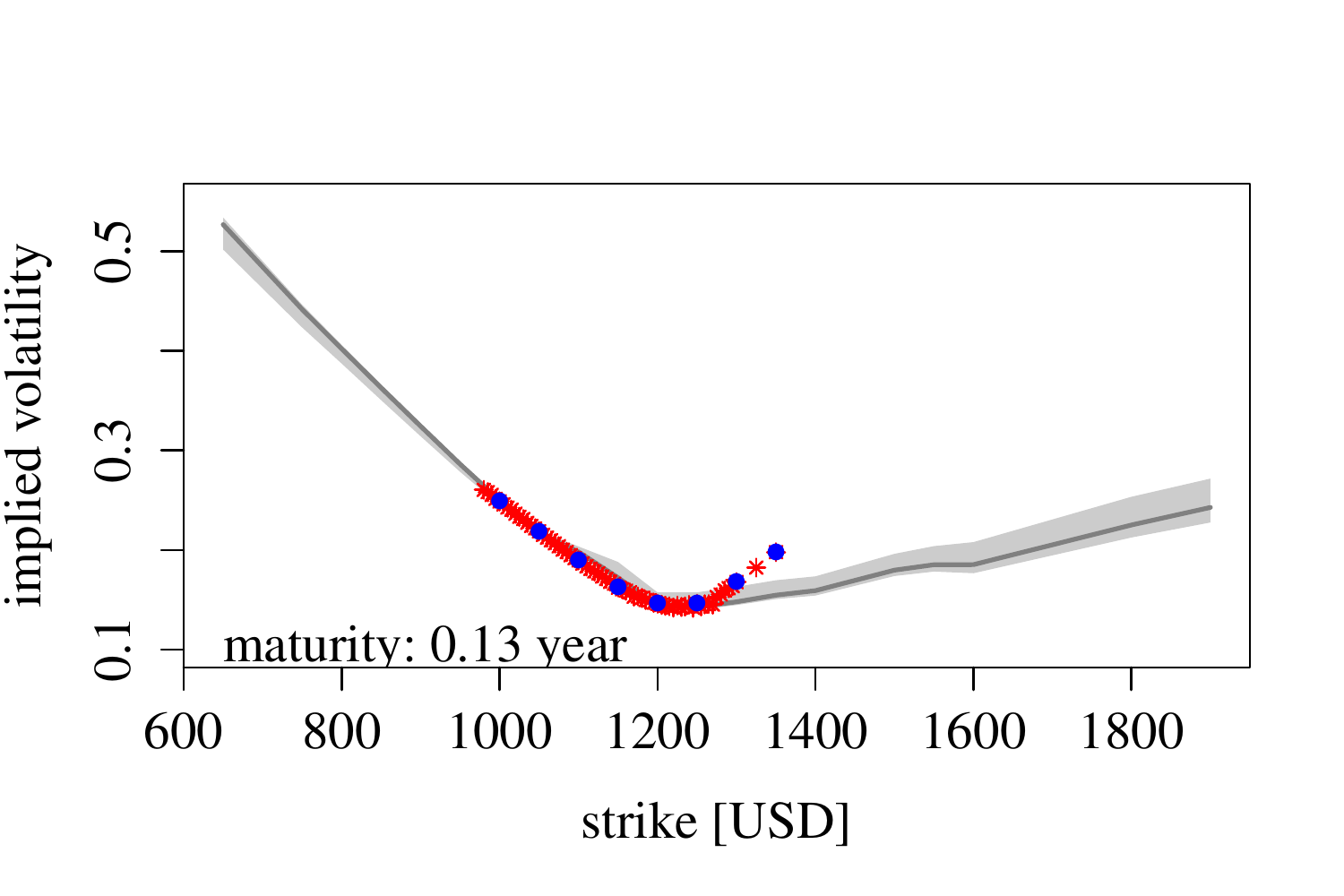} 
\includegraphics[scale=0.5,trim=46 65 25 50,clip]{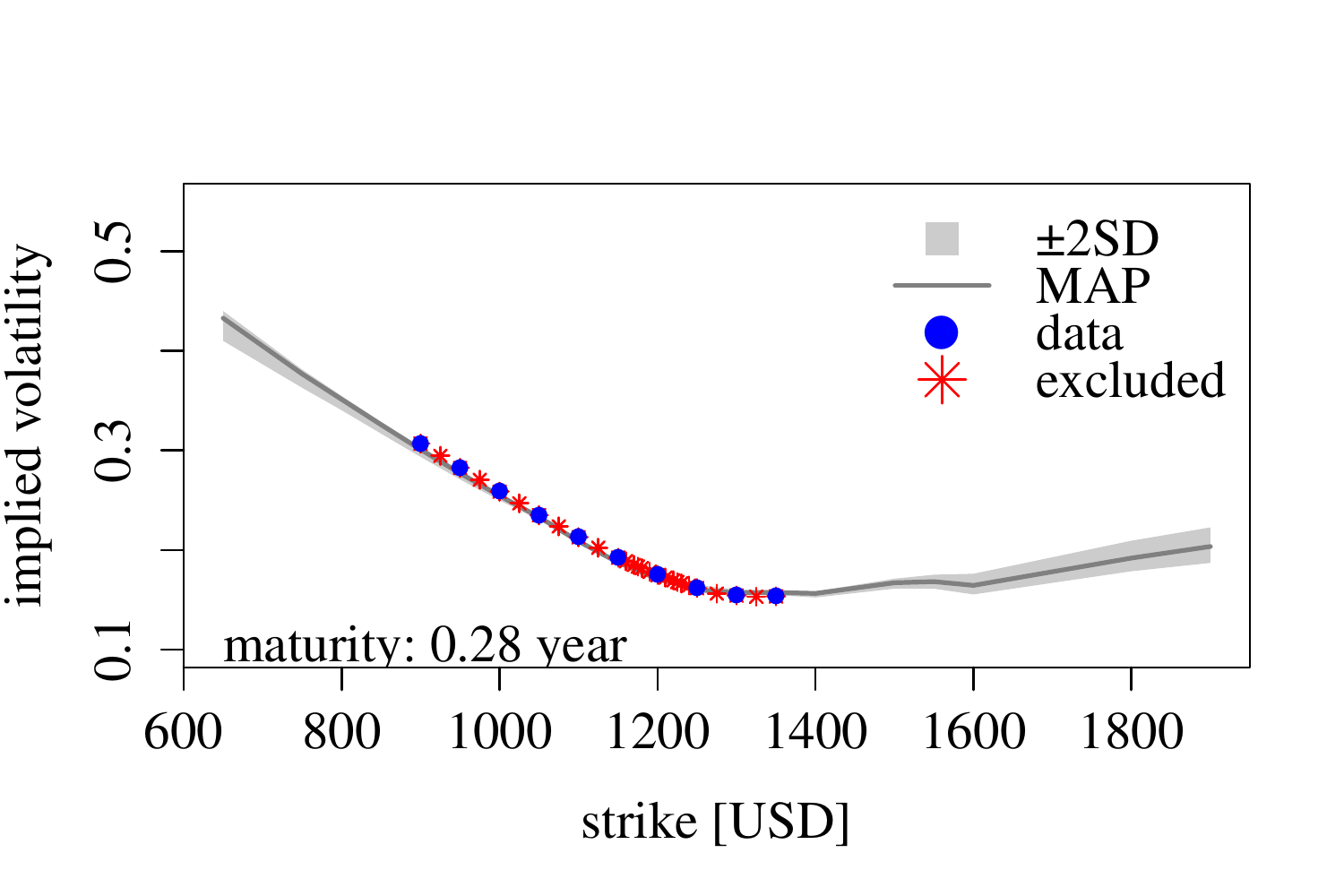} \\
\includegraphics[scale=0.5,trim=  0 65 25 50,clip]{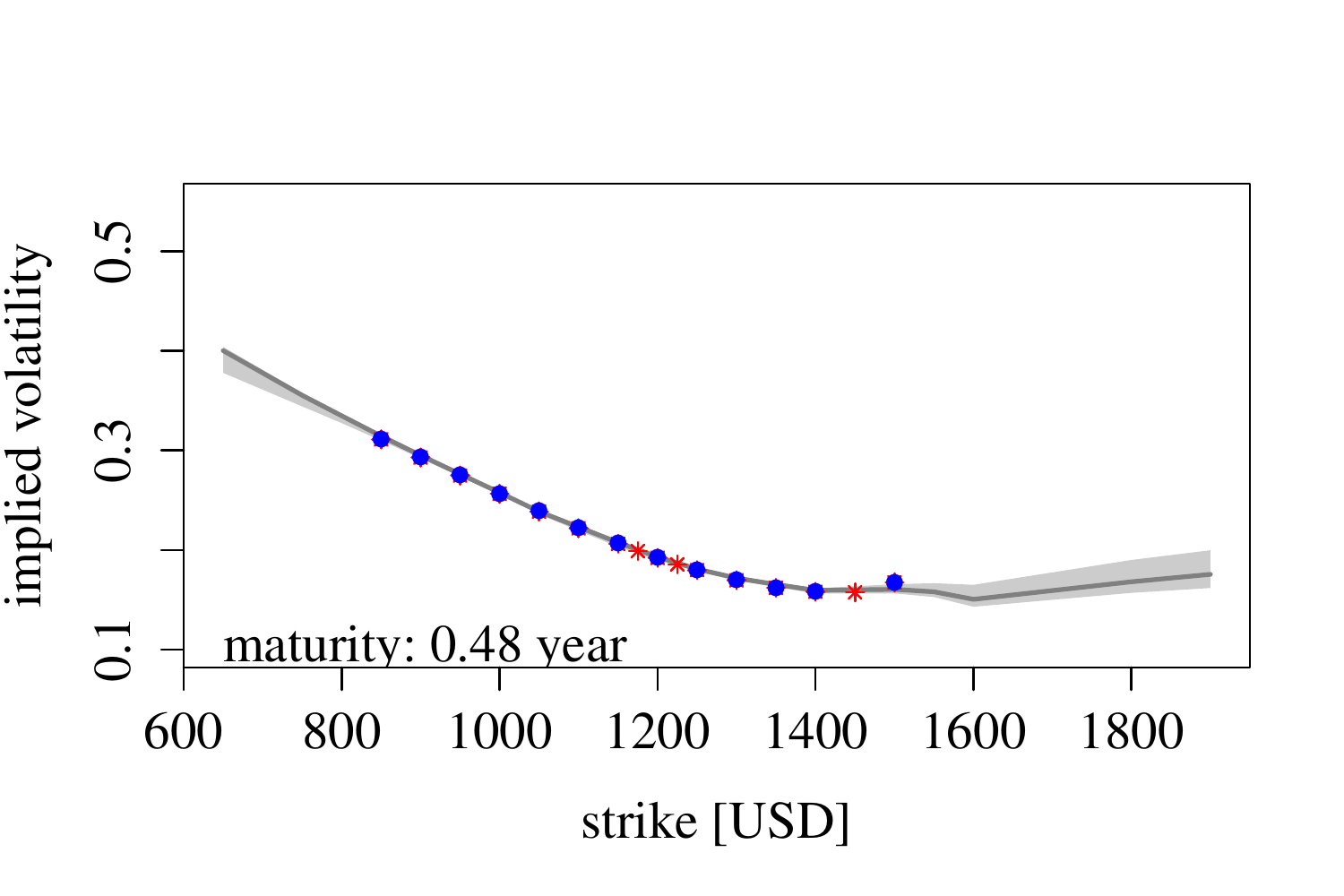} 
\includegraphics[scale=0.5,trim=46 65 25 50,clip]{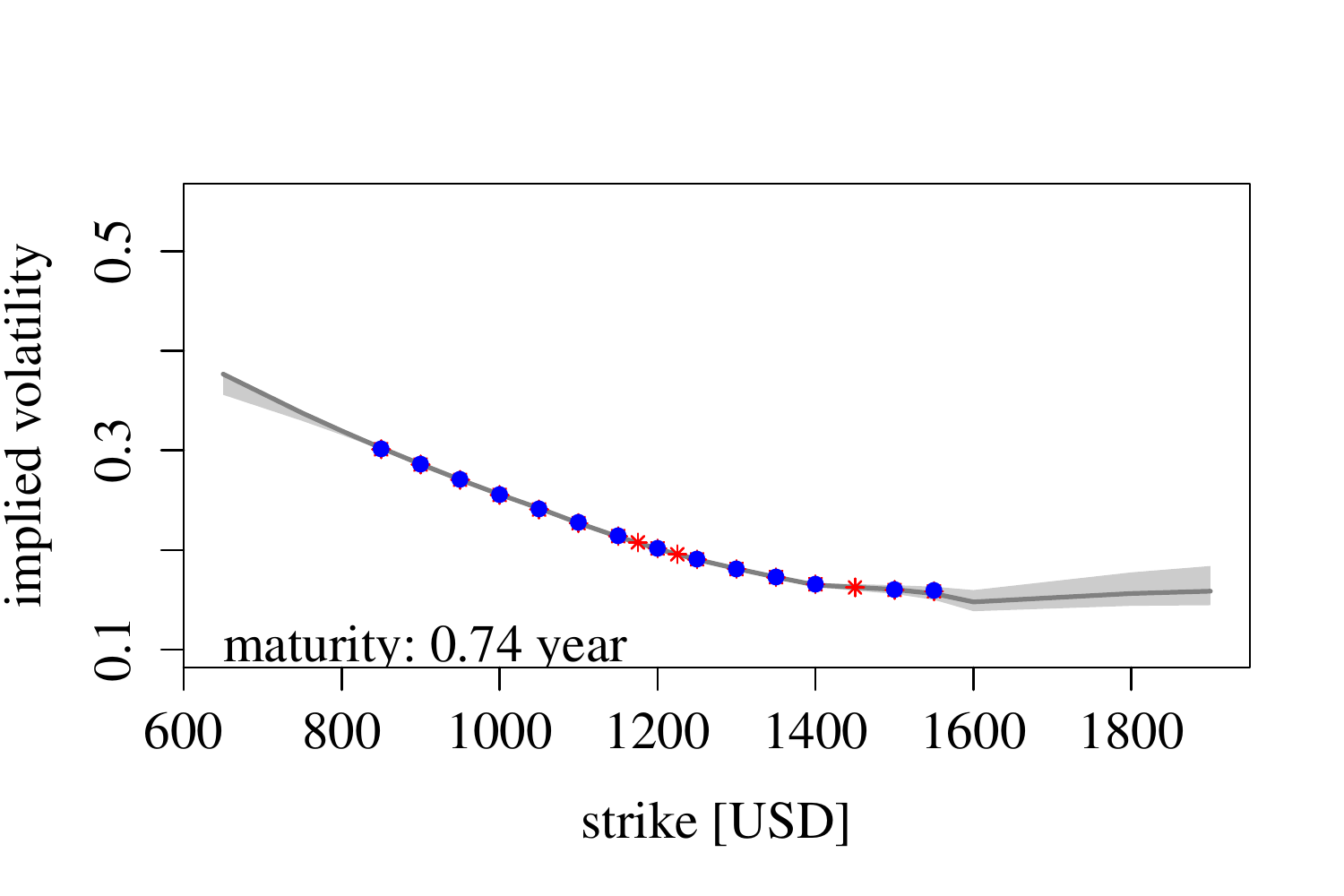} \\
\includegraphics[scale=0.5,trim=  0 65 25 50,clip]{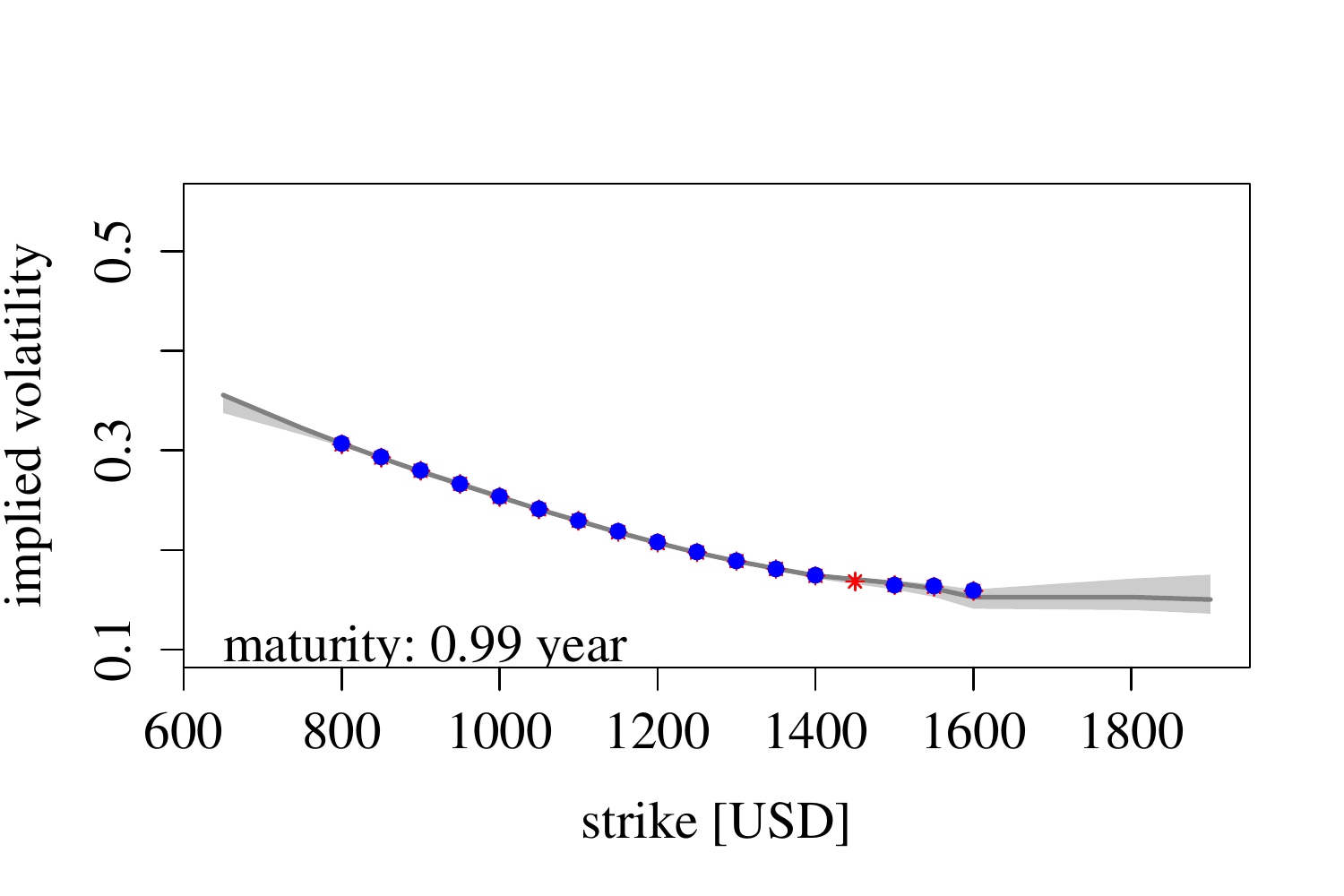} 
\includegraphics[scale=0.5,trim=46 65 25 50,clip]{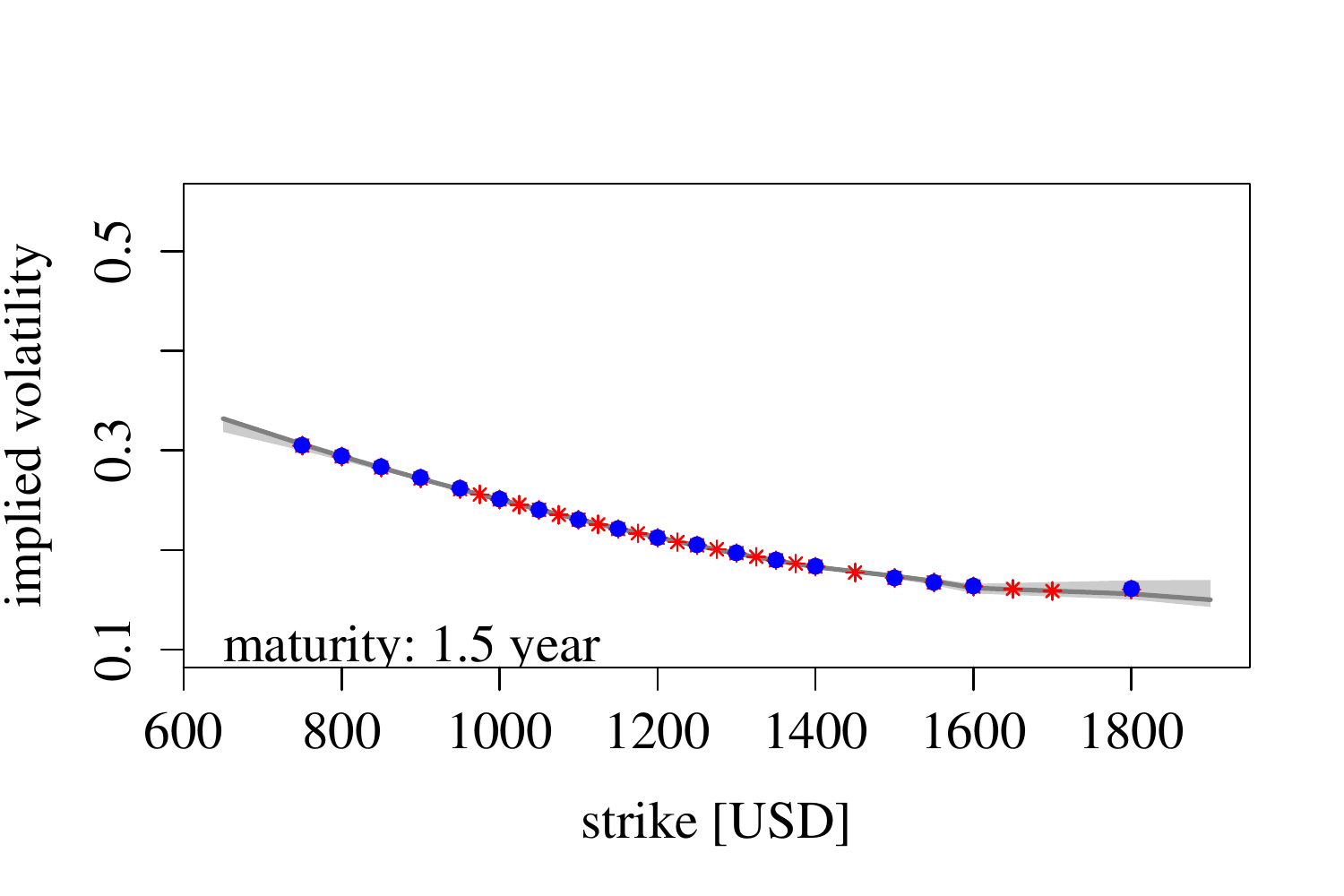} \\
\includegraphics[scale=0.5,trim=  0 10 25 50,clip]{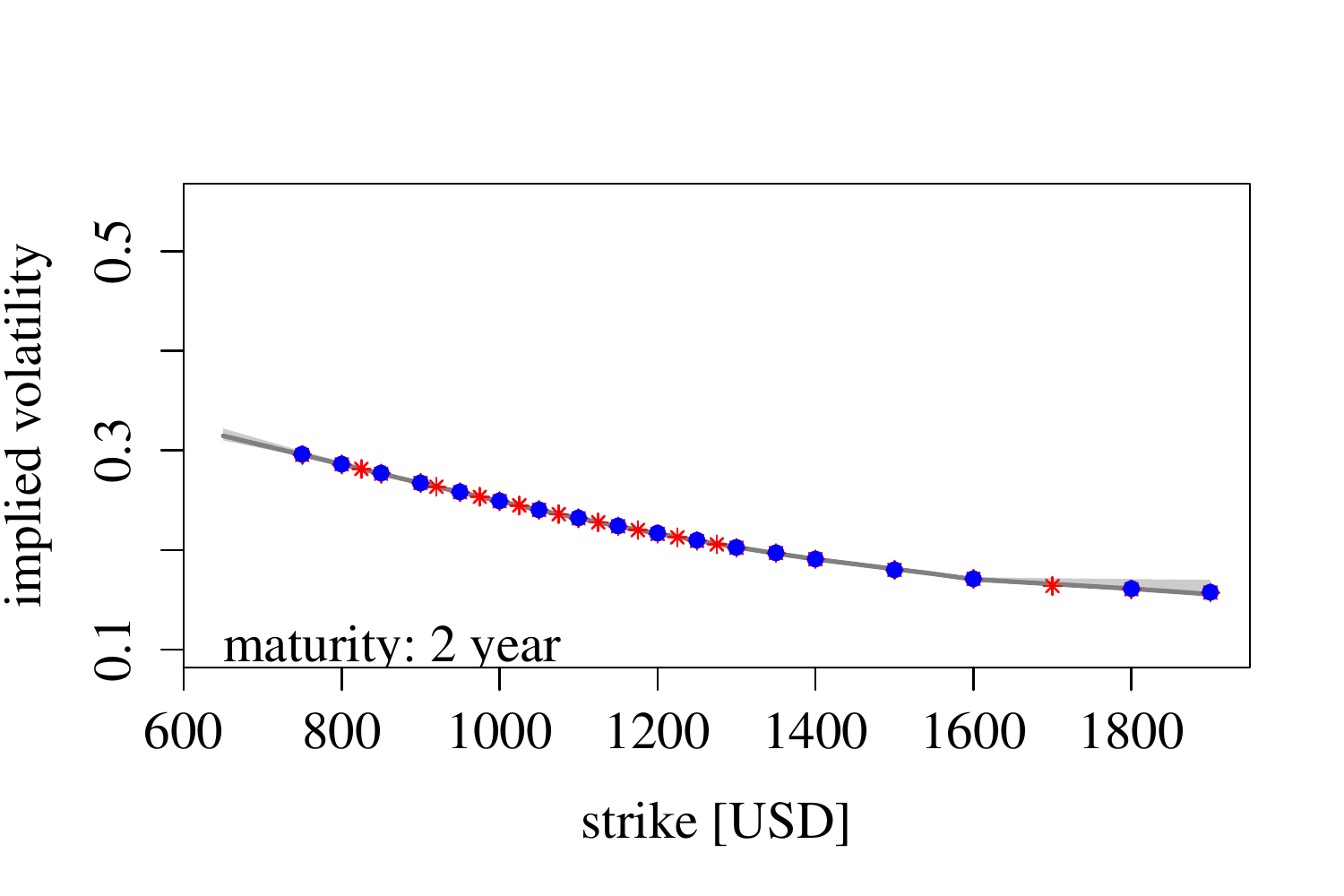} 
\includegraphics[scale=0.5,trim=46 10 25 50,clip]{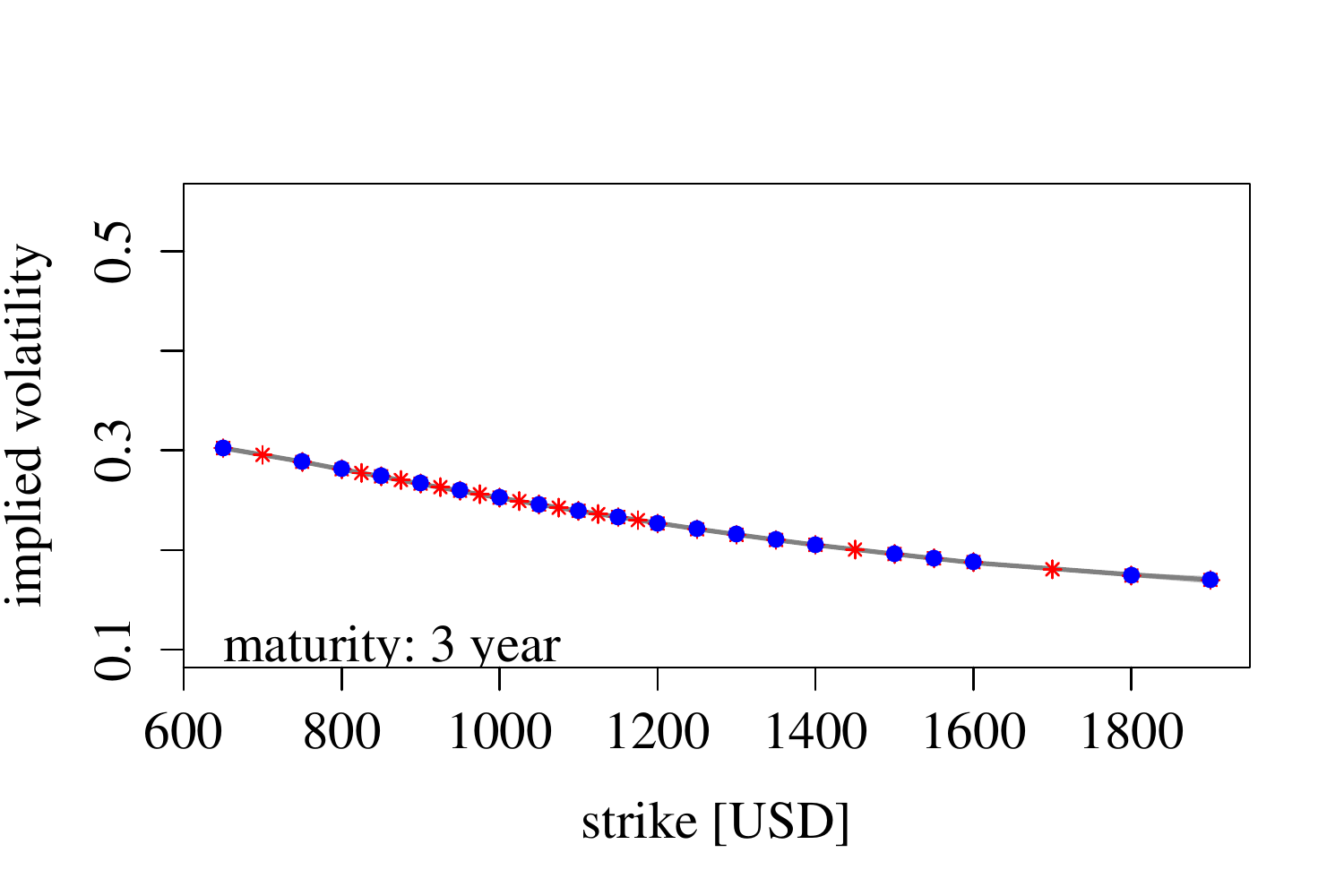} 
\caption{Posterior distribution over  implied volatilities: maximum a posteriori surface (solid grey line) and  $\pm2$ standard deviations around the  mean (light grey area). Estimates calculated from the transformed  sample of local volatility surfaces (Figure \ref{fig2}). Calibration data $\mathcal{D}$  plotted with blue dots  while red asterisks show the excluded market data (see Figure \ref{fig1}). }
\label{fig3}
\end{figure}

\paragraph{Re-pricing market options}
 To see how well the calibrated model explains market data, we transform the posterior sample over local volatility to samples over call prices and implied volatilities. The result is represented in Figure \ref{fig3} for implied volatilities. 
The figure show that the MAP surface is  close to the data    used for calibration (blue dots) and to the market data excluded from this set (red asterisks). The only exemptions are some out-of-money implied volatilities at the shortest maturity, as shown in the top left pane of Figure \ref{fig3}. This confirms the fact that  the local volatility model is often inconsistent with option market prices  at short maturities (\cite{gatheral2011volatility}). An explanation is that the underlying stock price is  a continuous process in the local volatility model, while the market anticipate sudden price changes, ``jumps'', which contribute to more pronounced implied-volatility skews (as observed in top-left Figure \ref{fig3}).

 Implied volatility MAP-to-market errors are within $-0.015\pm0.02$ and within  $-0.003\pm0.007$ if excluding the shortest maturity. Both indicate a good re-pricing performance of our approach (still, recall that the likelihood is based on errors of model and market prices, not implied volatilities) and competitive to what is reported in the literature: \cite{cont2004recovering} report  $\sim0\pm0.04$ for a set of DAX options, \cite{luo2010local}  a few percentage points. 
 
The uncertainty in the  local volatility estimate  propagate to only modest credible intervals for implied volatility, especially over the in-the-money region. The reason is, again,   model price/implied volatility  (in)sensitivity to the local volatility parameter---even if there is a high level of estimation uncertainty in local-volatility over some strike-maturity regions, it is not transferred back to uncertainty over  prices/implied volatilities. This robust behaviour of call option prices should, however, not be taken for granted for other derivatives, especially if they are designed to be heavily dependent on volatility.
 
\begin{figure}
\centerline{  
\includegraphics[scale=0.4,trim=0 0 0 50,clip]{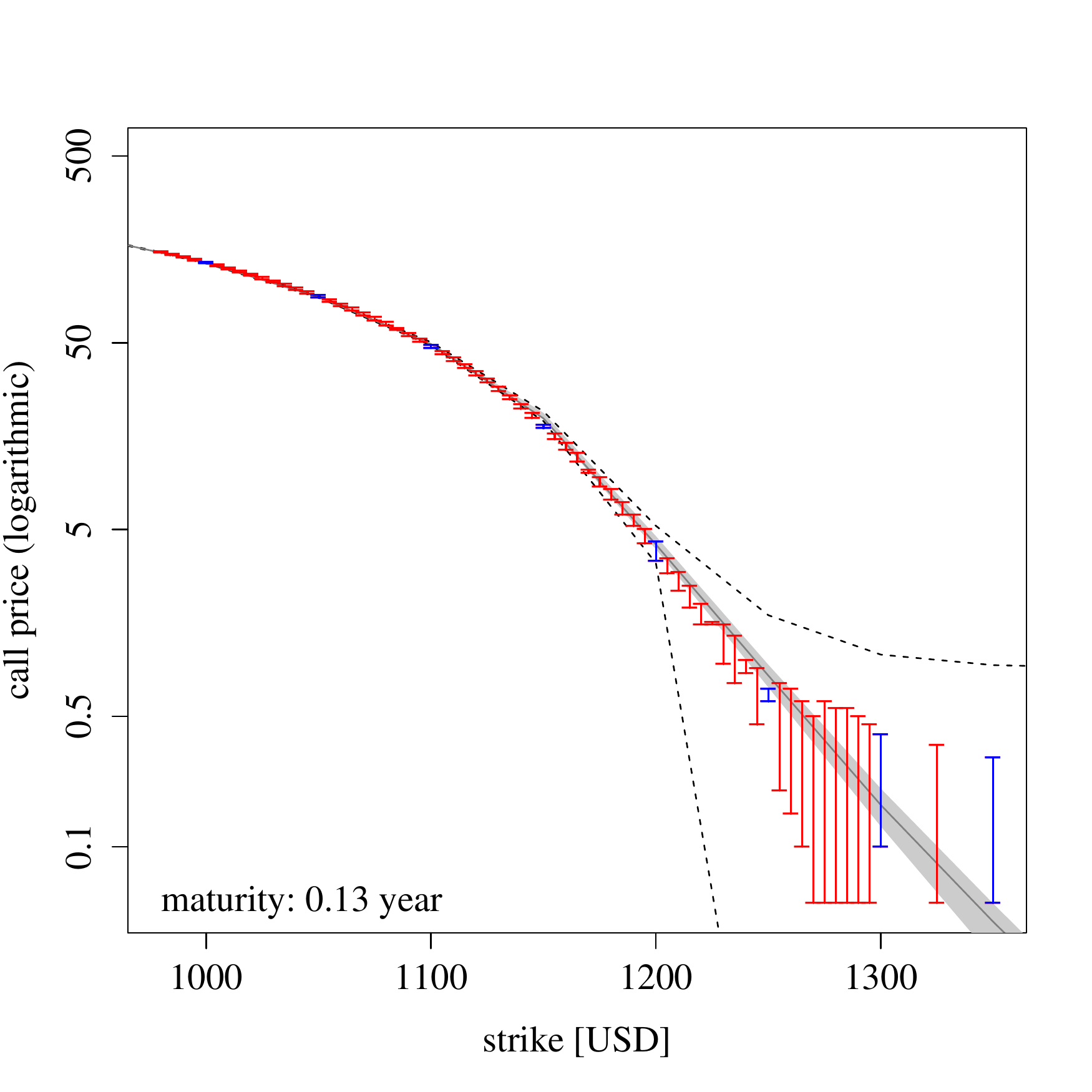}
\includegraphics[scale=0.4,trim=0 0 0 50,clip]{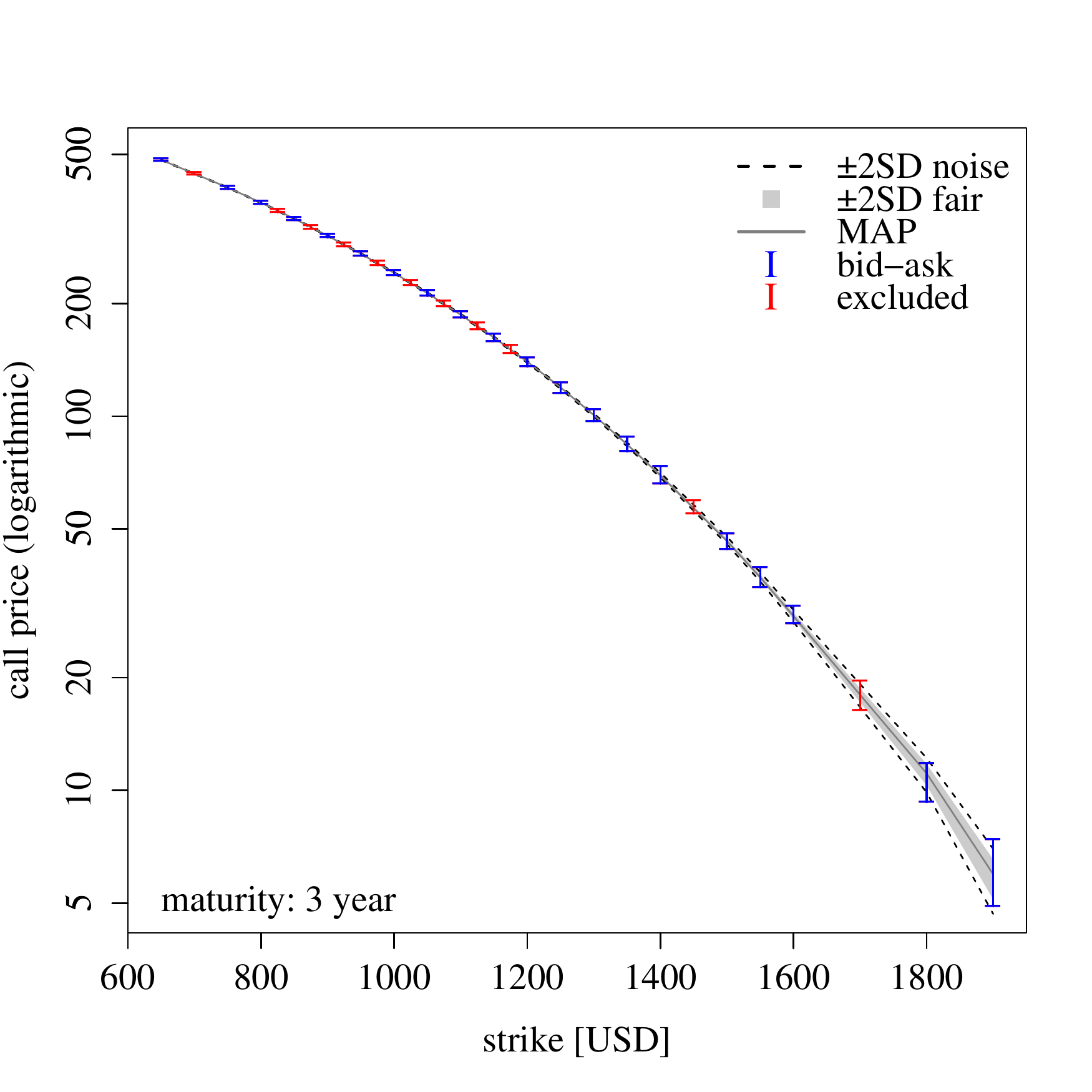}
}
\caption{The posterior distribution over call prices on a logarithmic scale, for the shortest (left) and longest (right) maturities of the market data used for calibration. Note that while the mid-market price is used for inference about the fair price, the figure shows quoted bid-ask spreads.	}
\label{fig3b}
\end{figure}

If we plot confidence regions over call prices they appear completely tight, with all market prices indistinguishable from the MAP estimate (figures not shown for brevity). Prices are  \$0--500 while their posterior standard deviations are \$$10^{-5}$--0.6. Figure \ref{fig3b} represents the posterior over call prices on a logarithmic scale together with bid-ask spreads, from which mid-prices are observed data. For the earliest maturity  in the left figure (with largest implied volatility errors), most bid-ask spreads---also for excluded data---enclose  the credible region over the model price even if the mid-price  sometimes fall outside. The fair price described by the posterior is hence permitted to be within the market spread. This is of course desirable since otherwise they would disagree on arbitrage.   For  out-of-money options  where prices are relatively small ($K>\$1200$), the noise  is dominating (the posterior mean of $\nsigma$ is 0.45) such that  mid-prices (and spreads) fall within their $\pm$2SD region. Even if the MAP estimate is not predicting these observations very well, the posterior is still consistent with observed data. 

For the latest maturity (Figure \ref{fig3b}, right) mid-prices fall more closely to the MAP estimate of the fair price while the noise  is relatively small: the credible region, also when including noise, is typically tight enough to be inside the bid-ask spread. This is the general picture also for remaining maturities (figures not included).

\begin{figure}
\makebox[\textwidth][c]{  
\includegraphics[scale=0.4,trim=60 50 60 50,clip]{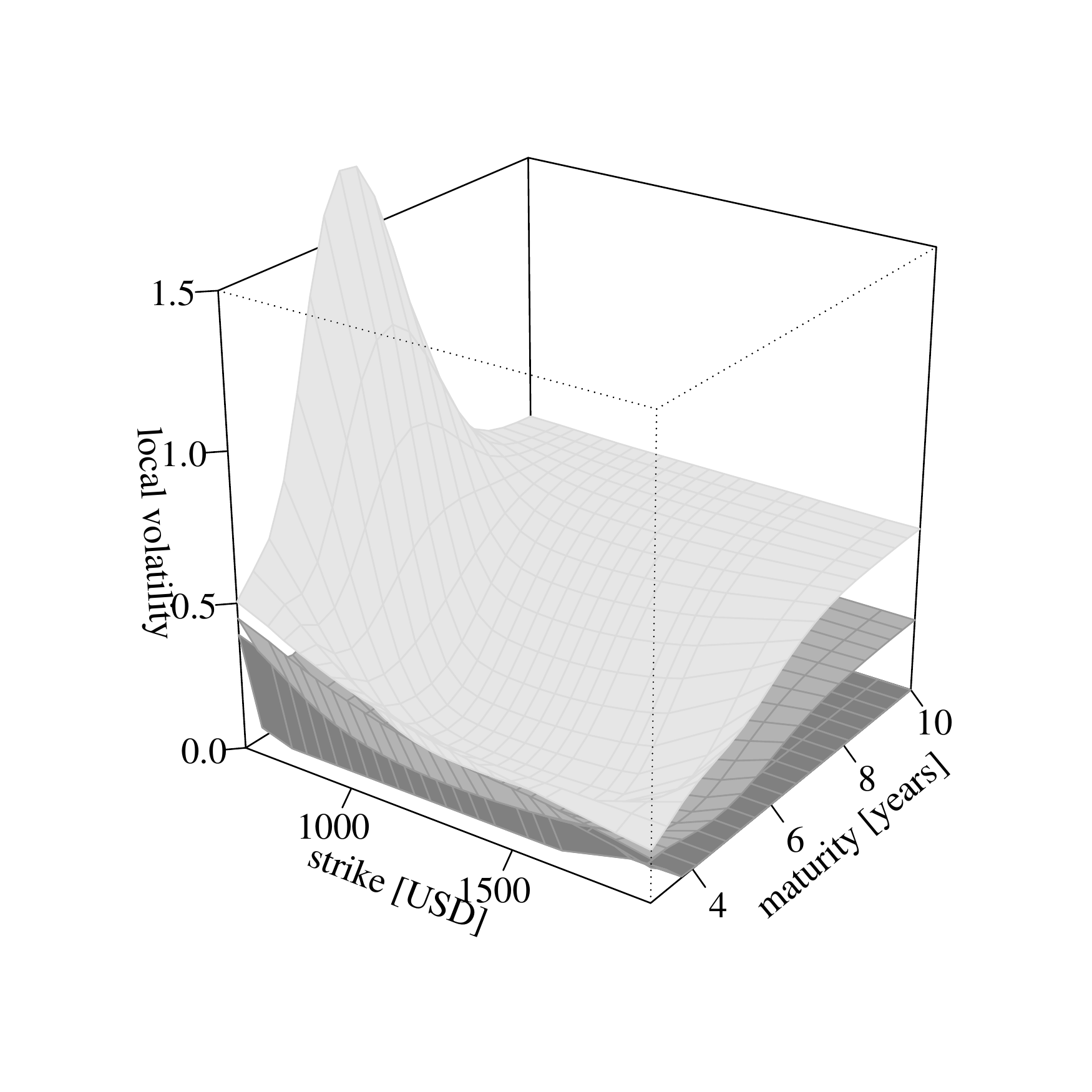}
\includegraphics[scale=0.4,trim=60 50 60 50,clip]{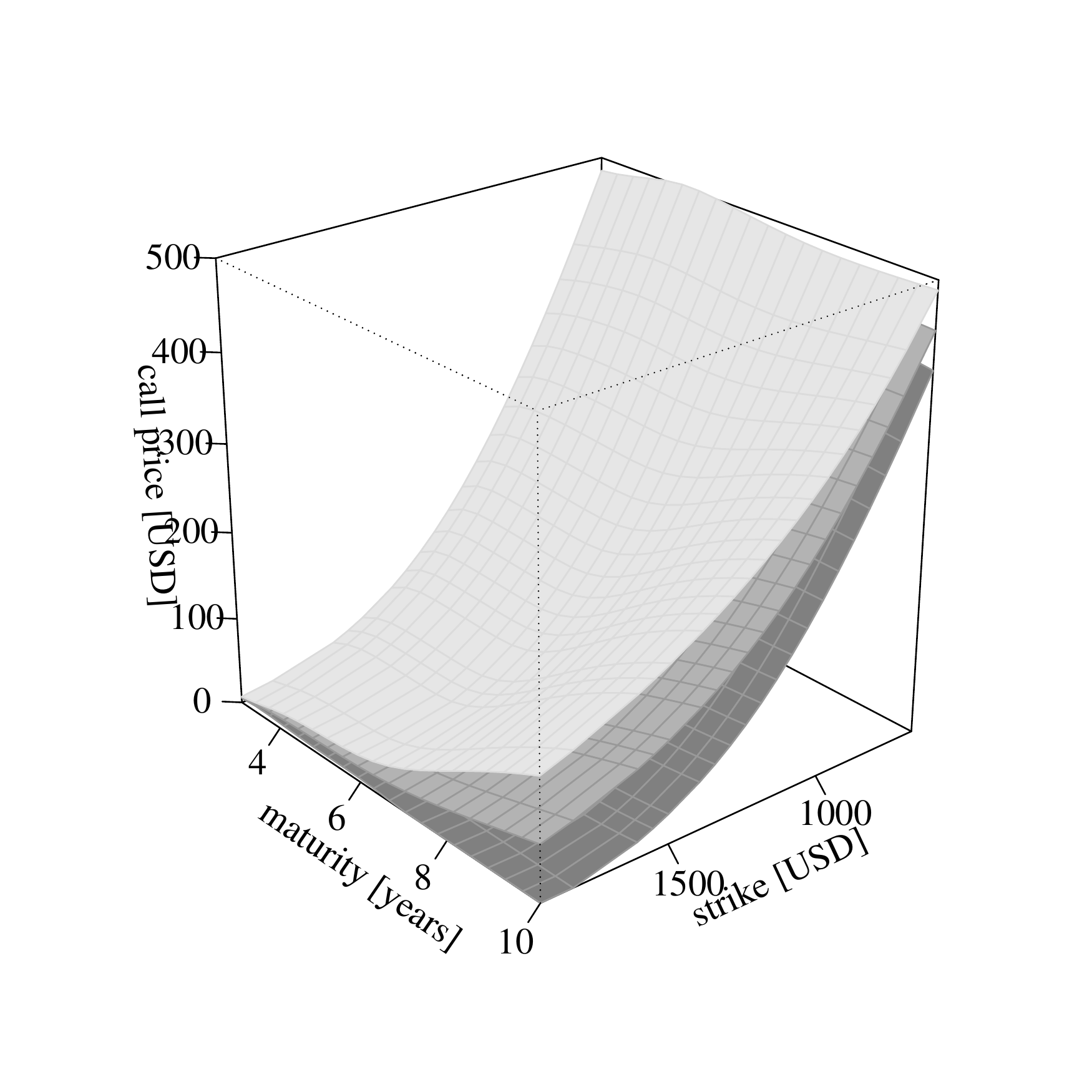}
\includegraphics[scale=0.4,trim=60 50 70 50,clip]{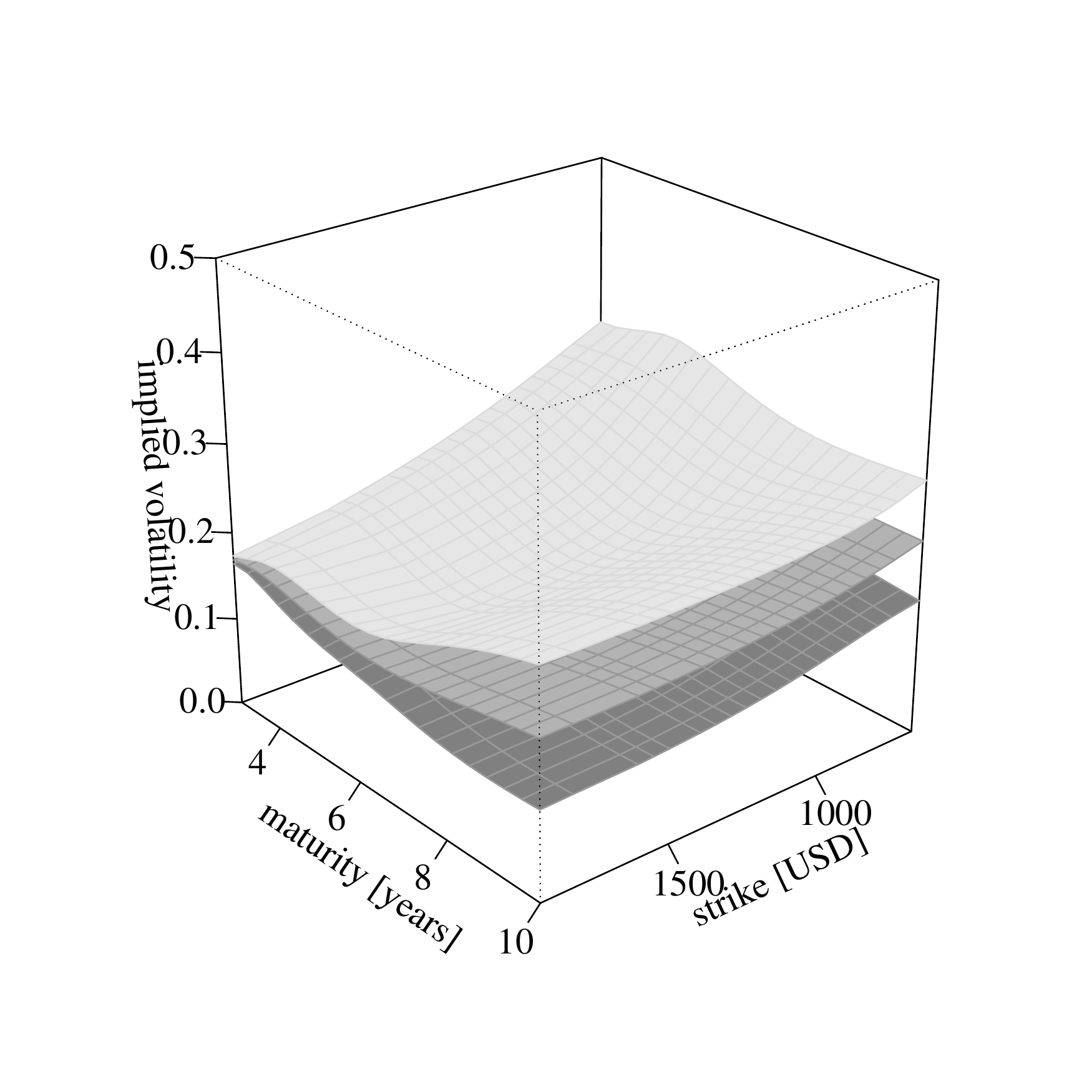}
}
\caption{\textbf{Left:}  Posterior predictive distribution over local volatility for  maturities of 4--10 years: credible interval of $\pm2$ standard deviations around the mean.  \textbf{Middle:} Predicted call prices. The figure is rotated to better show prediction uncertainty for  long-dated options. \textbf{Right:} Implied volatilities corresponding to predicted call prices. Compared to the calibration (Figure \ref{fig3}) the prediction uncertainty is generally higher   (same $z$-axis scale in both figures) and notably increasing in maturity.  } 
\label{fig4}
\end{figure}

\paragraph{Prediction} We apply the predictive distributions of Section \ref{secPred} to predict local volatility over unseen inputs. As an illustrative example, we consider long-dated options and extend the input grid to maturities  4--10 years. We generate a prediction sample of 1000 local volatility surfaces by direct simulation: each of 1000 randomly selected states $(\bf,\kappa,m)$ from the posterior sample are used to calculate the moments of the conditional prior in \eqref{eqnPredf}. We then generate  a local volatility surfaces 
 from every such set of moments. The result is shown in  Figure \ref{fig4} (left) as a credible envelope of $\pm2$ standard deviations around the  mean. Comparing with the calibration  (Figure \ref{fig2b}) the prediction uncertainty is substantial over small/large strikes and increasing in maturity up to $\sim$7 years. From this point, the surfaces level off to their unconditional (i.e. prior) confidence $\sim$$0.2\pm 0.3$. 
This is a consequence of the predictive properties of a Gaussian process. For points far away from the input grid there is not much predictive power from data as the conditional covariance declines, and the flat prior mean yields optimal predictions---see \eqref{eqnGPpred2} where the second term, in both moments, goes to zero.    

The local volatility sample is used to a predict call prices (Figure \ref{fig4}, middle) which in turn are transformed to a sample over implied volatilities (Figure \ref{fig4}, right). Both figures illustrate how prediction uncertainty associated with local volatility propagate to prices and implied volatilities: long-dated options  can only be consistently priced subject to an increasing  uncertainty, the range of which is  substantial for the long-dated maturities.
\begin{figure}
\centering
\includegraphics[scale=0.4,trim=  0 0 0 50,clip]{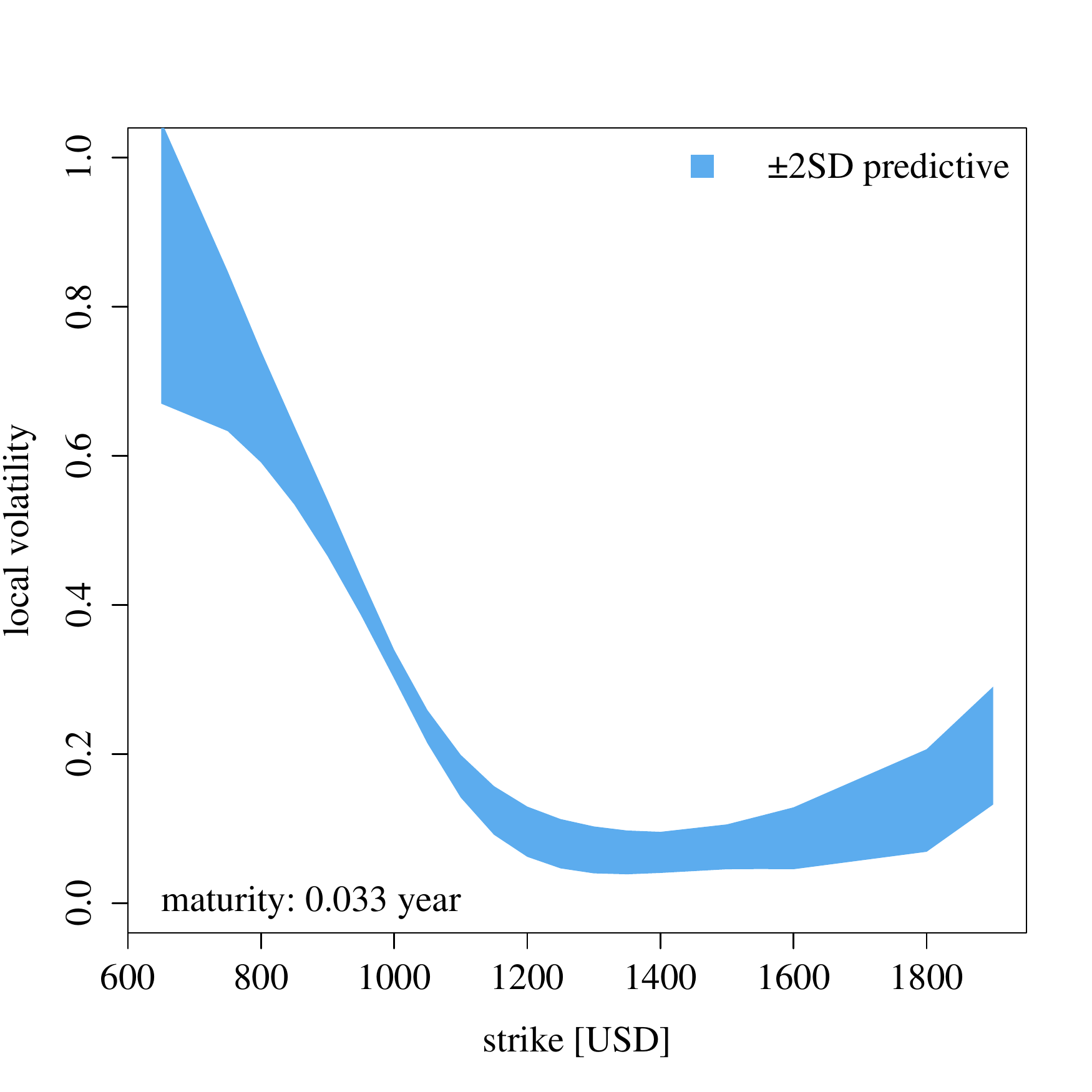} 
\includegraphics[scale=0.4,trim=  0 0 0 50,clip]{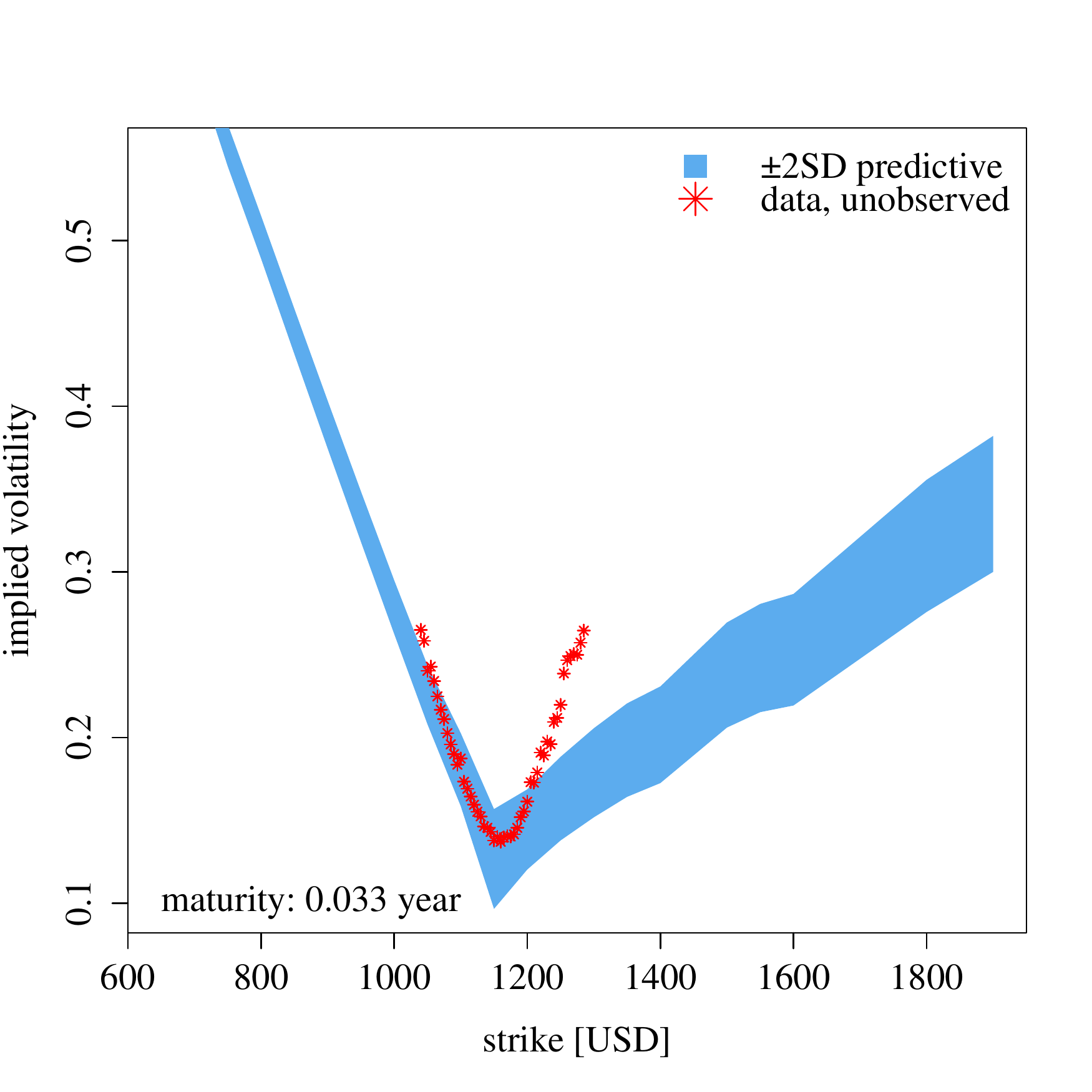} 
\caption{Predictive distribution at a maturity of 12 days (the shortest of the original data set). Credible interval of $\pm2$ standard deviations around the mean of local  (left) and implied  volatility (right).   }
\label{fig4b}
\end{figure} 
 
In Figure \ref{fig4b} we show predictions at a  maturity of 12 days. This is the shortest maturity of the original market data, but it was not included in the set $\{\bc,\bxh\}$ used for inference (see Figure \ref{fig1}). As we consider predictive inputs close to the input grid, we obtain a predictive interval similar in shape to that of the calibration (cf.  Figure \ref{fig2b}, top left). Transformed to call prices and implied volatilities, the predicted surfaces are again too flat in the out of money region,  confirming the fact that the local volatility model is incapable of capturing  market behaviour at very short maturities.

\begin{figure}
\makebox[\textwidth][c]{  
\includegraphics[scale=0.55,trim=50 50 50 50,clip]{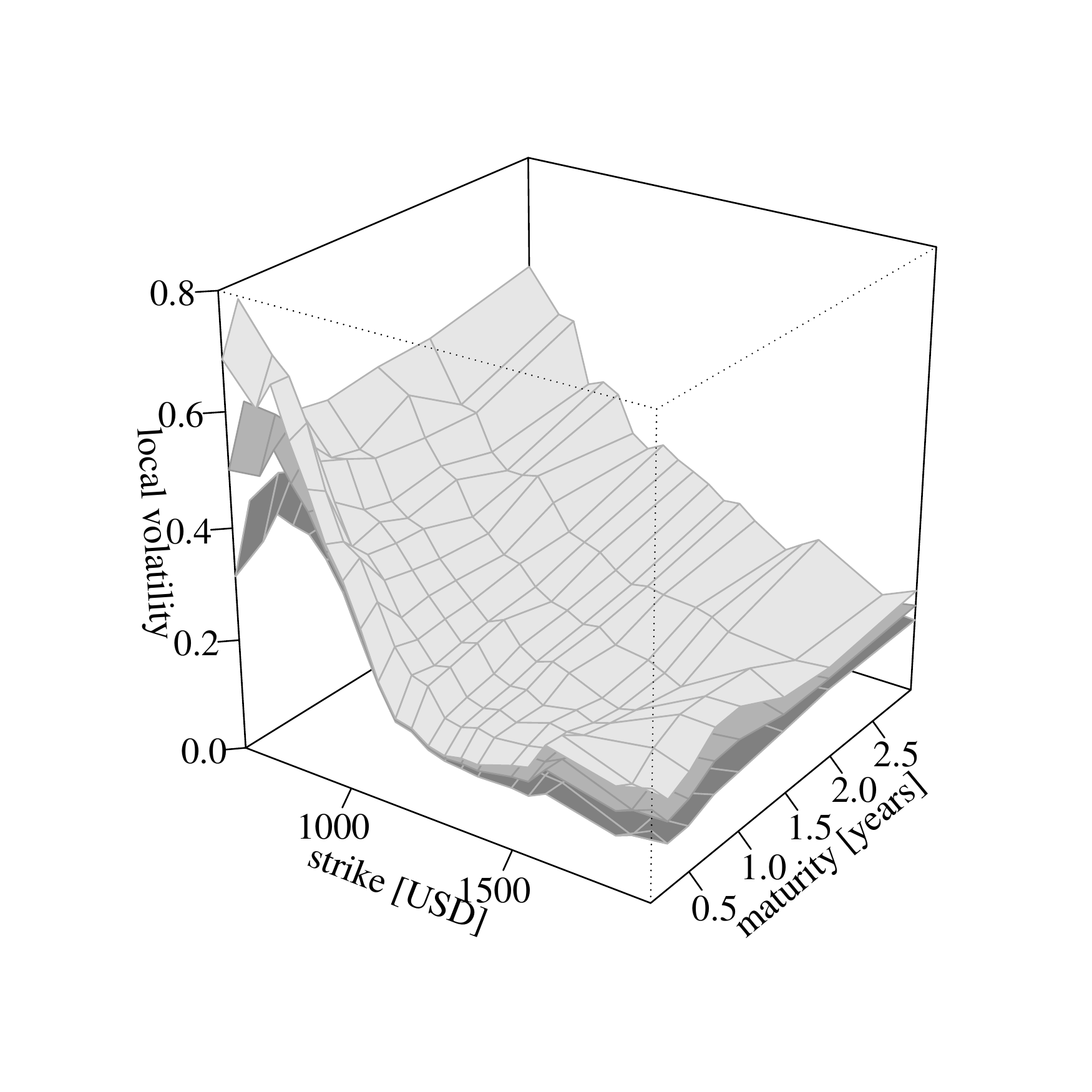}
\includegraphics[scale=0.45,trim=0 0 0 50,clip]{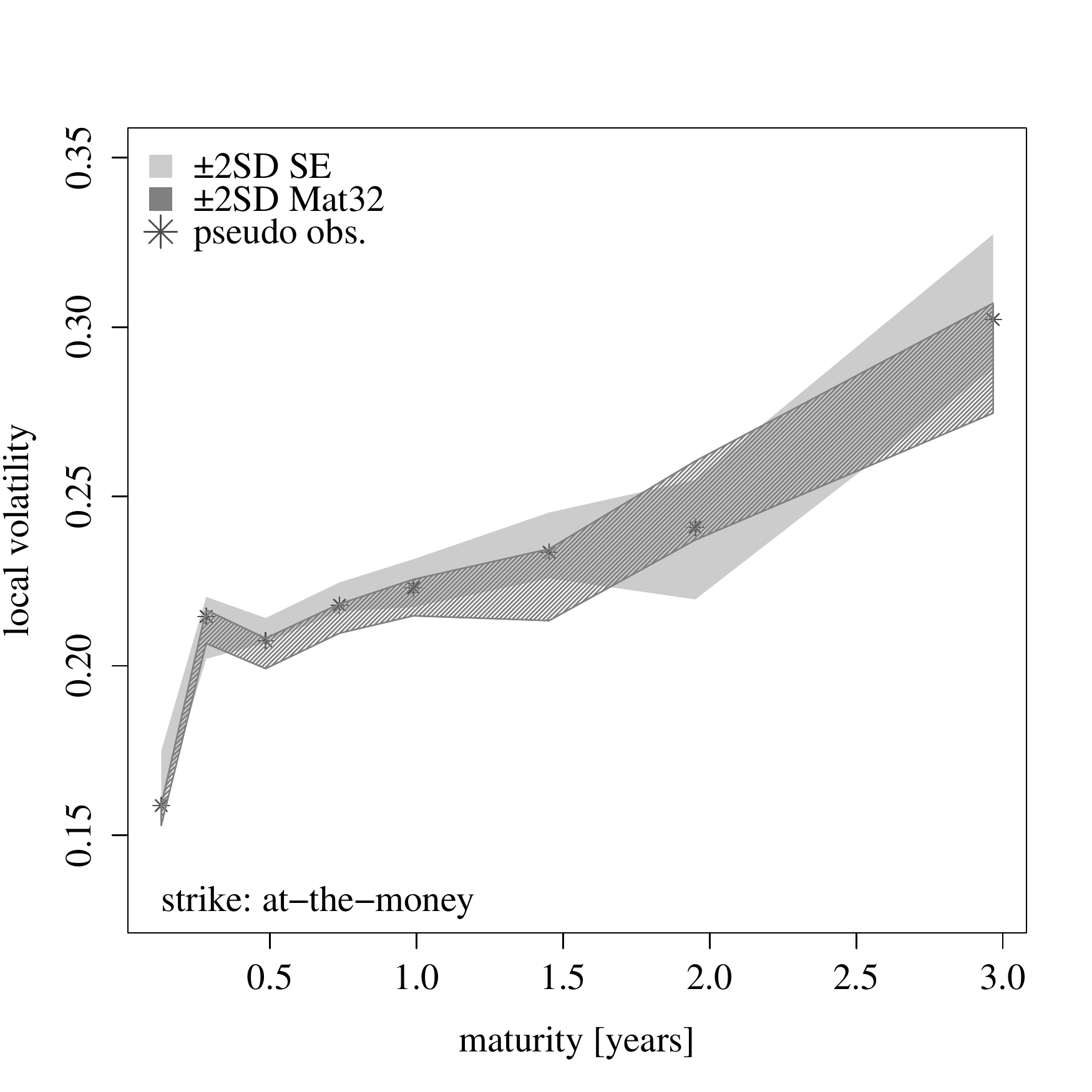}
}
\caption{\textbf{Left:} Local volatility calibrated to S\&P 500 call prices with a prior Mat\'{e}rn 3/2 covariance; credible envelope of $\pm2$SD around the mean along with the MAP surface. \textbf{Right:} Cross-section along maturity taken at the at-the-money strike level for the squared exponential (light grey) and Mat\'{e}rn (shaded dark grey).  }
\label{fig2bb}
\end{figure}

\paragraph{Comparison with a Mat\'{e}rn covariance}
Figure \ref{fig2bb} (left) shows the  credible envelope over  local volatility from the posterior-sample generated under a Mat\'{e}rn 3/2 covariance \eqref{eqnMat32}. As encoded by this prior, surfaces are  generally less smooth as compared to those of a squared exponential, cf. Figure \ref{fig2} (left). Still, credible regions over implied volatility from the Mat\'{e}rn  covariance  (not shown for brevity) are almost identical to those of the squared exponential shown in Figure \ref{fig3}. In terms of call prices, this covariances thus produce a similar distribution over the observed input set. What is slightly different is how the distribution is induced by the distribution over local volatility. The squared exponential assigns more probability to smooth surfaces while  it has to work with shorter length scales. Posterior sample averages are $\bar{l}_T = 0.3$, $\bar{l}_K = 0.2$ for the square exponential as compared to $\bar{l}_T = 0.7$, $\bar{l}_K = 0.3$ for the Mat\'{e}rn 3/2. The shorter maturity length-scale of the former give rise to a  somewhat more ``wiggly'' profile across maturity, as to preserve smoothness while still being able to fit observed prices---see Figure \ref{fig2bb} (right).

\begin{figure}
\makebox[\textwidth][c]{  
\includegraphics[scale=0.4,trim=60 50 60 50,clip]{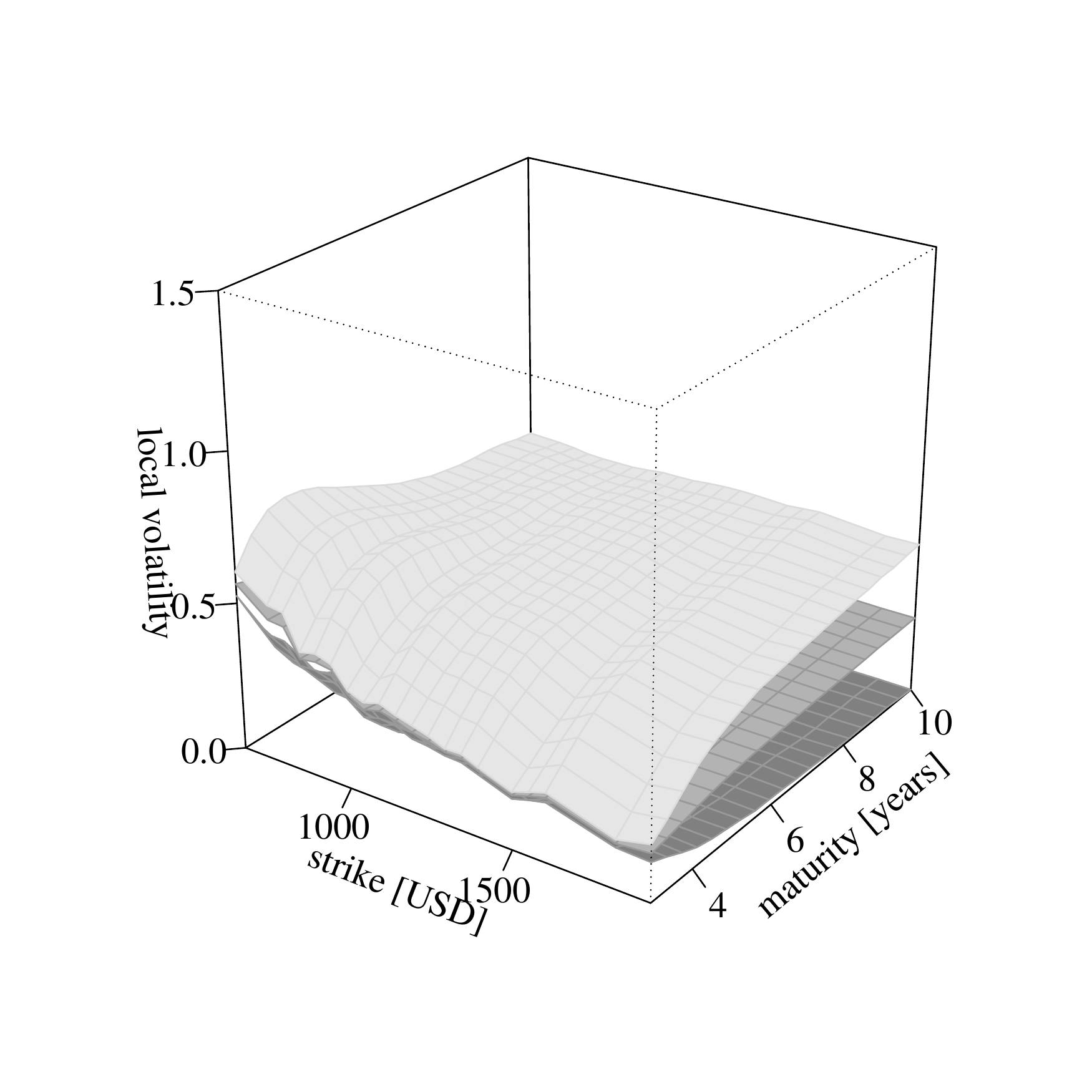}
\includegraphics[scale=0.4,trim=60 50 60 50,clip]{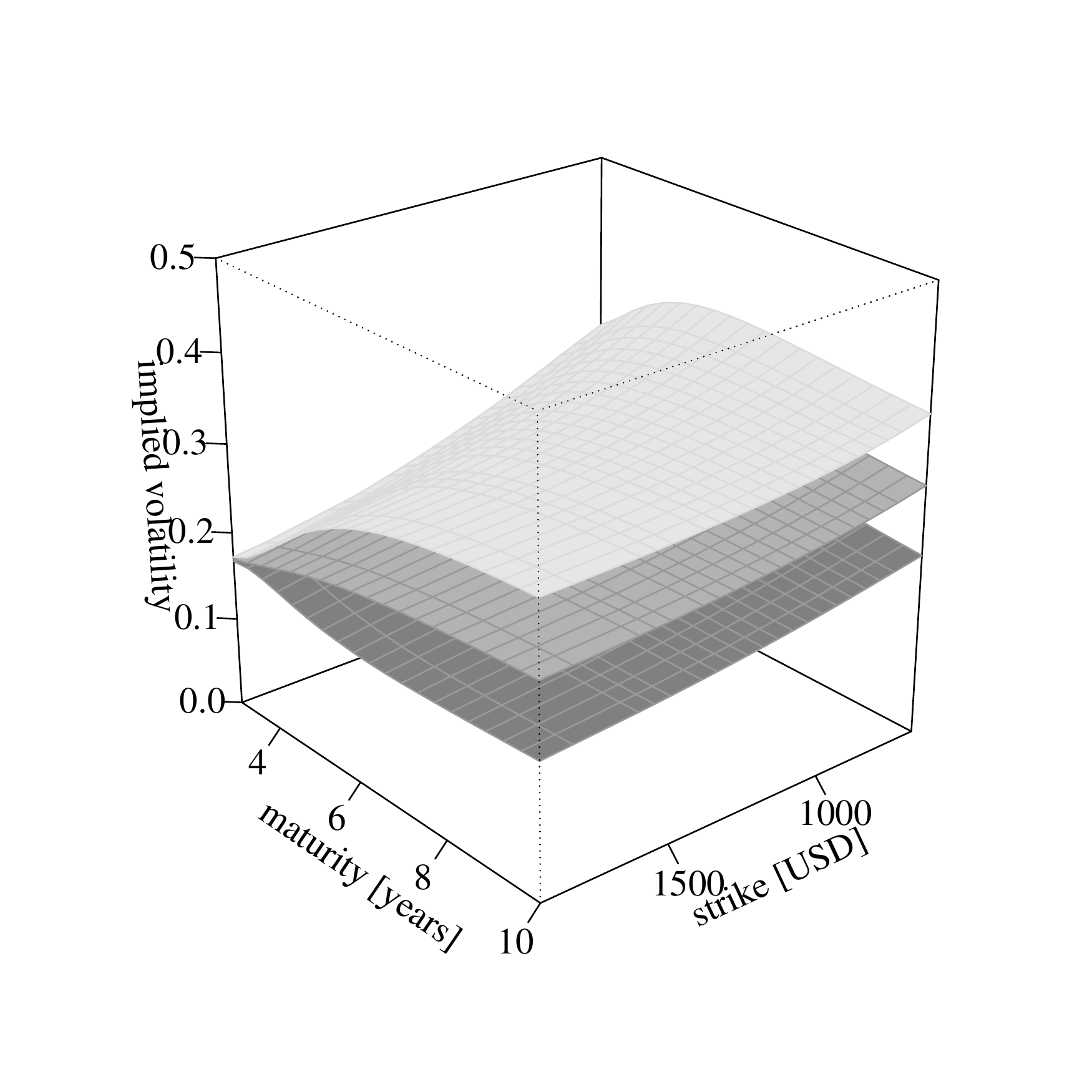}
\includegraphics[scale=0.3,trim=0 0 0 50,clip]{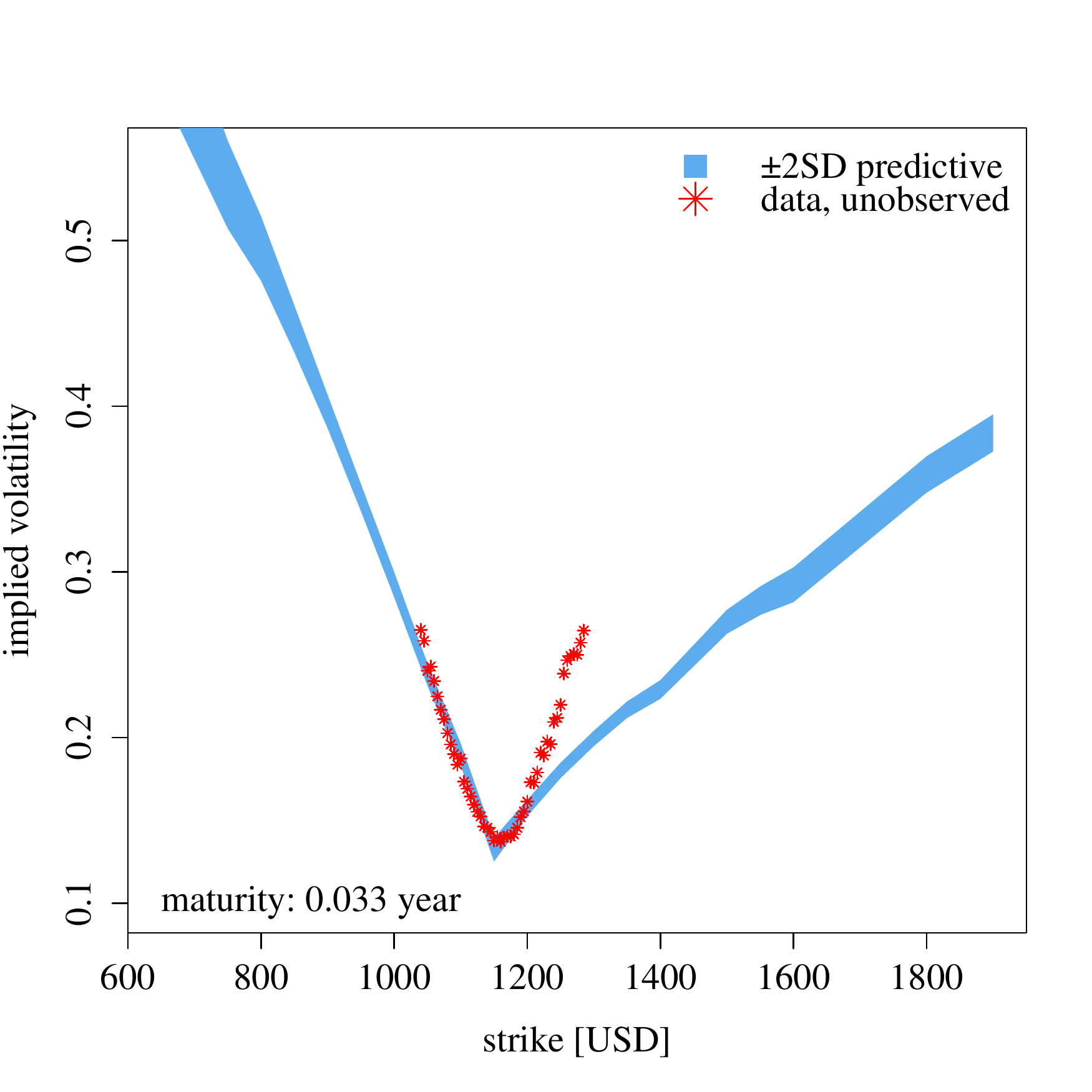}
}
\caption{Posterior predictive distribution under a Mat\'{e}rn 3/2 prior, illustrated by credible intervals of $\pm2$ standard deviations around the mean. Long-term local volatility (left) and corresponding  implied volatility (middle); short-term implied volatility (right).  }
\label{fig2bbb}
\end{figure}

Effects from the different covariances are more apparent in terms of predictions. While the squared exponential gives an overshot in local volatility when predicting long maturities (Figure \ref{fig4}, left)---again indicating that the model is specified with too short length-scales to preserve smoothness---predictions under Mat\'{e}rn behaves more nicely, see Figure \ref{fig2bbb} (left) (note that prior confidences are similar). When predicting at the shortest (unobserved) maturity of the original data, the Mat\'{e}rn seems to do a similar job: see Figure \ref{fig2bbb} (right) as compared to Figure \ref{fig4b} (right).

For a more principled comparison of the different kernels we approach their model evidences \eqref{eqME} by  Laplace's method (see \cite{mackay2003information} and \cite{moore2016fast})
\begin{equation*}
p(\bc|\mathcal{M}_i) \approx \underbrace{p(\bc|\bf^\text{map},\nsigma^\text{map})}_{\text{best fit likelihood}} \times \underbrace{ \frac{|\bK_{\bf|\bc}|^{\frac{1}{2}}}{|\bK_{\kappa^\text{map}}|^{\frac{1}{2}}} \times  \frac{|\bK_{\kappa,m,\nsigma|\bc}|^{\frac{1}{2}}}{|\ind|^{\frac{1}{2}}} }_{\text{Occam factors, }\bf\text{ and parameters} }.
\end{equation*}
This is the best fit likelihood achieved by the model times an Occam factor which penalises the model for having  parameters.  The Occam factor is a ratio of volumes of the parameter space; the posterior accessible volume  to the prior accessible volume. The denominator thus represents the prior range of possible parameter values while the nominator represents the posterior range. Prior model complexity is penalised by the former while models which do not have to be fine-tuned to data are favoured by the latter.

The log-evidence for the squared exponential covariance is $-32.4$, of which the Occam factors are 0.45 for the Gaussian process and 0.52 for hyperparameters. The log-evidence is $-26.3$ for the Mat\'{e}rn 3/2 with  Occam factors 0.38 for the Gaussian process and 0.53 for hyperparameters. As for the results, Bayesian model selection will ultimately favour the Mat\'{e}rn covariance in view of the observed  data.

\section{Time evolution of  local volatility }\label{secLVdynamics}

Up to this point, we have considered local volatility calibration based on data as observed on a single time point $t=0$, i.e. in view of ``today's" call prices.
While the local volatility model is able to consistently and effectively fit the option  market in this static sense---see the discussion of equation \eqref{eqn1}---it is also known to have poor properties in that the optimal calibration as of today will not yield an optimal calibration as of tomorrow. For this reason, the local volatility model has to be recalibrated, say daily,  to be useful in practice (see the discussion in \cite{hull2002methodology}). 

This motivates us to introduce a third input dimension in our prior model that represents  time-dependency of local volatility
\begin{equation}\label{eqBerra1}
\sigma: (t,T,K) \mapsto  \sigma(t,T,K)
\end{equation}
where we use a variable $(t,x)\equiv(t,T,K)$ in the augmented input space $\bar{\mathcal{X}} = \R_+\times\mathcal{X}$. 
Our Bayesian framework with Gaussian processes  readily  generalises to this setting such that we may draw posterior inference about local volatility  over (calender) time. 

\subsection{Augmented probabilistic model}

To accommodate \eqref{eqBerra1} we take 
$\bar{\mathcal{X}}$ to be the input space of our Gaussian process prior and introduce a time component in the covariance function with a separate length-scale. For the Mat\'{e}rn 3/2 covariance function
\begin{equation}\label{eqBerra4}
k(t_i,x_i;t_j,x_j;\kappa) = \sigma_f^2\,k_{\text{Mat32}}(T_i,T_j;l_T)\,k_{\text{Mat32}}(K_i,K_j;l_K)\,k_{\text{Mat32}}(t_i,t_j;l_t)
\end{equation}
where $\kappa = (l_T,l_K,l_t,\sigma_f)$ is the vector of covariance parameters. 
For the observation model, we assume independent and identically distributed noise across time 
\begin{equation}\label{eqBerrall}
\log p(\bc|\bf,\nsigma) = -\frac{1}{2\nsigma^2}\sum_{t=1}^{\tau}\sum_{i=1}^{n_t} (C(\obs{x}_{t,i},\bf_t)-\obs{c}_{t,i})^2 - \frac{n_{\text{tot}}}{2}\log\left(2\pi \nsigma^2\right).
\end{equation}
We introduce a separate time index $t\in\{1,2,\dots,\tau\}$ such that for a  set of $n_{\text{tot}}$ 
observations $\bc = \{\bc_t\}_{t=1}^{\tau}$,  
 each $\bc_t$ collects a surface $\bc_t = \{\obs{c}_{t,i}\}_{i=1}^{n_t}$ of $n_t$ mid-prices observed at a single date $t$. The index $i$ refers to the corresponding  strike-maturity $\obs{x}_{t,i}=(t,T_i,K_i)$ of the option at that date. Similarly, we use $\bx_t$ to denote an input set of $N_t$ strike-maturities as of date $t$ and $\bx = \{\bx_t\}_{t=1}^{\tau}$ for the full input set over all dates (and likewise for market sets $\bxh_t$ and $\bxh$). We follow the same convention for functional evaluations: $\bf_t = f(\bx_t)$ for the local volatility surface at $t$ and $\bf = \{\bf_t\}_{t=1}^{\tau}$ for the collection of local volatility surfaces over all $\tau$ dates, or, equivalently, of $f$ evaluated at every input $x$ of the full input set $\bx$.

Together with the hyper-prior over $(\kappa,m,\nsigma)$ where $l_t$ is assumed independently distributed according to \eqref{eqnSSG}, we  obtain a joint posterior of functional values and hyperparameters of the same form as  (\ref{eqnJointPost}). With shorthand notation $\bx_{1:\tau}\equiv \{\bx_t\}_{t=1}^{\tau}$, 
\begin{equation}\label{eqBerra2}
p(\bf_{1:\tau},\kappa,m,\nsigma|\bc_{1:\tau}) = \frac{1}{p(\bc)} \underbrace{ p(\bc_{1:\tau}|\bf_{1:\tau},\nsigma) }_{\text{likelihood}} \underbrace{ p(\bf_{1:\tau}|\kappa,m) }_{\bf\text{-prior}} \underbrace{ p(\kappa,m,\nsigma) }_{\text{hyperprior}}.
\end{equation}

\paragraph{Inference}

Sampling the full posterior \eqref{eqBerra2} with the strategy outlined in Section \ref{secNum} quickly becomes prohibitive as the computational cost for making proposals scales as $\mathcal{O}(N_{\text{tot}}^3) = \mathcal{O}(\tau^3 N^3)$. Critically, observed strikes and maturities changes over time such that $\bx$ no longer has Kronecker structure. For this reason, we exploit the time-structure of the problem and  approach the sampling of \eqref{eqBerra2} by a sequential method proposed by \cite{mt_ssgp}. 

To this end, consider  sampling $\bf_t$ when  a new set of market data $\{\bc_t,\bxh_t\}$ arrives at time $t$. For given hyperparameters and past values $\bf_{1:t-1}$, this means targeting the conditional
\begin{equation}\label{eqBerra3}
p(\bf_t | \bf_{1:t-1},\kappa,m,\nsigma,\bc_{1:t})\propto p(\bc_t|\bf_t,\nsigma)p(\bf_t|\bf_{1:t-1},\kappa,m)
\end{equation}
where we  use that all terms of the full likelihood 
drop out except the most recent $p(\bc_t|\bf_t,\nsigma)$ due to the factorising form  of \eqref{eqBerrall}. This is blocked sampling of $\bf = \{\bf_t,\bf_{1:t-1}\}$ with $\bf_{1:t-1}$ fixed.  We can therefore apply the updating step for functional values described in Section \ref{secNum} by treating the predictive prior in \eqref{eqBerra3}---defined by the predictive equations (\ref{eqnGPpred})---as the Gaussian variable of \eqref{eqESS}. The cost for this is, however,  increasing cubically with $t$ and as a remit we apply  a simple form of  data selection
\begin{equation}\label{eqBerra5}
p(\bf_t|\bf_{1:t-1},\kappa,m) \approx p(\bf_t|\bf_{t-k:t-1},\kappa,m).
\end{equation}
Limiting the conditional dependency to the $k\geq1$ most recent time steps is reasonable for isotropic covariances, as with \eqref{eqBerra4}, since the dependency on distant variables is weak and of less importance for the predictive distribution, see \cite{osbornebayesian}. 
In effect, the approximation \eqref{eqBerra5} captures the computational cost to $\mathcal{O}(k^3N^3)$ for a fixed $k$. It (re)introduces more variance in the conditional posterior over $\bf_t$ through its approximative ``acting prior" \eqref{eqBerra5}, which is the proposal distribution of the elliptical slice sampler. This can potentially have beneficial effects for the sampling efficiency---see e.g. \cite{Jacob:2017aa} for related ideas. 

While sequential sampling of functional values is relatively straightforward, the computational challenge is updating  hyperparameters from their posterior conditional on the growing sequence of functional values. At time $t$, the target is
\begin{equation}\label{eqBerra7}
p(\kappa,m,\nsigma|\bf_{1:t},\bc_{1:t}) \propto  p(\bc_t|\bf_t,\nsigma)p(\bf_t|\bf_{1:t-1},\kappa,m) p(\bf_{1:t-1},\kappa,m,\nsigma|\bc_{1:t-1})
\end{equation}
where the rightmost distribution---the posterior from the previous time-step---is the challenging factor. We 
approach \eqref{eqBerra7} by breaking the dependency between parameters and functional values of the previous posterior, and approximate the parameter marginal based on the hyperprior assumption \eqref{eqnSSG}. We thus target the vector $\bz$ of the transformation $(\kappa,m,\nsigma) = \text{ssg}(\bz)$, and take
\begin{equation}\label{eqBerra6}
p(\bf_{1:t-1},\bz|\bc_{1:t-1}) \approx \mathcal{N}(\bz|\vect{m}_{\bz,t},\bK_{\bz,t})q(\bf_{1:t-1})
\end{equation}
where $\vect{m}_{\bz,t}$ and $\bK_{\bz,t}$ are the sample mean-vector and covariance matrix estimated from the parameter states generated at the previous sequential step $t-1$. Combining \eqref{eqBerra6} and \eqref{eqBerra5} with \eqref{eqBerra7}, we can  again apply the updating steps for hyperparameters of Section \ref{secNum} at a cost captured by a fixed $k=1$. Here we note that the marginal term with functional values in \eqref{eqBerra6} is irrelevant; so is \eqref{eqBerra5} when updating $(m,\nsigma)$. When updating $\kappa$, the conditioning in \eqref{eqBerra7}  prescribes a strong informativeness about $\bf_t$ relative the likelihood, such that an update based on \eqref{eqnWhite} may be preferred to the more involved surrogate data slice sampling. 

In all, we arrive at a sequential procedure: given the $t-1$ sample $\{ \bf_{1:t-1}^{(l)},\kappa^{(l)},m^{(l)},\nsigma^{(l)} \}_{l=1}^M$, we continue to build up the vector of functional values by adding a sampled $\bf_t^{(l)}$ based on $\bf_{t-k:t-1}^{(l)}$ and then updating $\kappa^{(l)}$, $m^{(l)}$ and $\nsigma^{(l)}$ for  $l=1,\dots,M$. For the initial time step $t=1$, we use the prior methodology of Section \ref{secNum} (i.e. the generated sample of Section \ref{seqBerra}).

\begin{figure}
\makebox[\textwidth][c]{  
\includegraphics[scale=0.37,trim=0 0 20 0,clip]{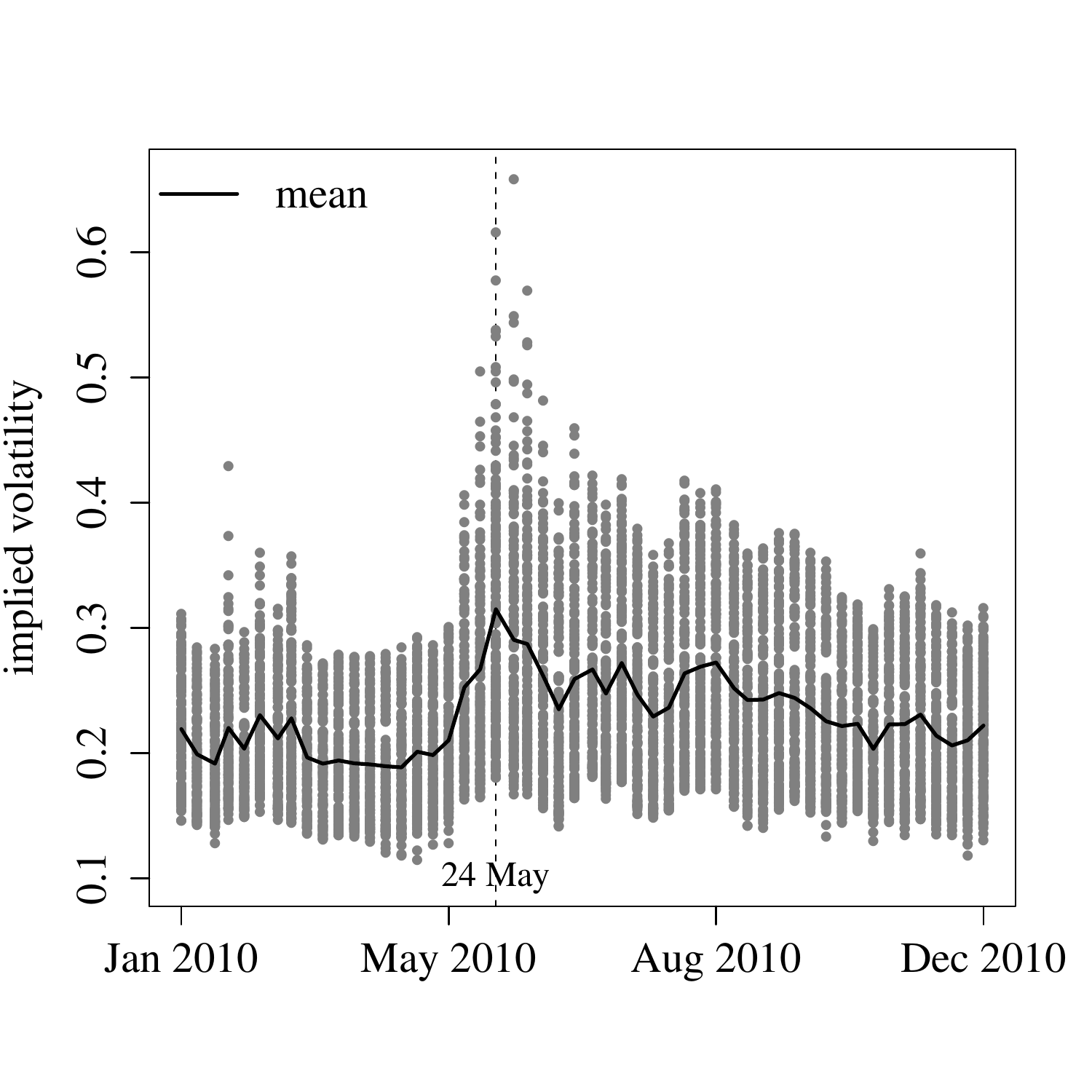} 
\includegraphics[scale=0.37,trim=50 50 50 50,clip]{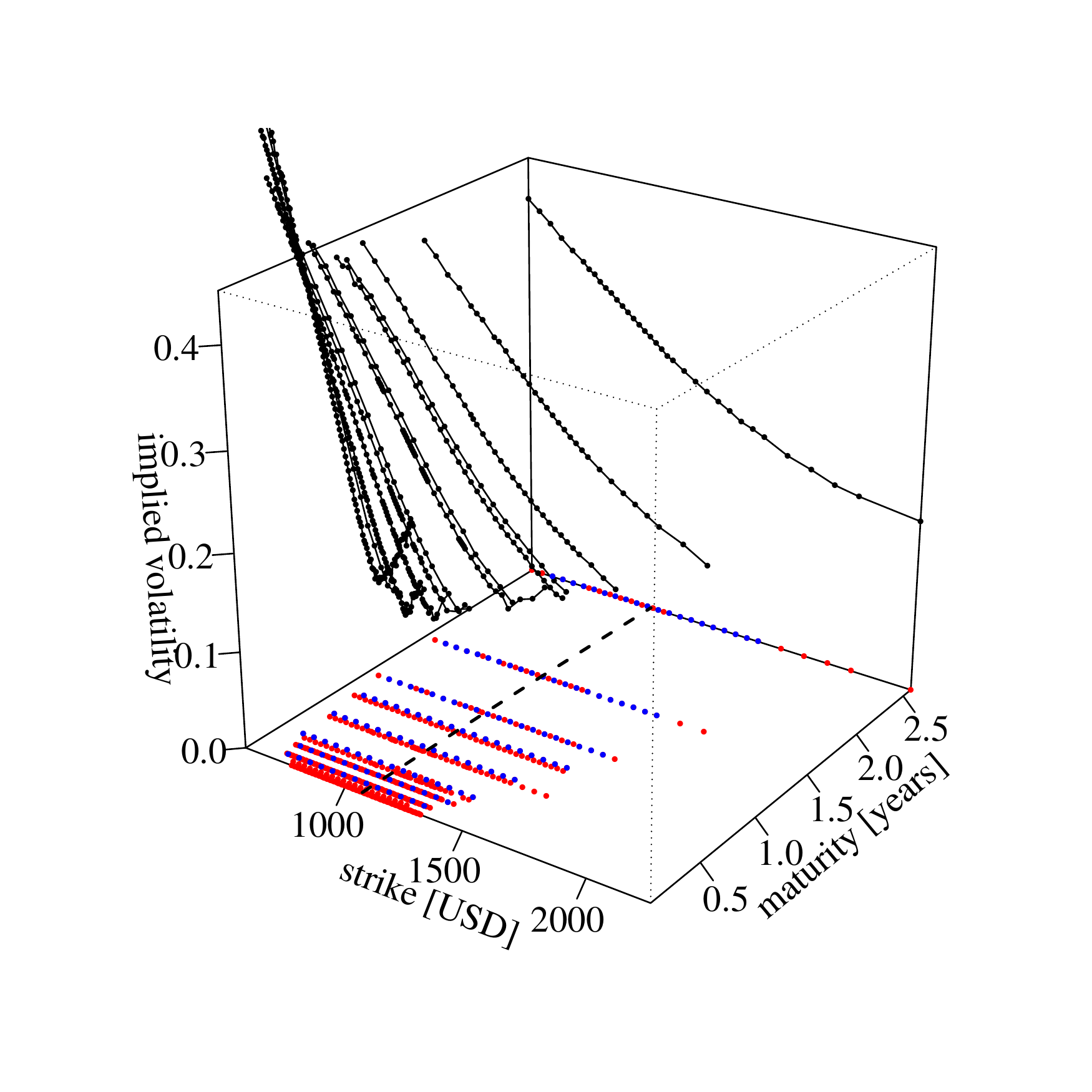}
\includegraphics[scale=0.37,trim=50 50 50 50,clip]{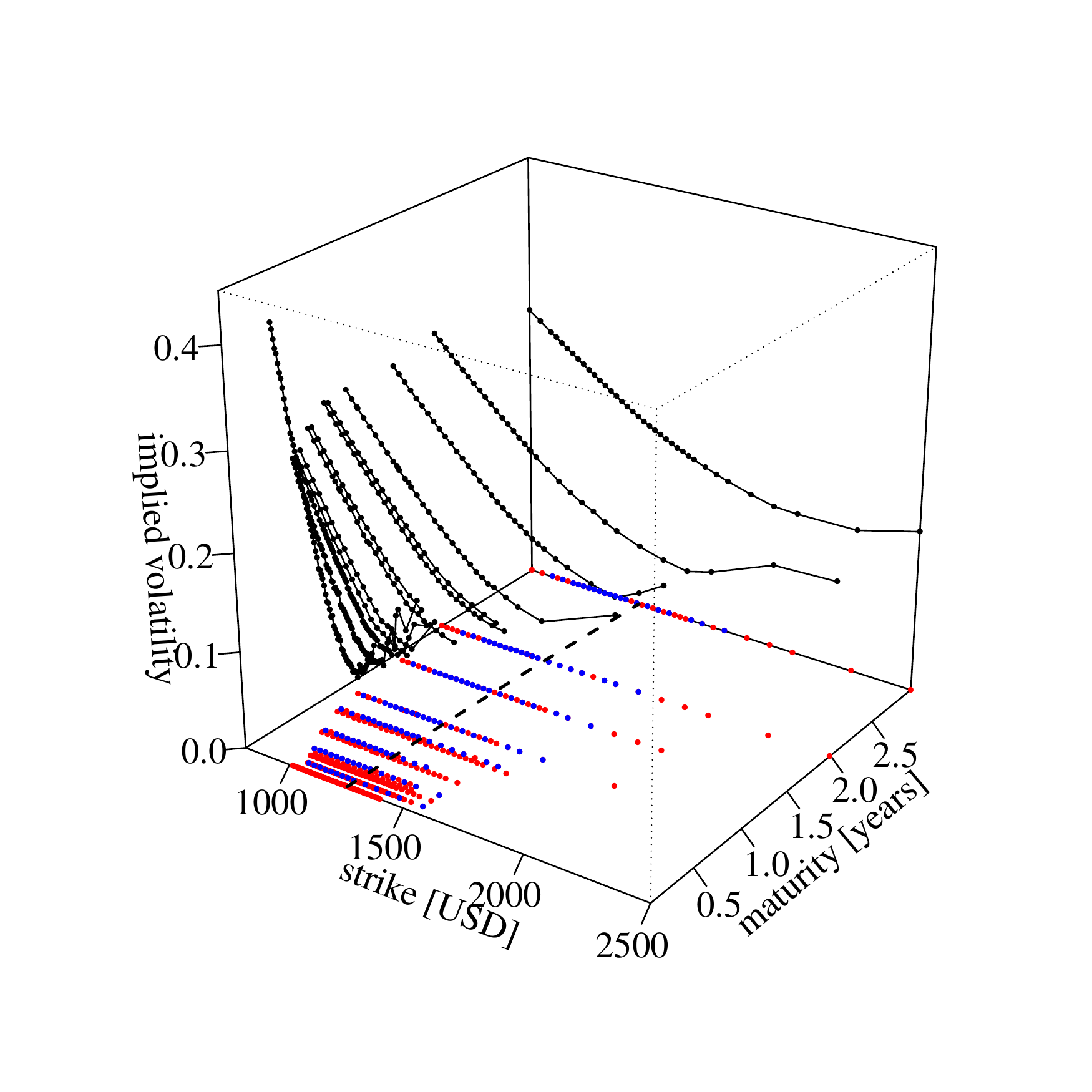}
}
\caption{Weekly data of S\&P 500 options from  4 January 2010 to 31 December 2010. \textbf{Left:} Implied volatilities from the market data $\{{\bc}_t\}_{t=1}^{\tau}$, each surface grouped together  by its observation time $t$. The black line shows the mean of each implied volatility surface. \textbf{Middle:} Implied volatility surface of 24 May 2015. This surface has the highest mean value. \textbf{Right:} Implied volatility surface of 27 December 2015 (the last date of the dataset).}
\label{fig5b}
\end{figure}

\subsection{Experiments}

\paragraph{Data and set-up}
We use market data of S\&P 500 call options from the period 4 January 2010 to 31 December 2010. We consider  weekly observations, i.e. 52 surfaces of call prices where each set   is the result from the data-preparation procedure of Section \ref{secBerra2}. For every observation time $t=1,\dots,52$, we construct a calibration set  $\{\bc_t,\bxh_t\}$ and input set $\bx_t$ to include 20 different strikes  and 8 maturities of the original  data. These are illustrated in Figure \ref{fig5b} in terms of  implied volatilities. In terms of strikes and maturities, the distribution of market observations $\bxh_t$ over the input set $\bx_t$ is fairly constant across time---similar to what is observed and discussed with Figure \ref{fig1} (left)---and the coverage, \#observations out of \#input-points, ranges from 60\% to 90\%. To set-up the model, we keep the hyperprior assumptions of Section \ref{secBerra2} along with $l_{t,\max}=1$ for the calendar-time length-scale. We thus  scale the observation times to lie in the unit interval.

\begin{figure}
\makebox[\textwidth][c]{  
\includegraphics[scale=0.4,trim=60 50 60 20,clip]{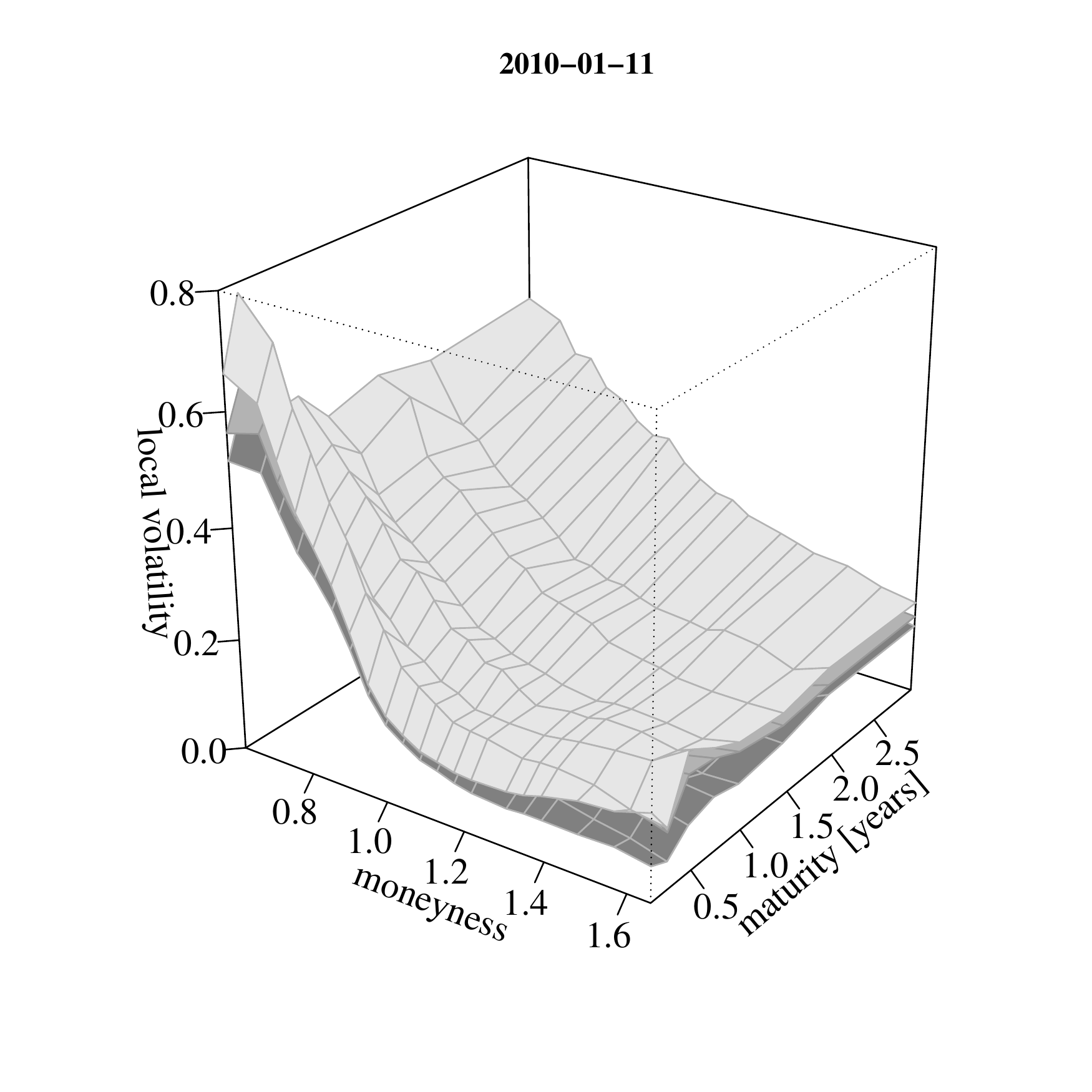}
\includegraphics[scale=0.4,trim=60 50 60 20,clip]{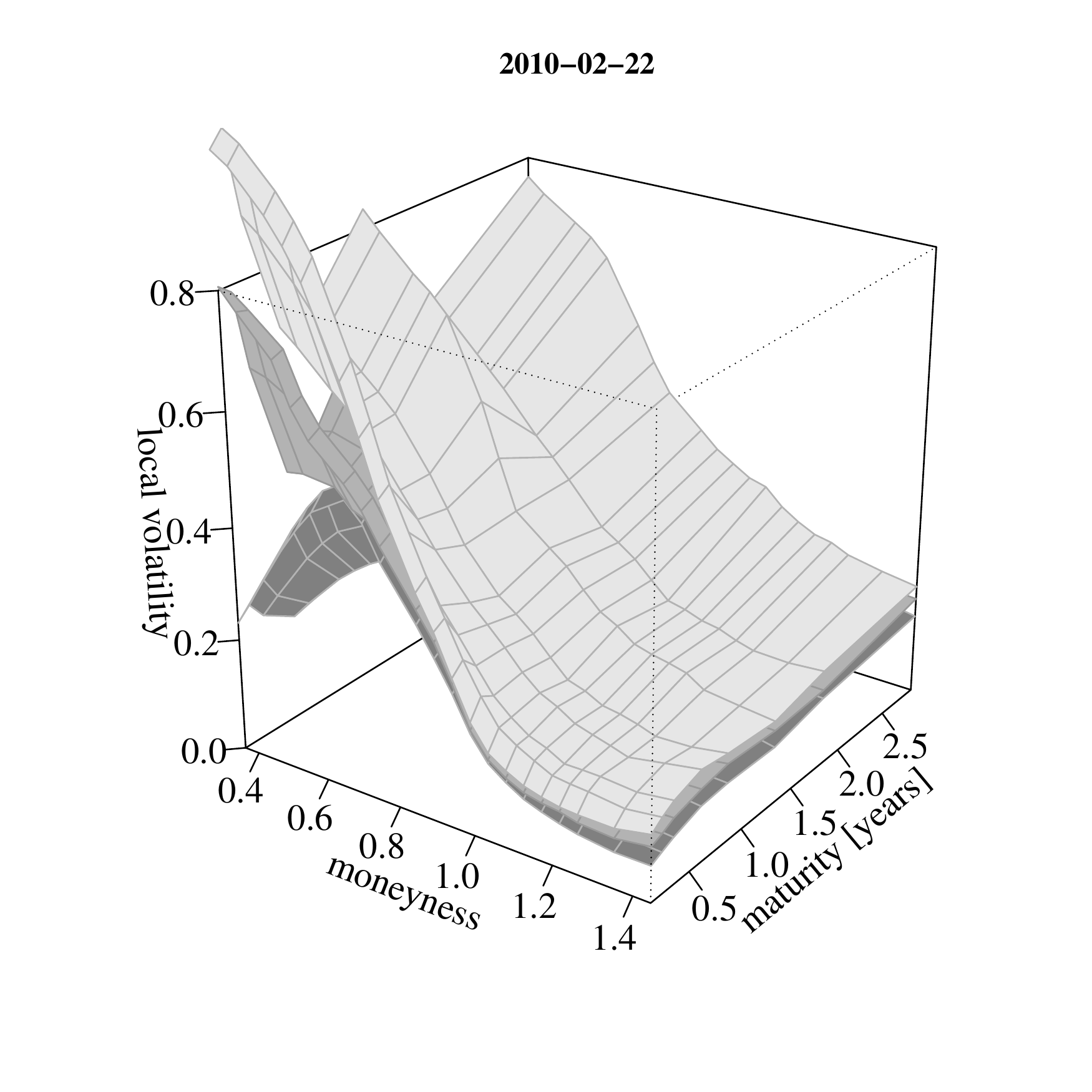}
\includegraphics[scale=0.4,trim=60 50 60 20,clip]{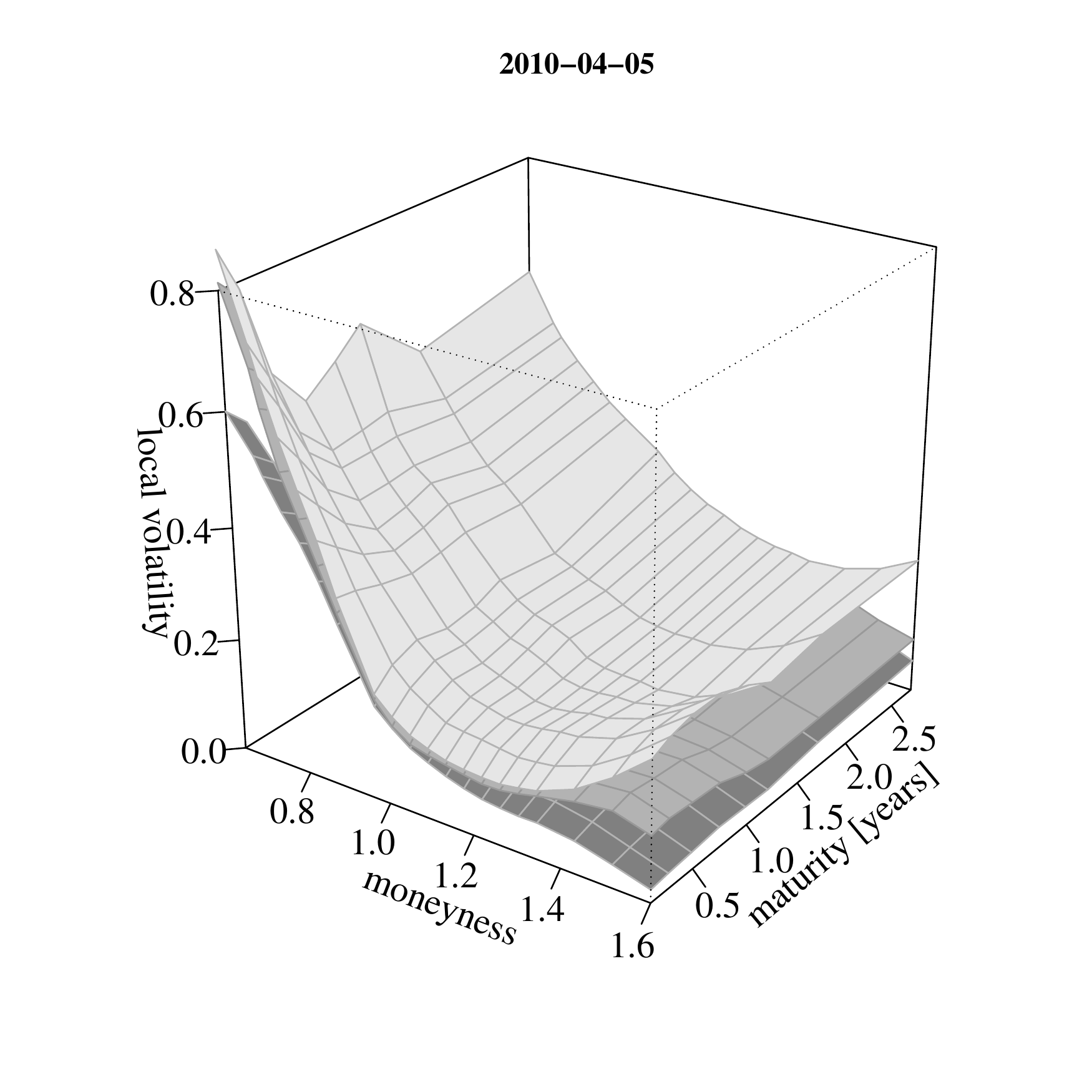}
}
\makebox[\textwidth][c]{  
\includegraphics[scale=0.4,trim=60 50 60 20,clip]{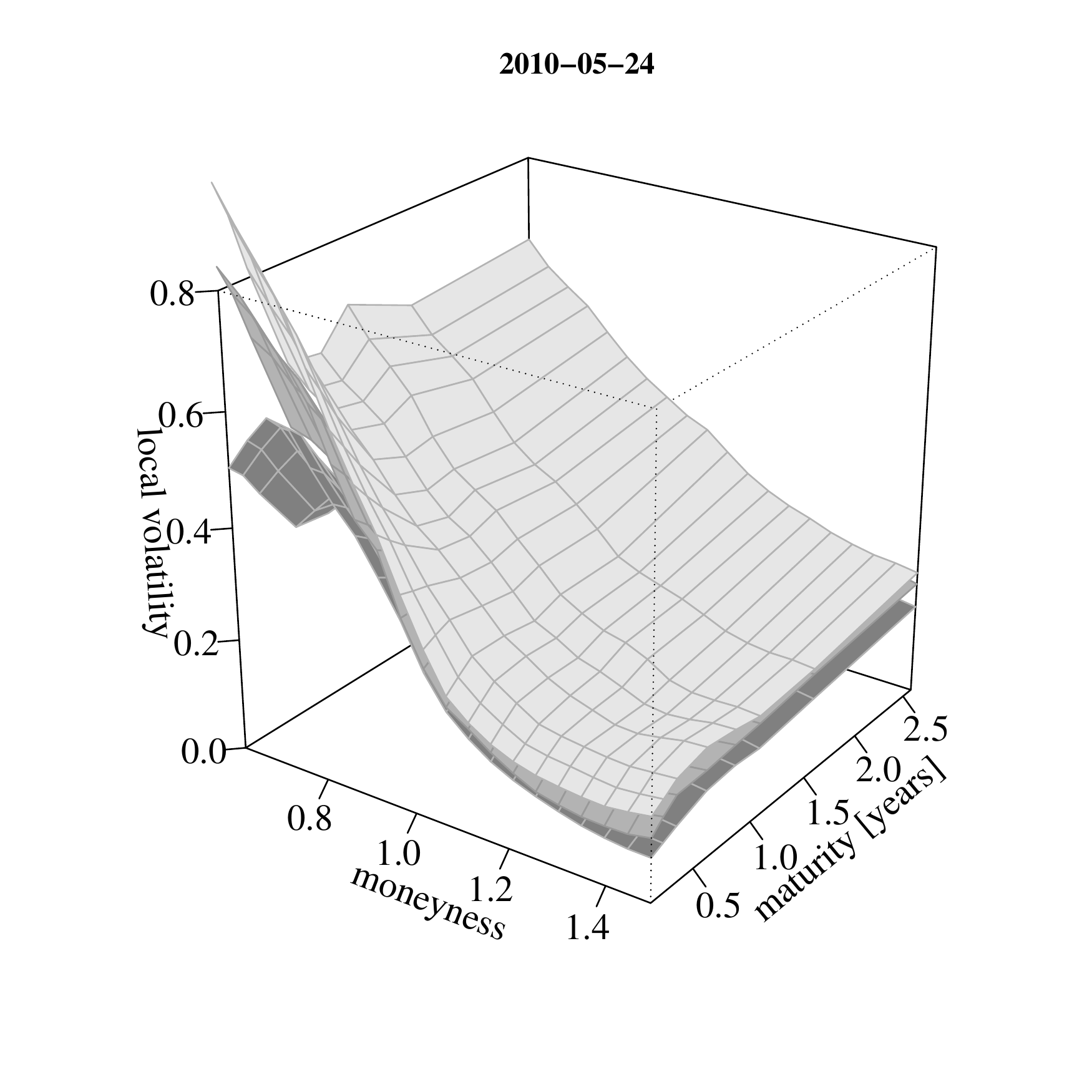}
\includegraphics[scale=0.4,trim=60 50 60 20,clip]{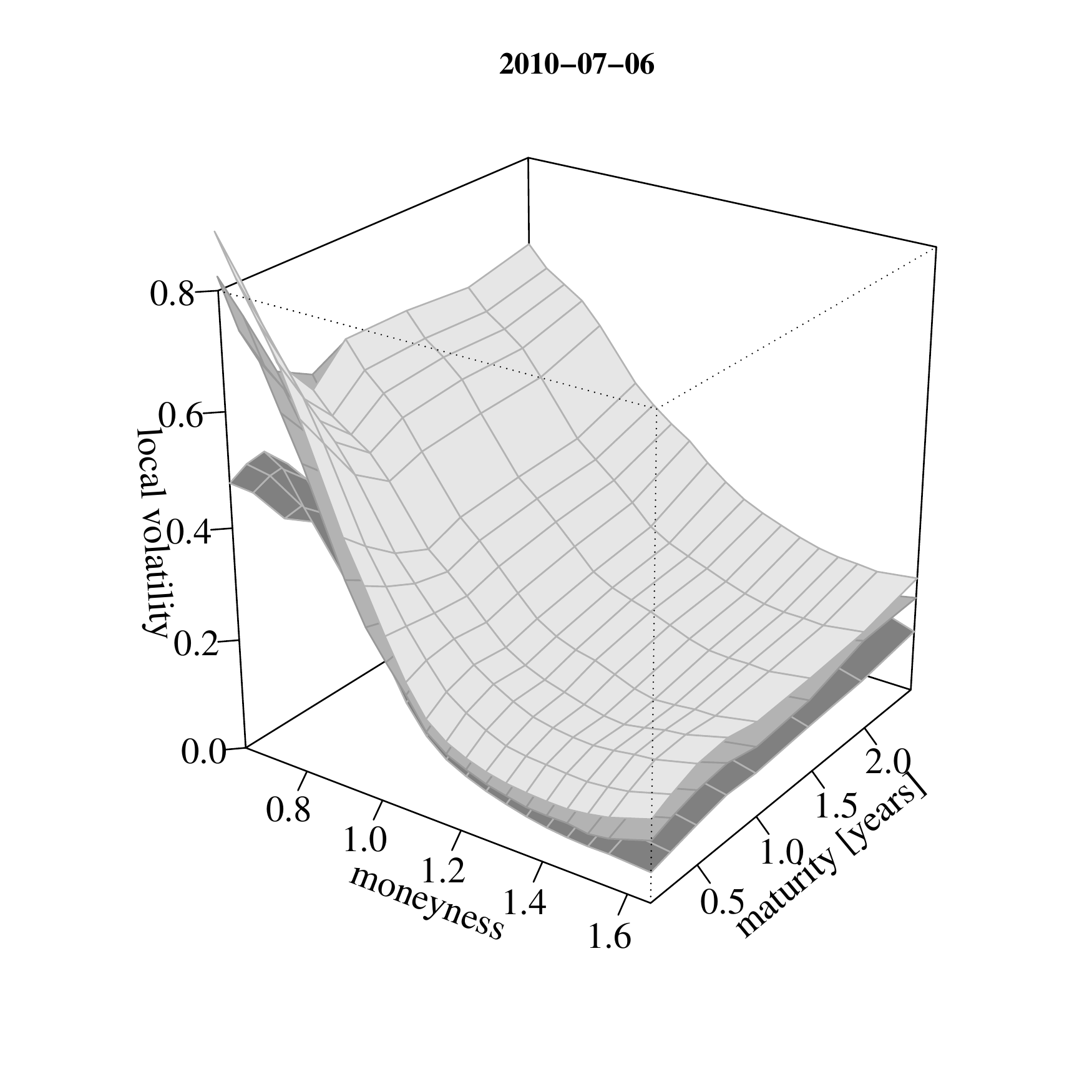}
\includegraphics[scale=0.4,trim=60 50 60 20,clip]{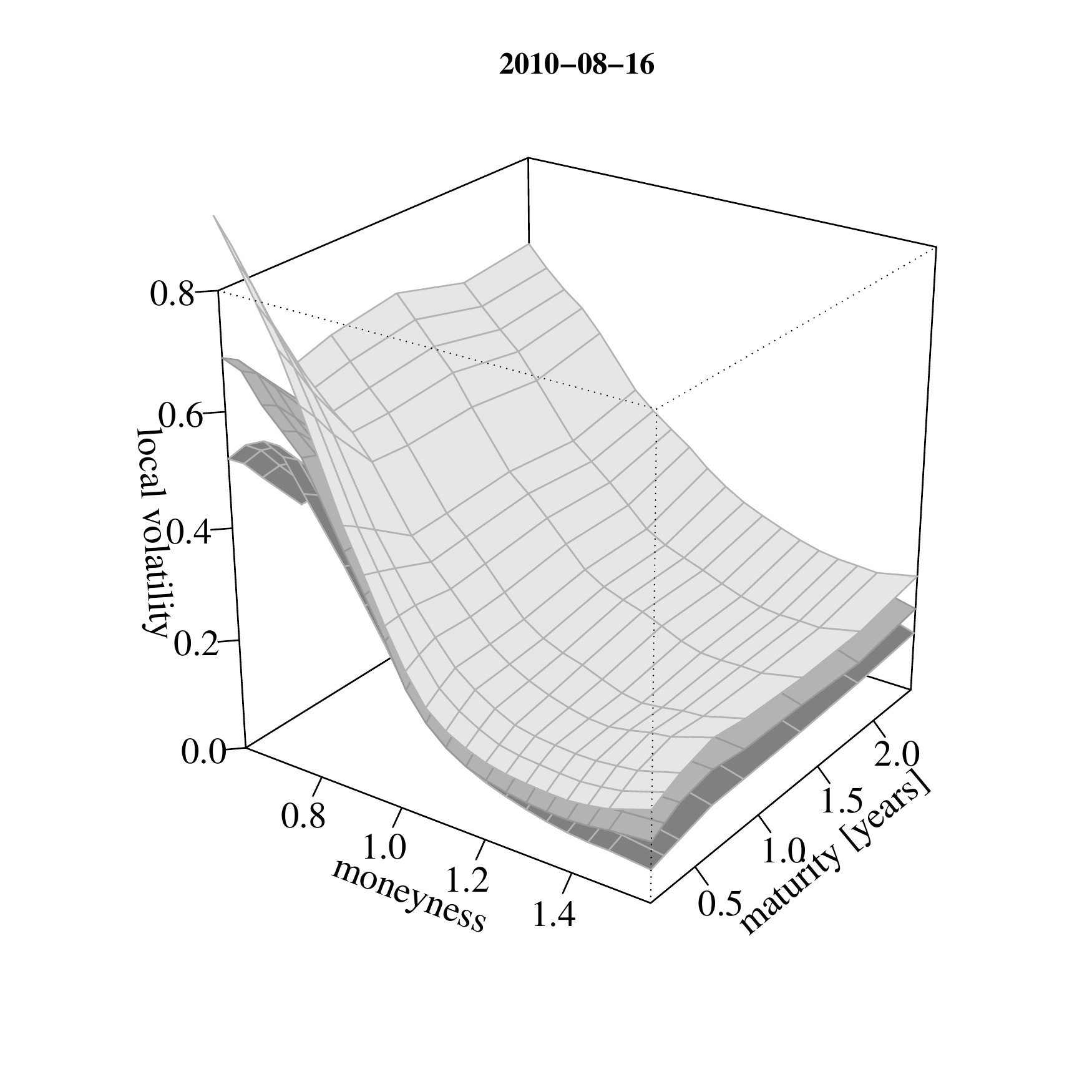}
}
\makebox[\textwidth][c]{  
\includegraphics[scale=0.4,trim=60 50 60 20,clip]{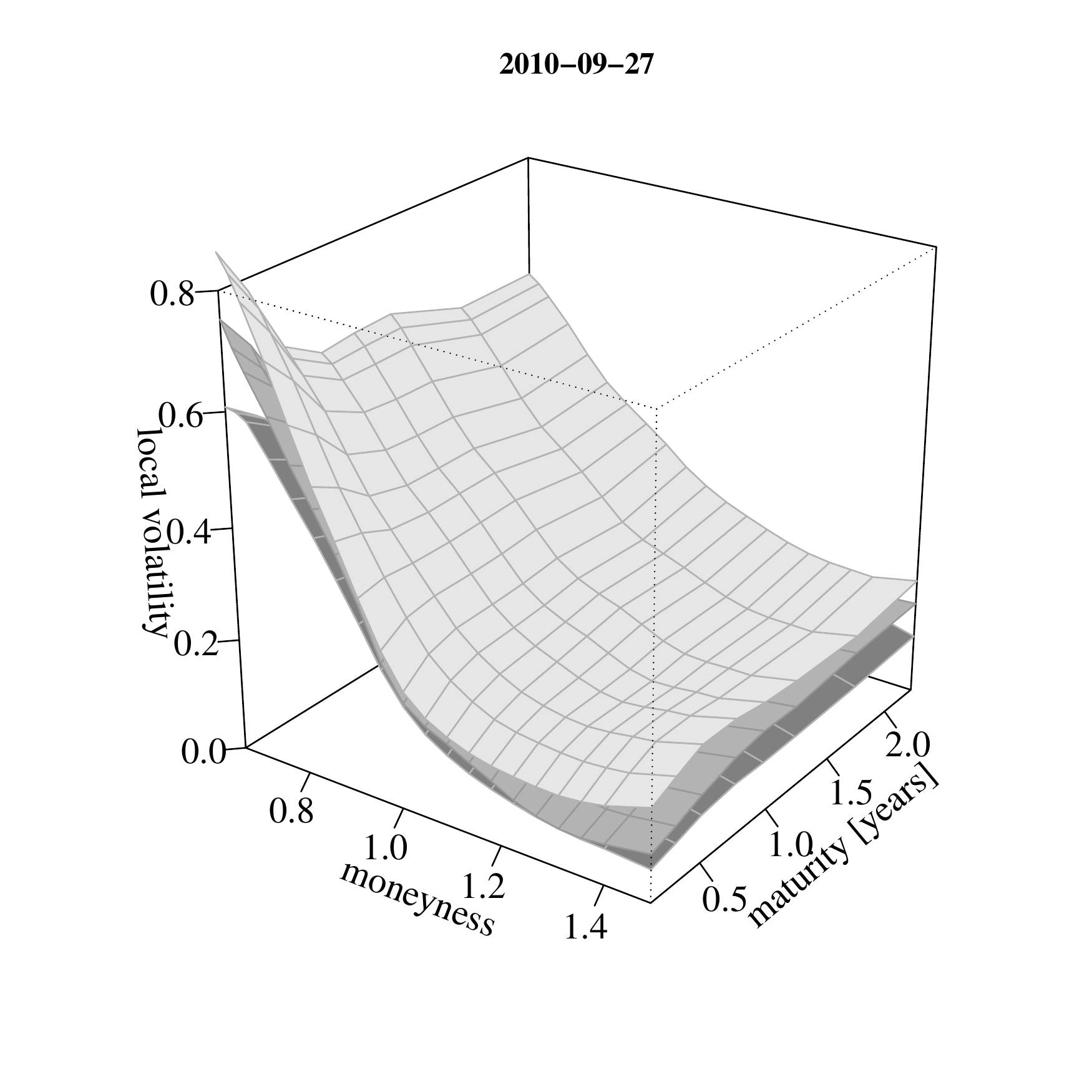}
\includegraphics[scale=0.4,trim=60 50 60 20,clip]{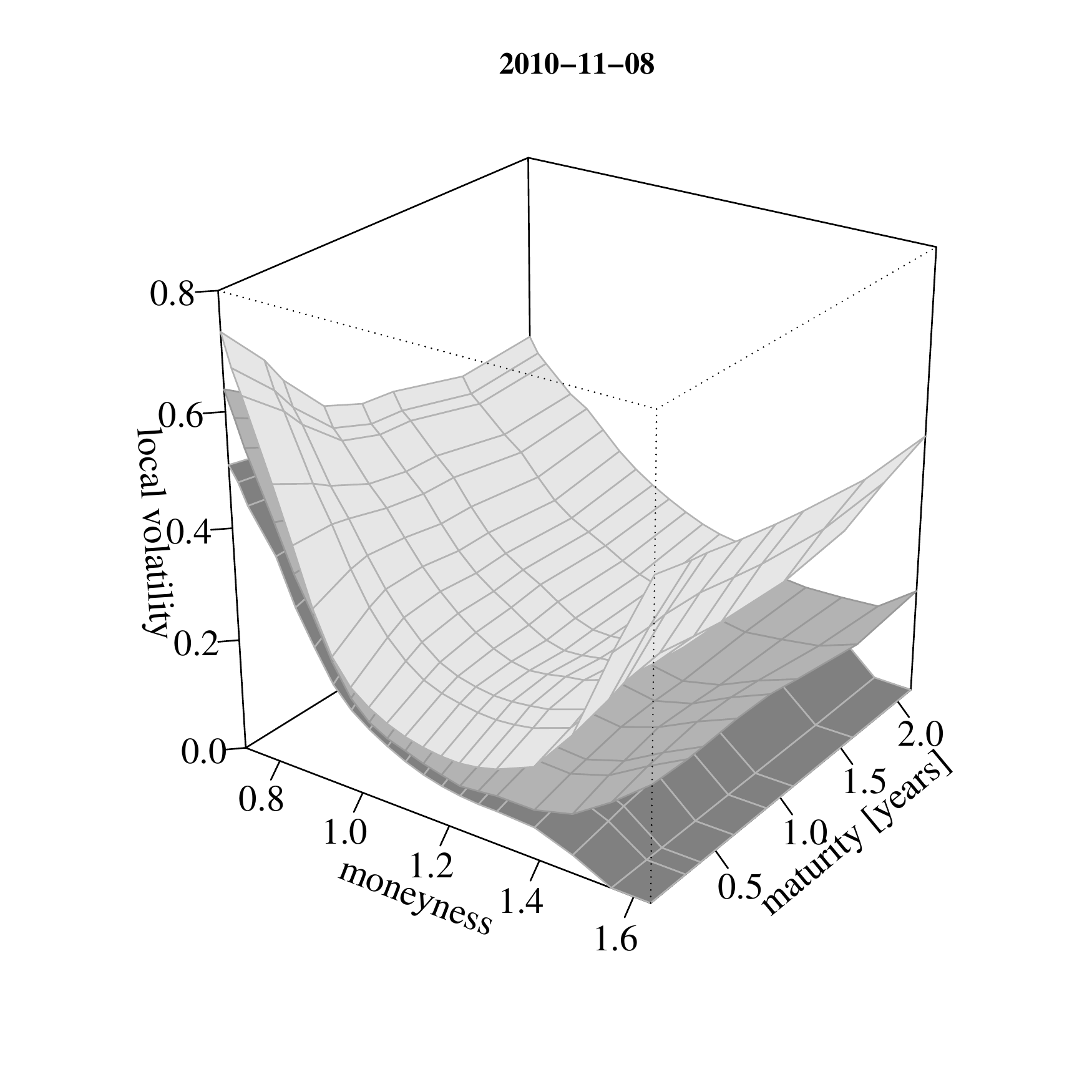}
\includegraphics[scale=0.4,trim=60 50 60 20,clip]{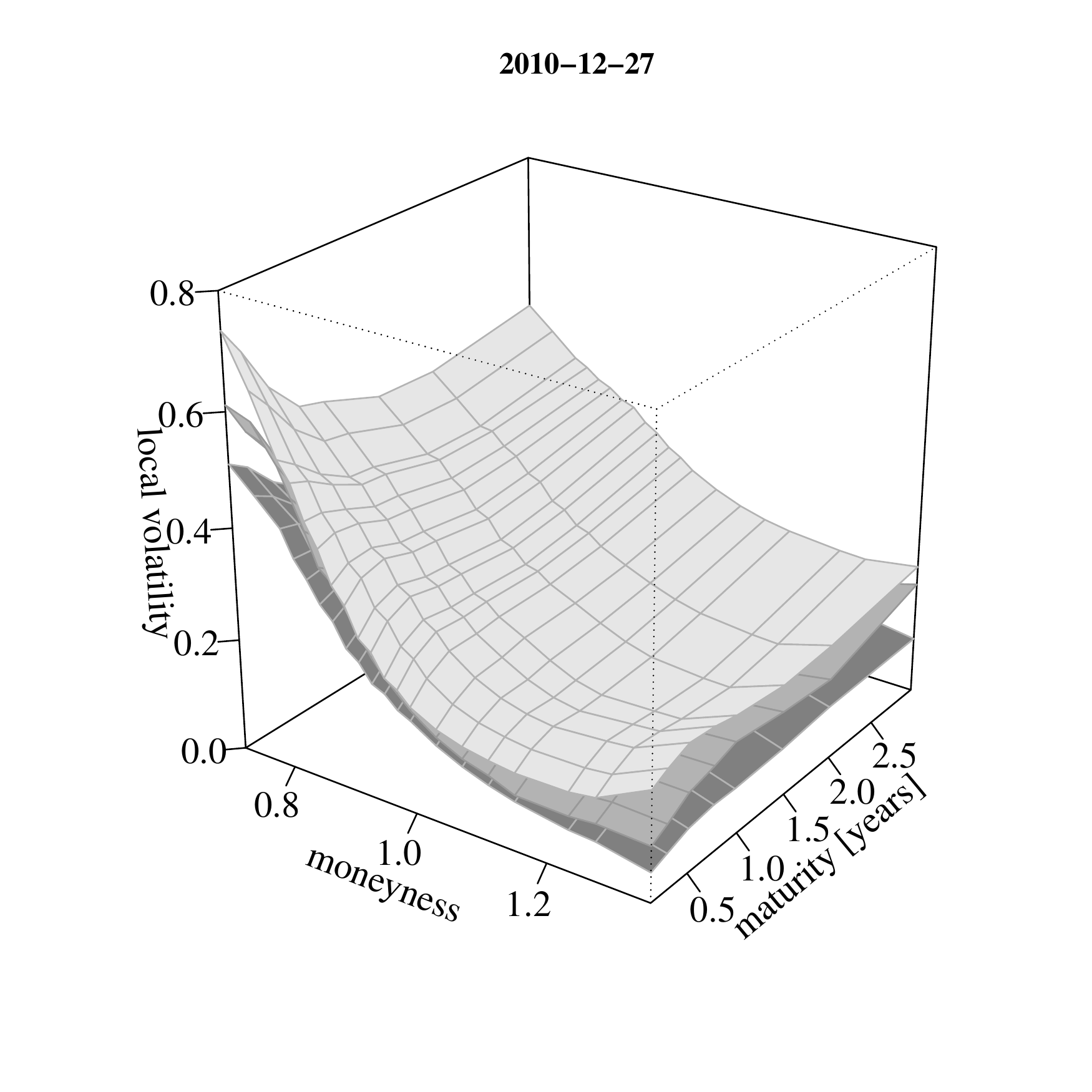}
}
\caption{Local volatility calibrated to S\&P 500 call prices from different dates during 2010. The figures show credible envelopes and MAP estimates as extracted from the posterior sample of each time point. }
\label{fig6b}
\end{figure}

\paragraph{Calibration} We run the sequential method of the previous section to obtain a sample $\{\bf_{1:52}^{(l)}\}_{l=1}^M$ where a set of surfaces $\bf_{1:52}^{(l)} = \{\bf_{1}^{(l)},\dots,\bf_{52}^{(l)}\}$ is the $l^{\text{th}}$ generated state of a local volatility time-series. Similarly, for a single time $t$  the sample $\{\bf^{(l)}_t\}_{l=1}^M$  represents the calibration---the posterior over local volatility---as of that date. Figure \ref{fig6b} illustrates posteriors at nine different time points as represented by their credible envelopes. 

\begin{figure}
\makebox[\textwidth][c]{  
\includegraphics[scale=0.5,trim=0 0 10 0,clip]{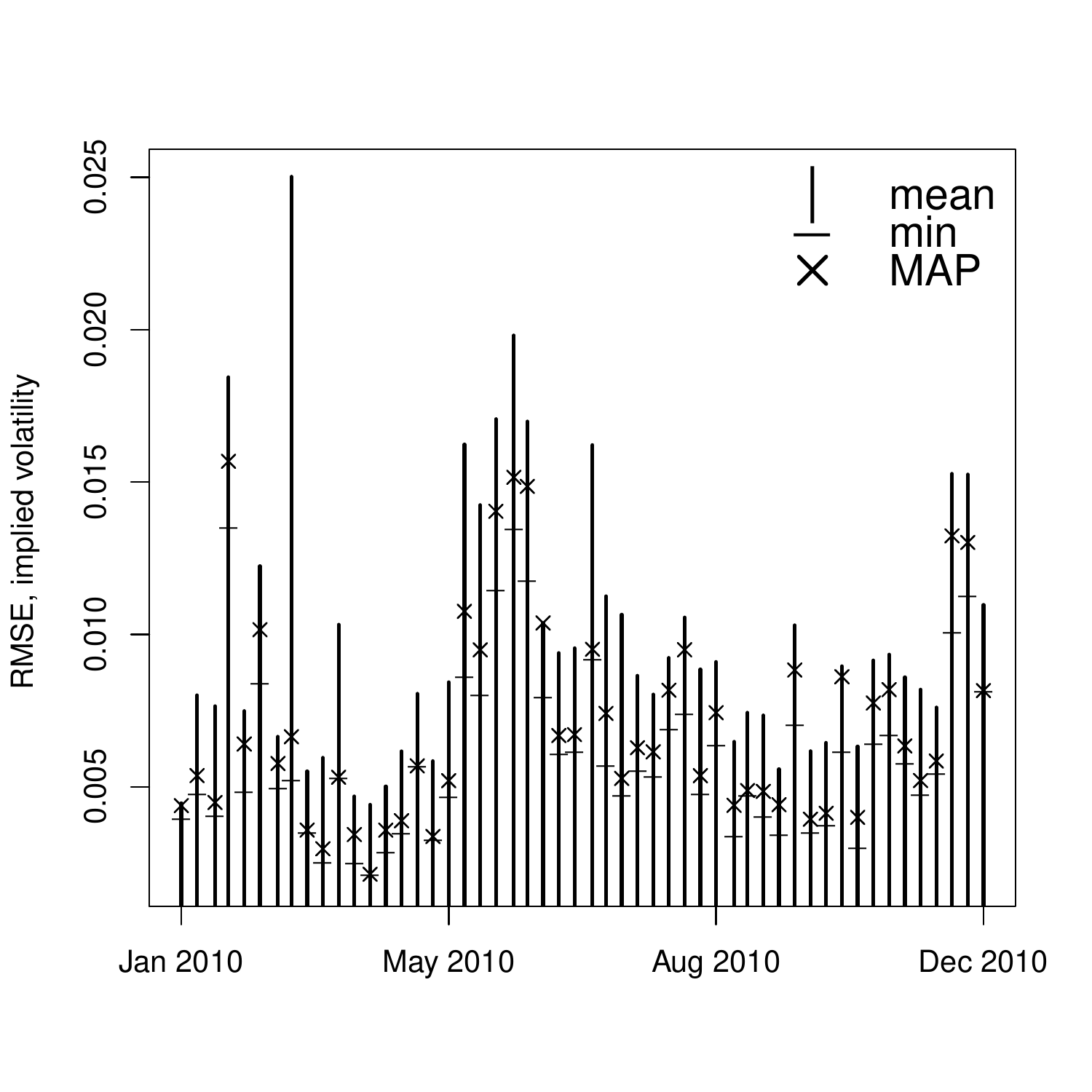} 
\includegraphics[scale=0.5,trim=0 0 10 0,clip]{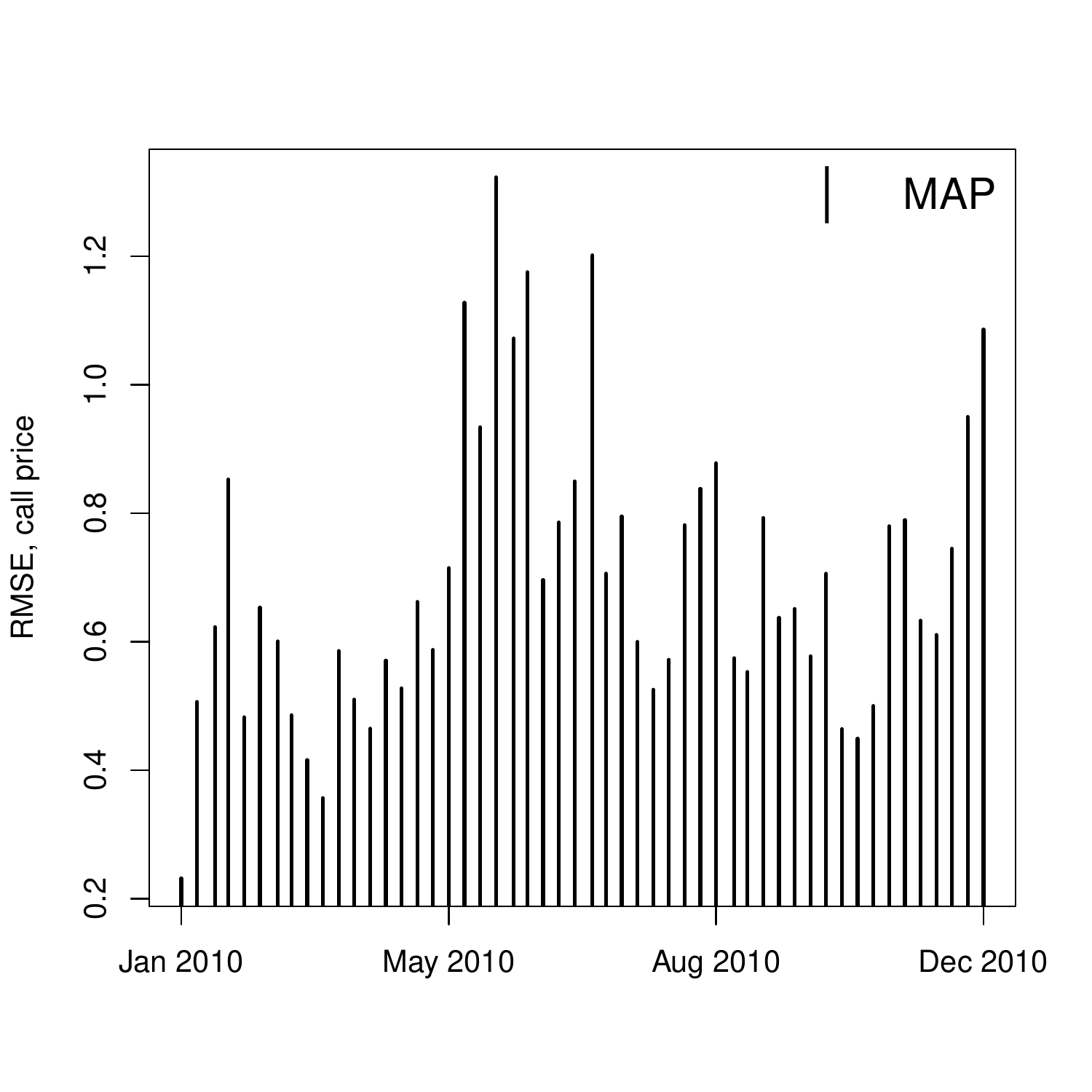} 
}
\caption{Root mean squared errors from the local volatility sample at each time $t$. \textbf{Left:} Errors in terms of implied volatility from the MAP surface along with the mean and minimum across the sample. \textbf{Right:} Errors in terms of call prices of the MAP local volatility.}
\label{fig5bbb}
\end{figure}

To assess the calibration in term of re-pricing performance, we consider the  root mean squared error from each sampled local volatility surface 
\begin{equation}\label{eqrmse}
{RMSE}(\bf_t^{(l)}) = \sqrt{ \frac{1}{n_t}\sum_{i=1}^{n_t} (\sigma_{\text{BS}}(\hat{x}_{t,i},\bf_t^{(l)})-\hat{\sigma}_{\text{BS},t,i})^2 }
\end{equation}
where $\sigma_{\text{BS}}(\hat{x}_i,\bf)$ and $\hat{\sigma}_{\text{BS},i}$ are the volatilities implied by $C(\hat{x}_i,\bf)$ and $\hat{c}_i$, respectively. Figure \ref{fig5bbb} (left) shows the root mean squared error obtained from the MAP surface for each $t$ along with the mean and minimum across each sample (the latter not necessarily the same as the error achieved by the MAP estimate). The minimum measure is fairly stable over time which indicates that the sequential sampling method yields satisfying results. In particular, note that minimal errors of the entire period do not appear for the $t=1$ sample.  

Figure \ref{fig5bbb} (right) shows the corresponding errors in terms of call prices (for which MAP and minimum coincides). Here, the initial sample $t=1$ produces the smallest (minimum) error since it results from a Markov chain with many more states (50,000 vs. 1000) and as it has a much less informative prior over both $\bf$ and its hyperparameters while consecutive samples draw on information from their predecessors. After all, we are interested in inferring each individual local volatility surface as well as their dependency over time. To be able to do the latter, especially from a computational point of view, we must expect to sacrifice precision in the former.

\begin{figure}
\centering
\includegraphics[scale=0.5,trim=  0 65 25 50,clip]{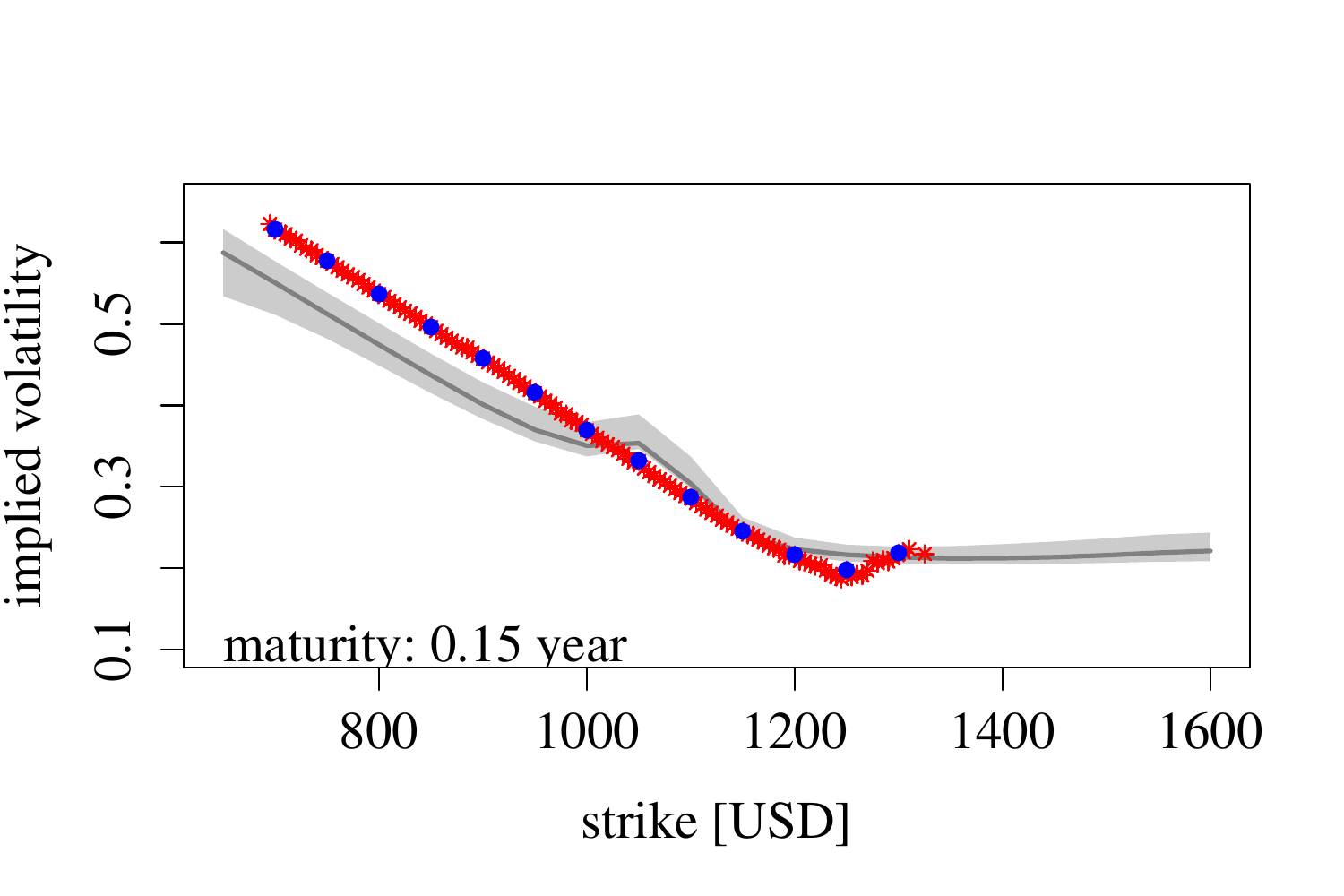} 
\includegraphics[scale=0.5,trim=46 65 25 50,clip]{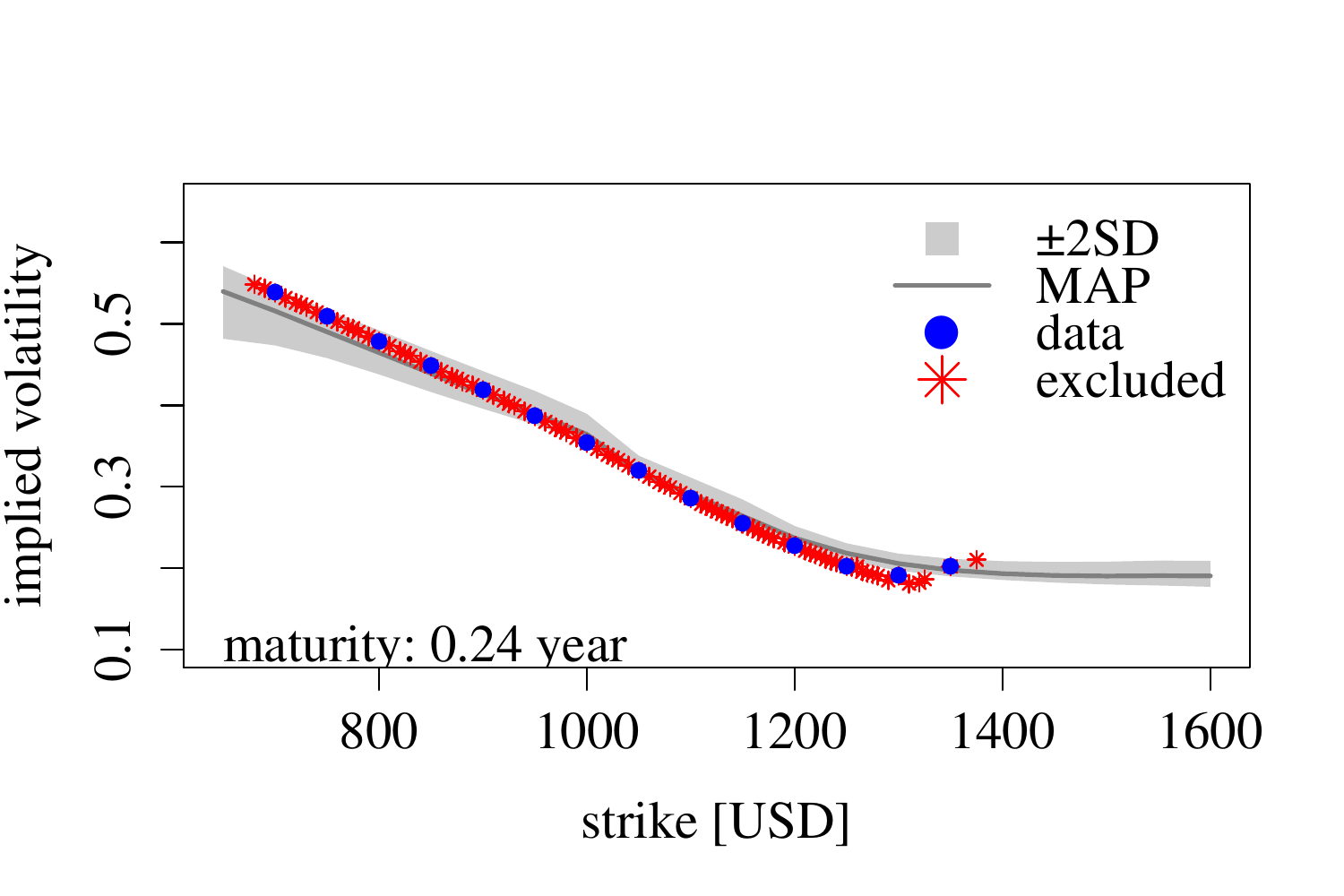} \\
\includegraphics[scale=0.5,trim=  0 65 25 50,clip]{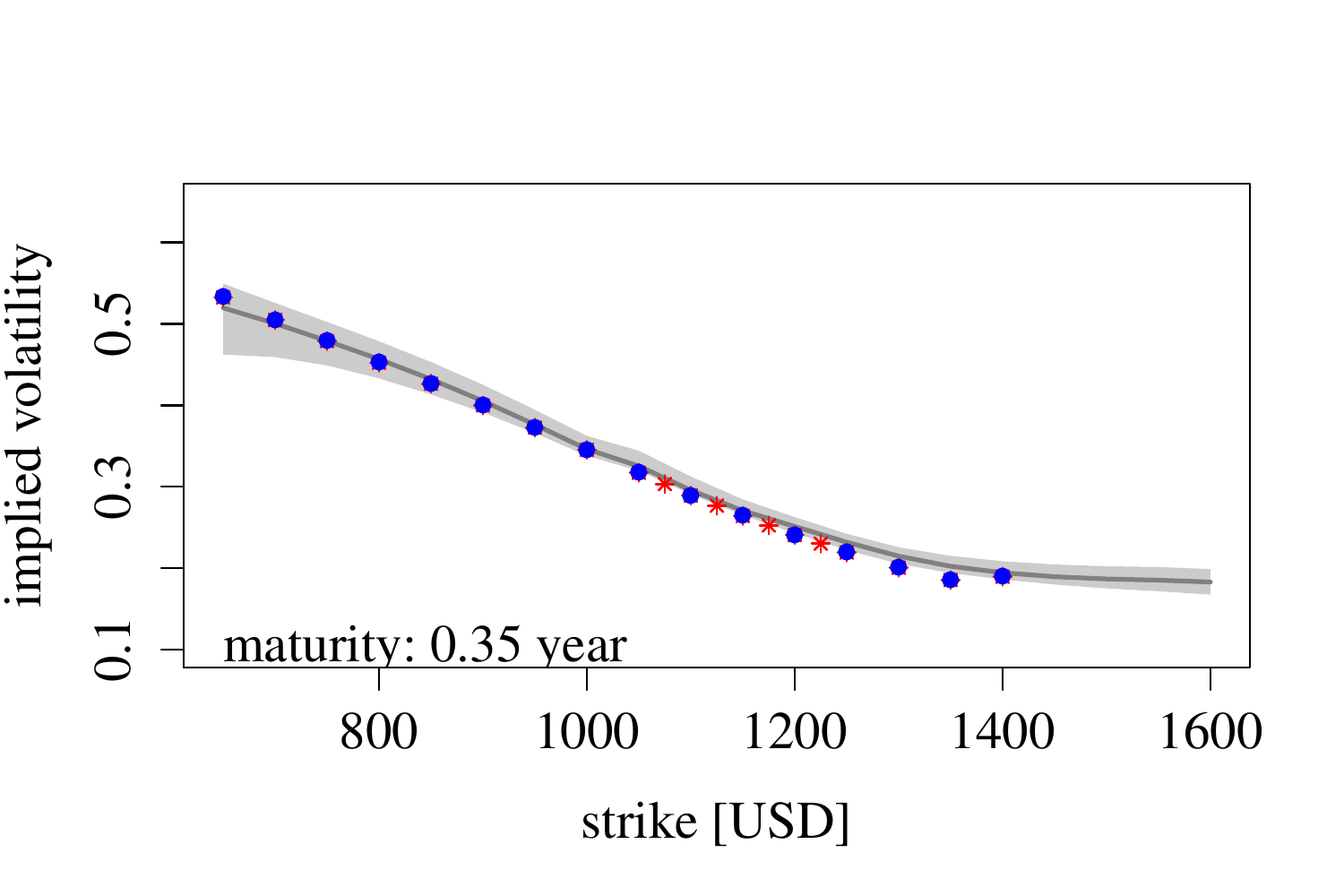} 
\includegraphics[scale=0.5,trim=46 65 25 50,clip]{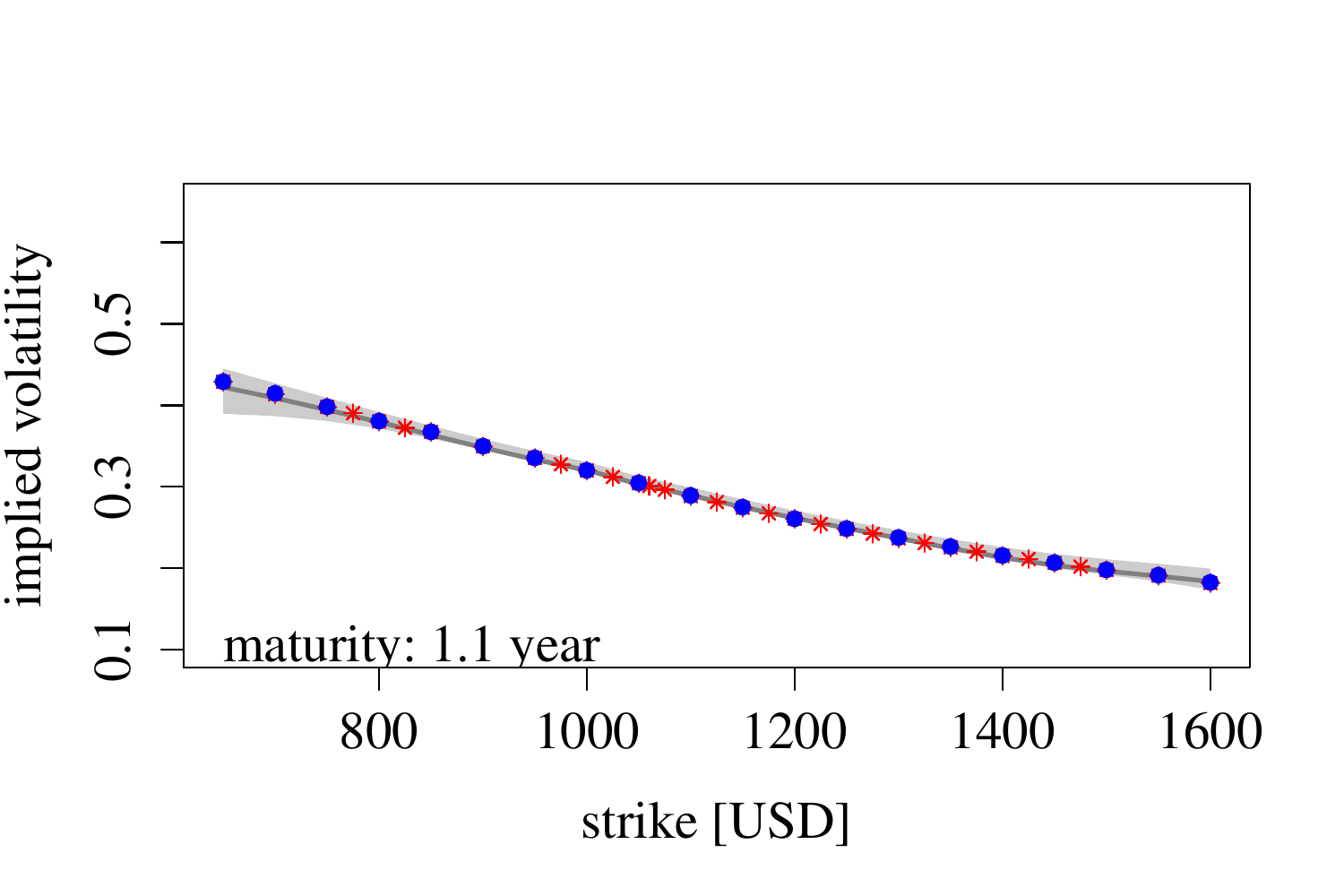} \\
\includegraphics[scale=0.5,trim=  0 10 25 50,clip]{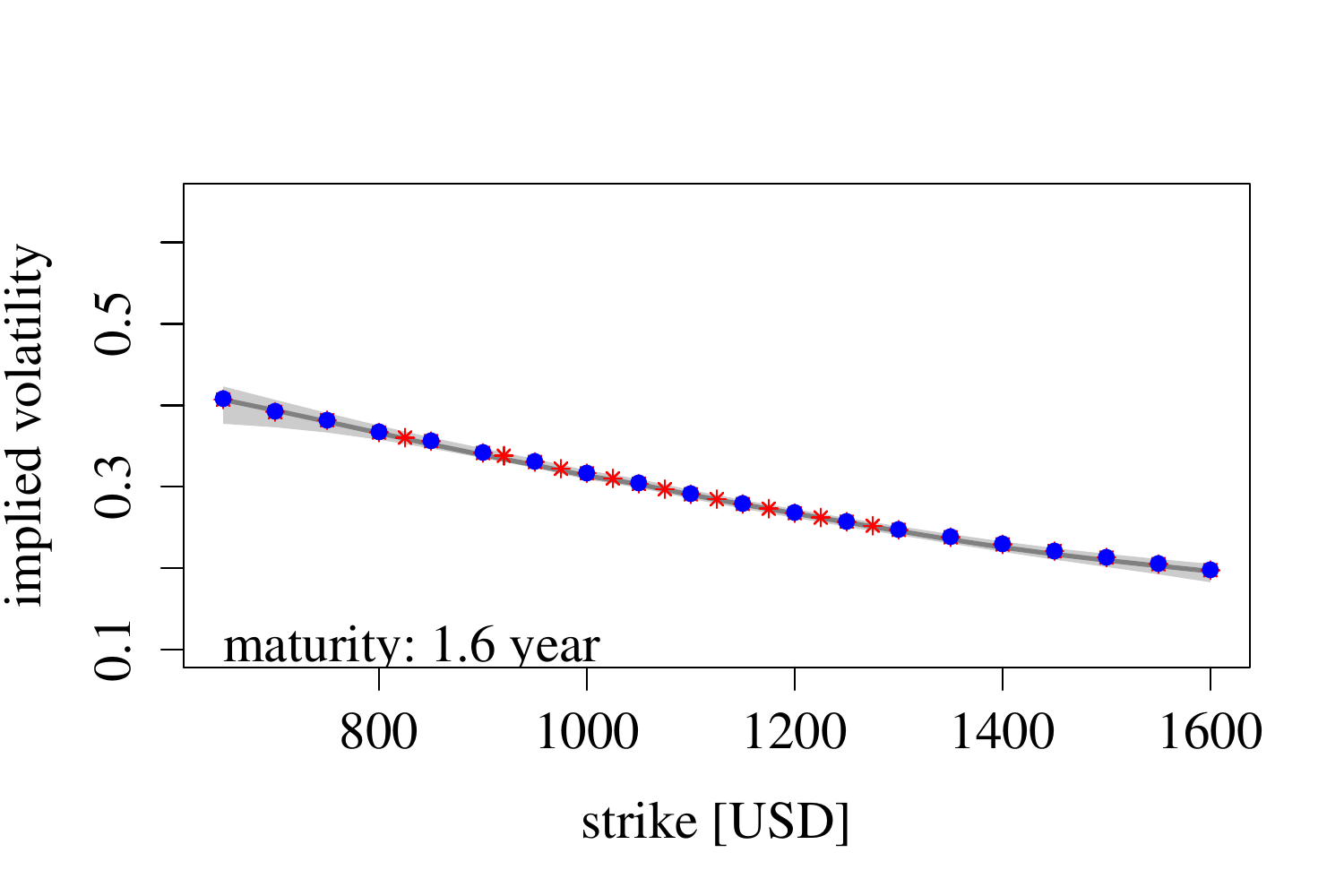} 
\includegraphics[scale=0.5,trim=46 10 25 50,clip]{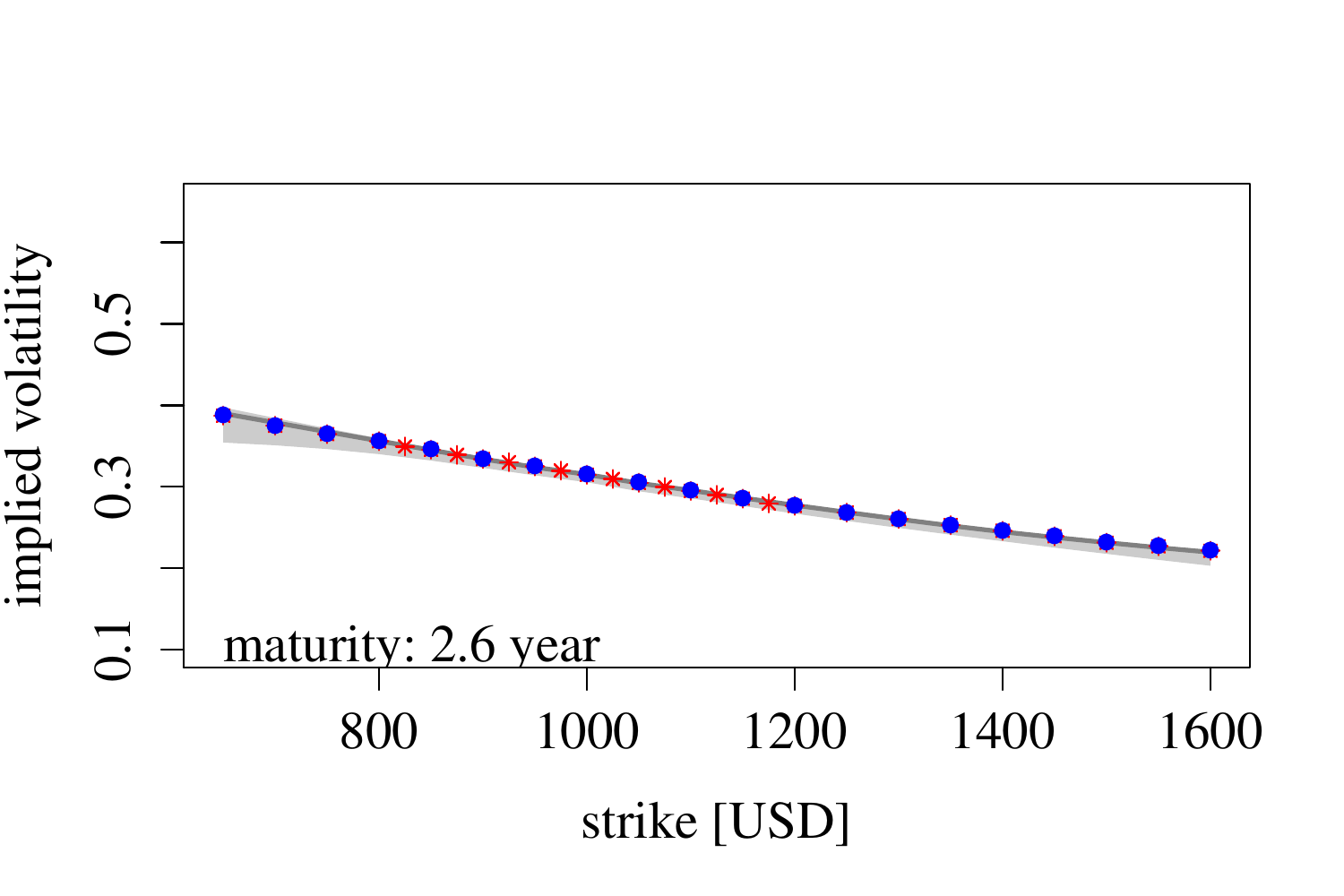} 
\caption{Posterior distribution over  implied volatilities on  24 May 2010.}
\label{fig333}
\end{figure}

The poorest calibration in terms of the largest call-price RMSE (at MAP) appears on  24 May 2010. This coincides with the largest observed average implied volatility, see left and middle panes of Figure \ref{fig5b}. The latter followed a rapid rise in volatilities after the  \textit{Flash Crash} of 6 May 2010. The overall calibration error mostly stem from errors at the shorter end of  maturities (as also observed in Section \ref{seqBerra})---see Figure \ref{fig333} for implied volatilities. 

\paragraph{Prediction at a future time}
The sequential calibration approach can be viewed as day-to-day (in our case weekly) re-calibration aided by previous date's estimate acting as a regularisation term. What more is, we  infer dependency over time in our probabilistic model. We can therefore consider 
posterior predictions at future time-points of the local volatility surface along with associated uncertainty measures. To this end, the predictive distribution is obtained by marginalisation. 

Suppose we have  observations up to and including time  $\tau$, i.e. a time-series  of market data $\{\bc_t,\bxh_{t}\}_{t=1}^{\tau}$. The $s$-step forward prediction of local volatility   at time $\tau+s$ and unseen inputs  $\bxs_{\tau+s}$ is then  given by
\begin{equation}\label{eqPRED}
p(\bfs_{\tau+s}|\bc_{1:\tau}) = \int \underbrace{	  p(\bfs_{\tau+s}|\bf_{1:\tau},\kappa,m) }_{\text{conditional prior}} \underbrace{p(\bf_{1:\tau},\kappa,m,\nsigma|\bc_{1:\tau})}_{\text{time-}\tau\text{ posterior}} d\kappa \, dm \, d\nsigma \, d\bf_{1:\tau}.
\end{equation}
In \eqref{eqPRED} we  use  that $p(\bfs_{\tau+s}|\bf_{1:\tau},\kappa,m,\nsigma,\bc_{1:\tau}) = p(\bfs_{\tau+s}|\bf_{1:\tau},\kappa,m)$, due to the likelihood \eqref{eqBerrall} factorising over time. 

\begin{figure}[!t]
\centering 
\includegraphics[scale=0.5,trim=  0 65 25 50,clip]{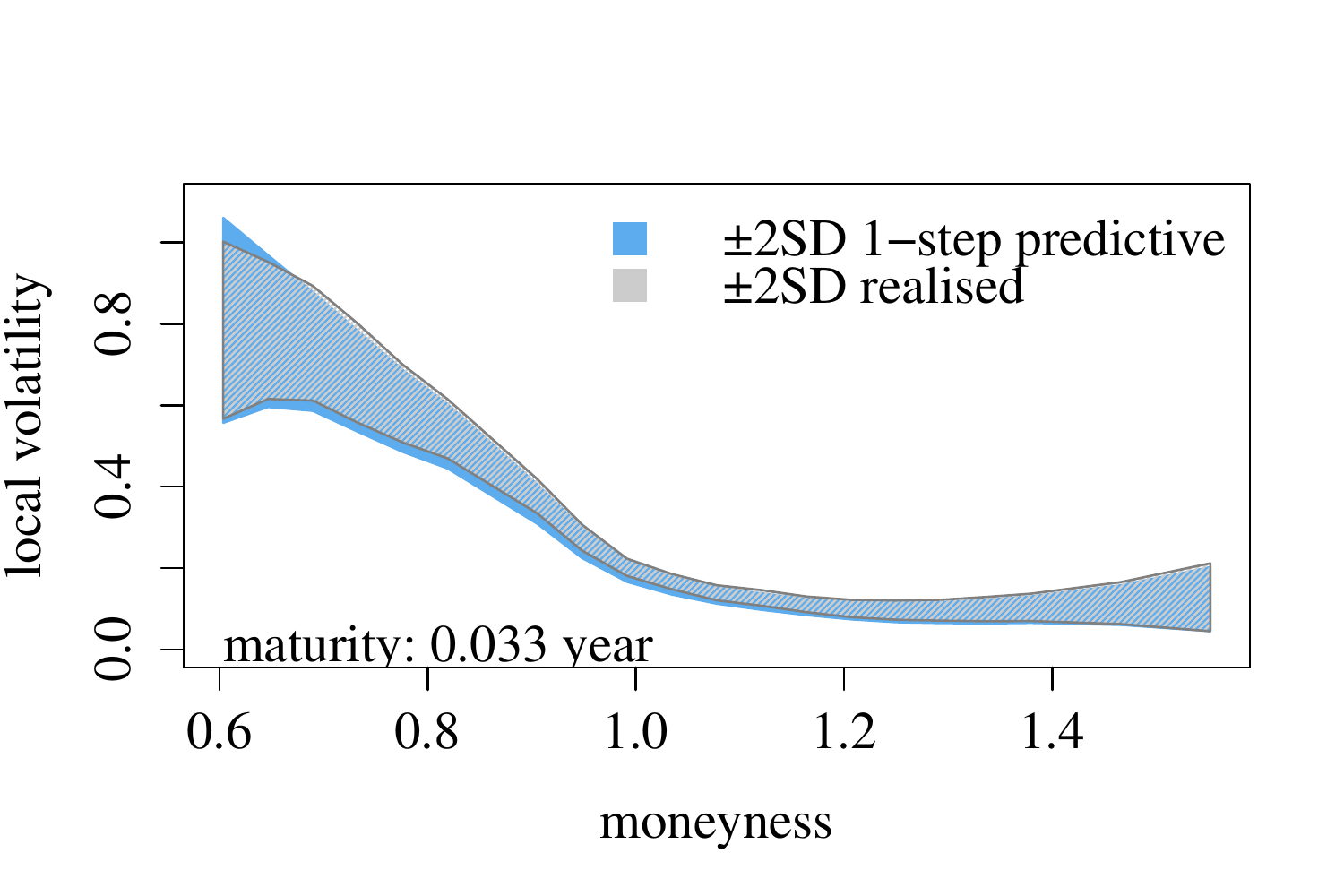} 
\includegraphics[scale=0.5,trim=  0 65 25 50,clip]{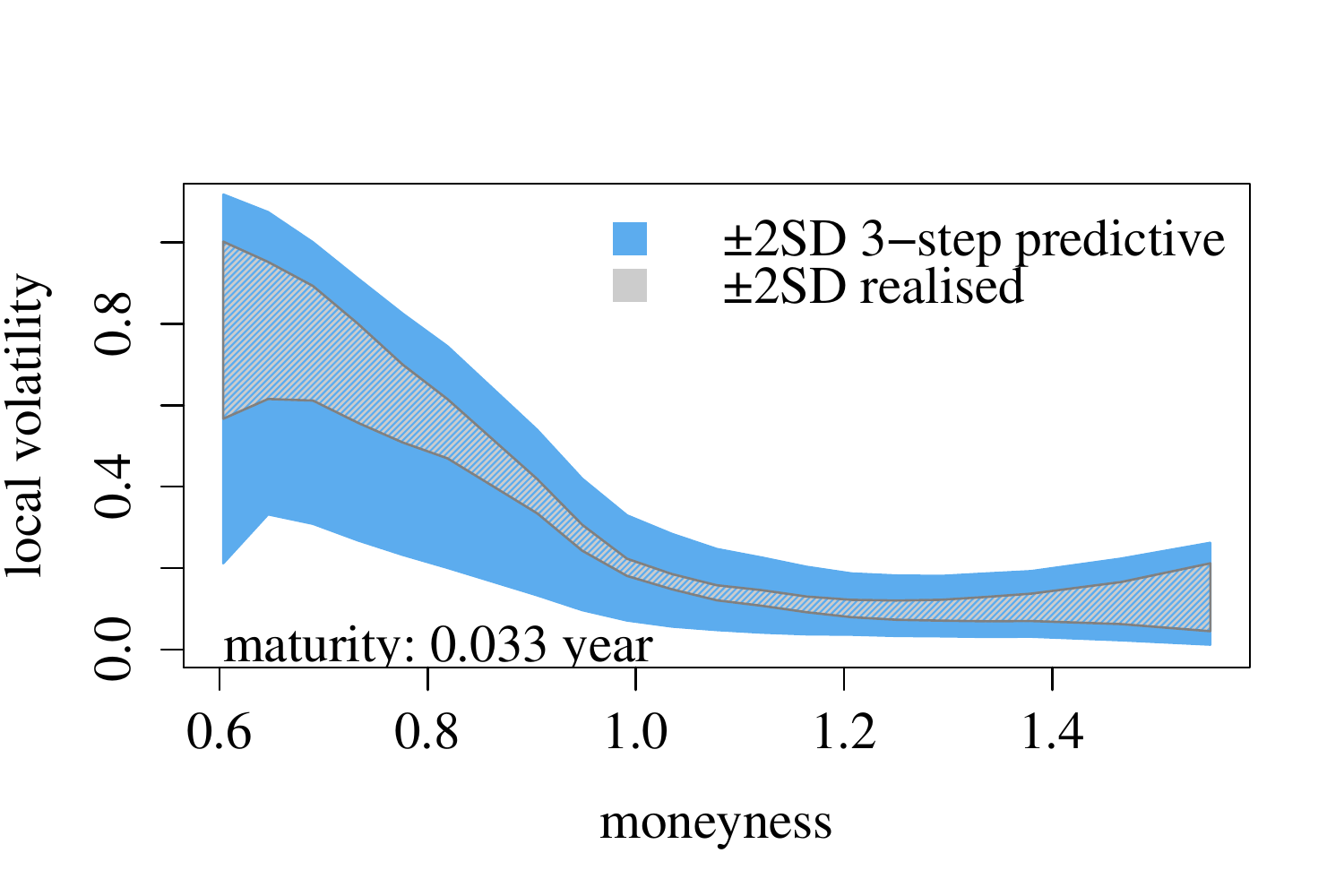} \\
\includegraphics[scale=0.5,trim=  0 65 25 50,clip]{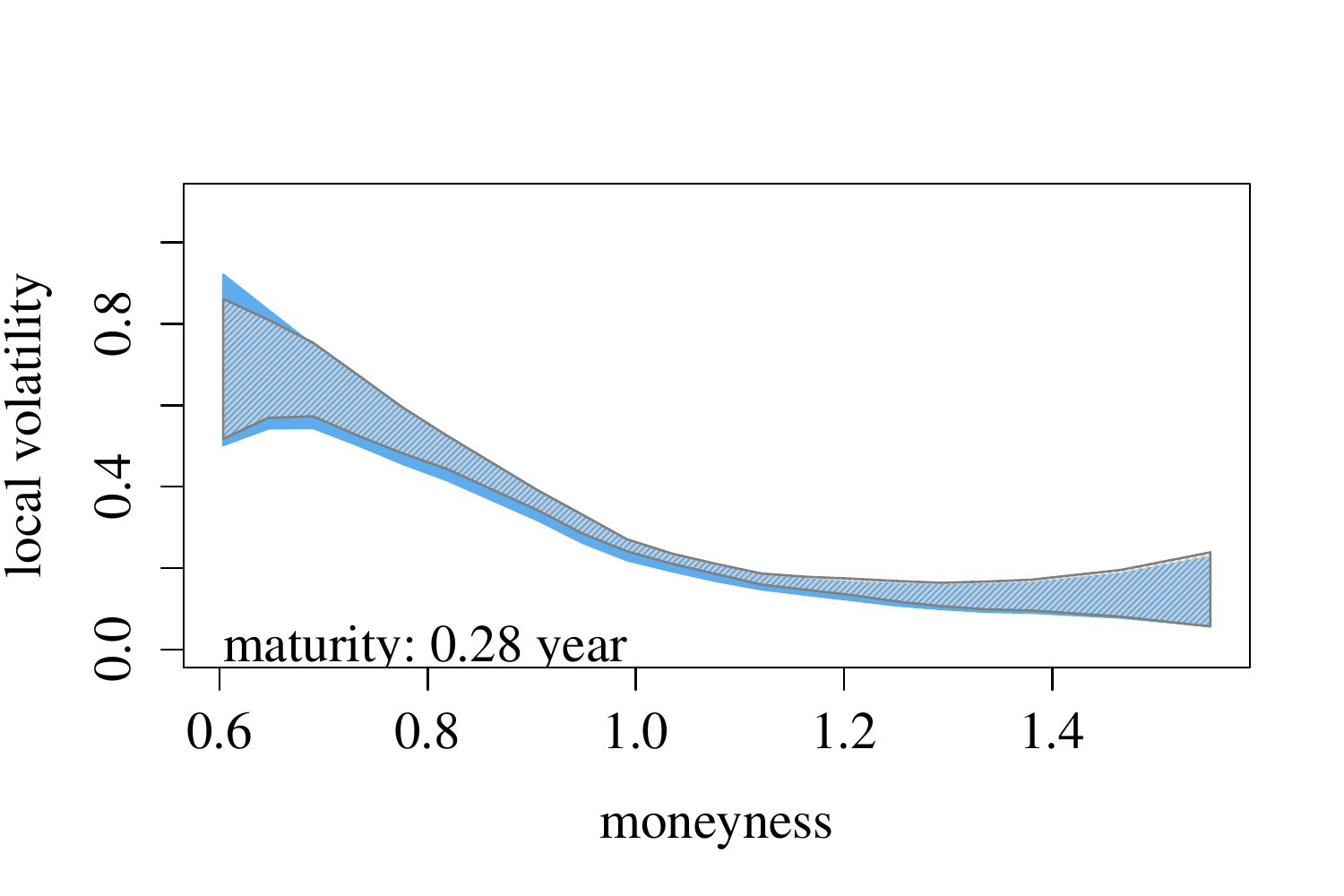} 
\includegraphics[scale=0.5,trim=  0 65 25 50,clip]{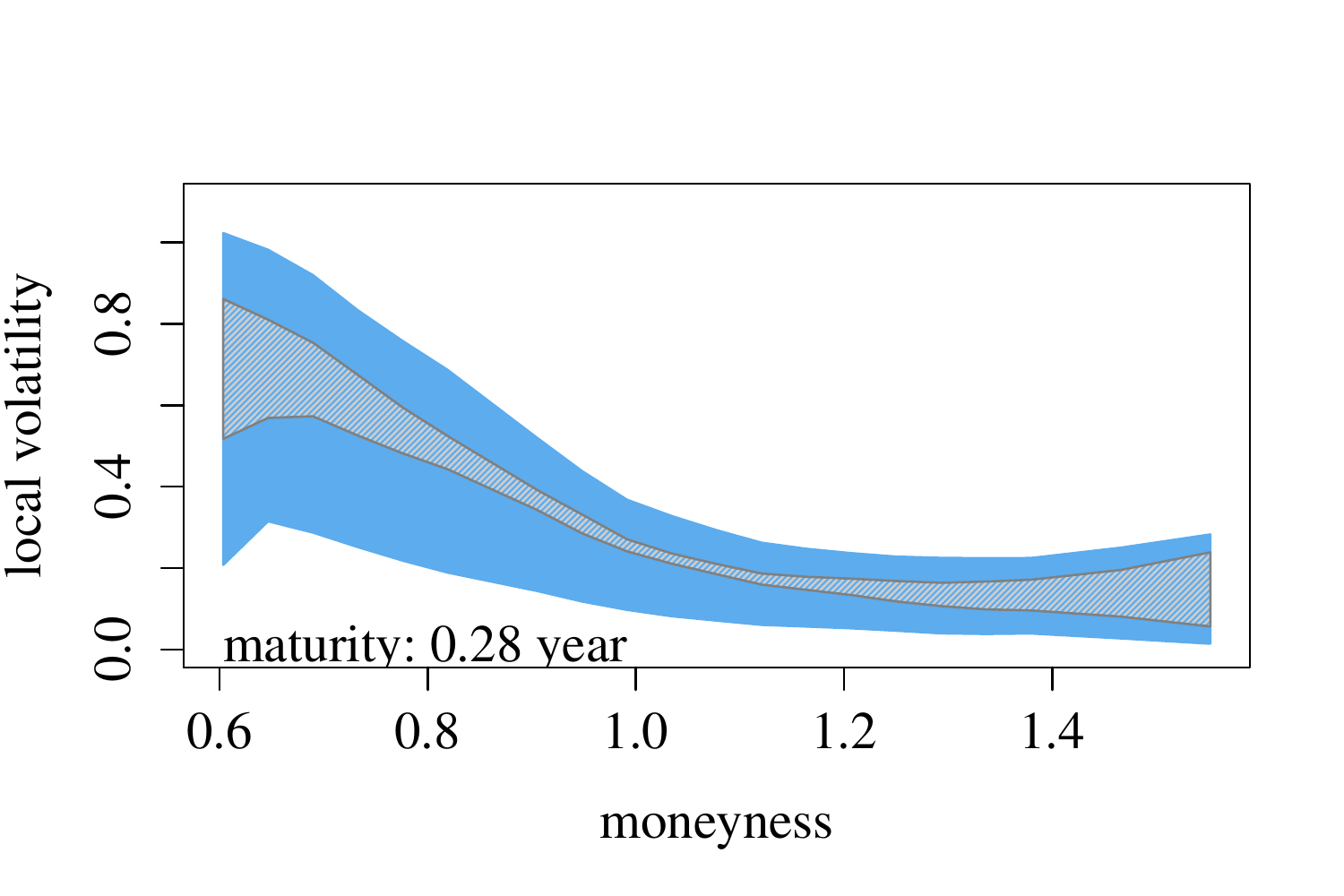} \\
\includegraphics[scale=0.5,trim=  0 65 25 50,clip]{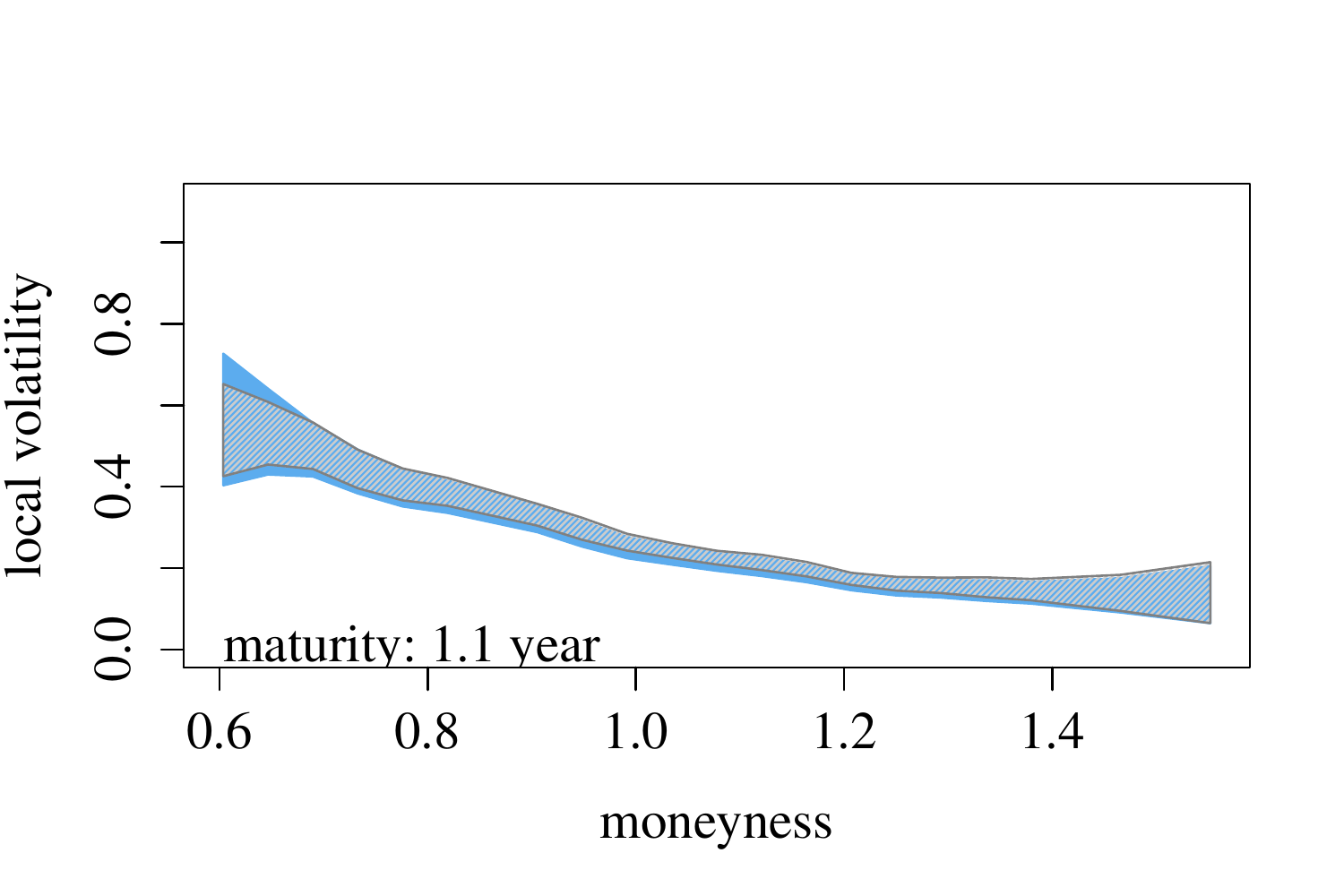} 
\includegraphics[scale=0.5,trim=  0 65 25 50,clip]{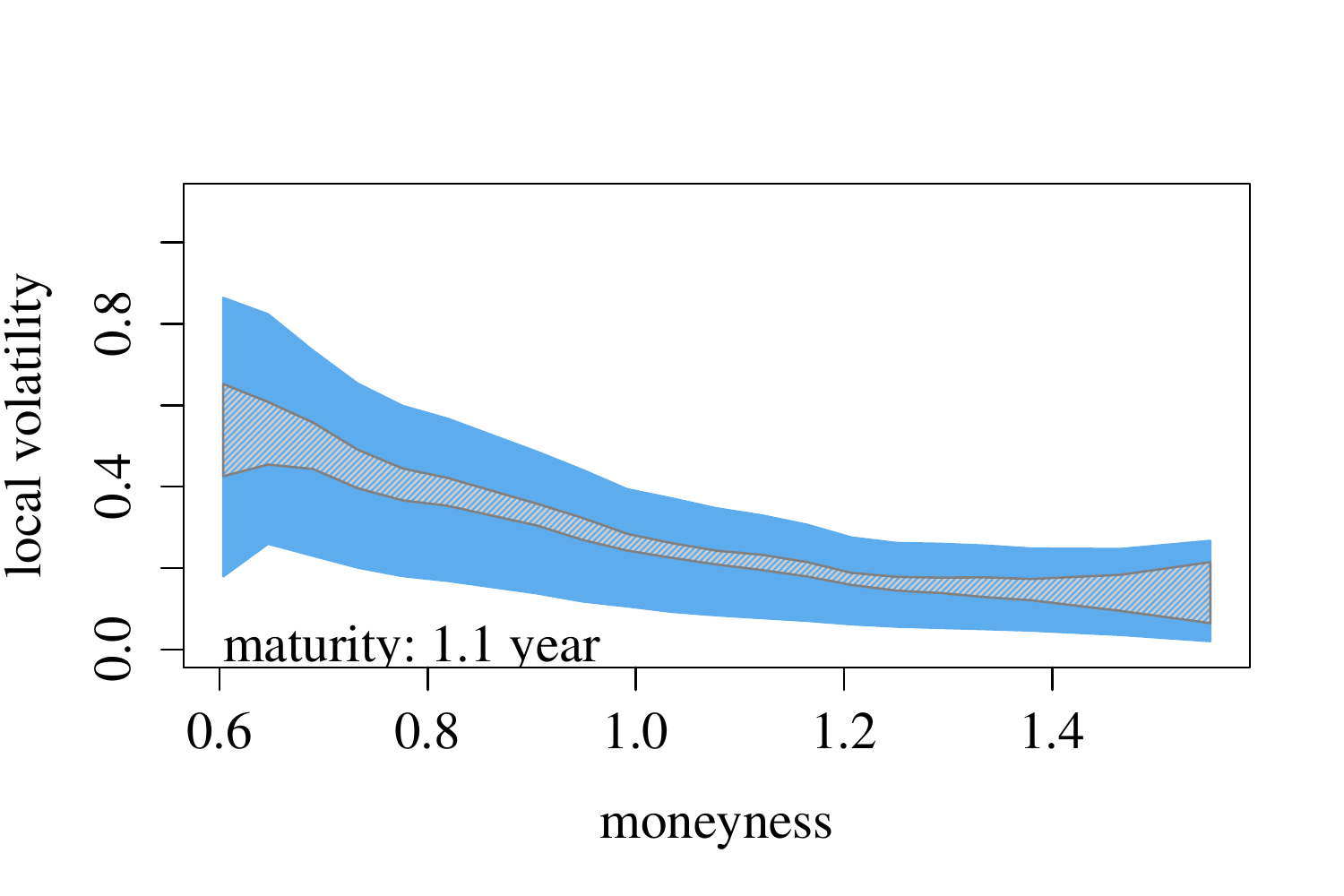} \\
\includegraphics[scale=0.5,trim=  0 10 25 50,clip]{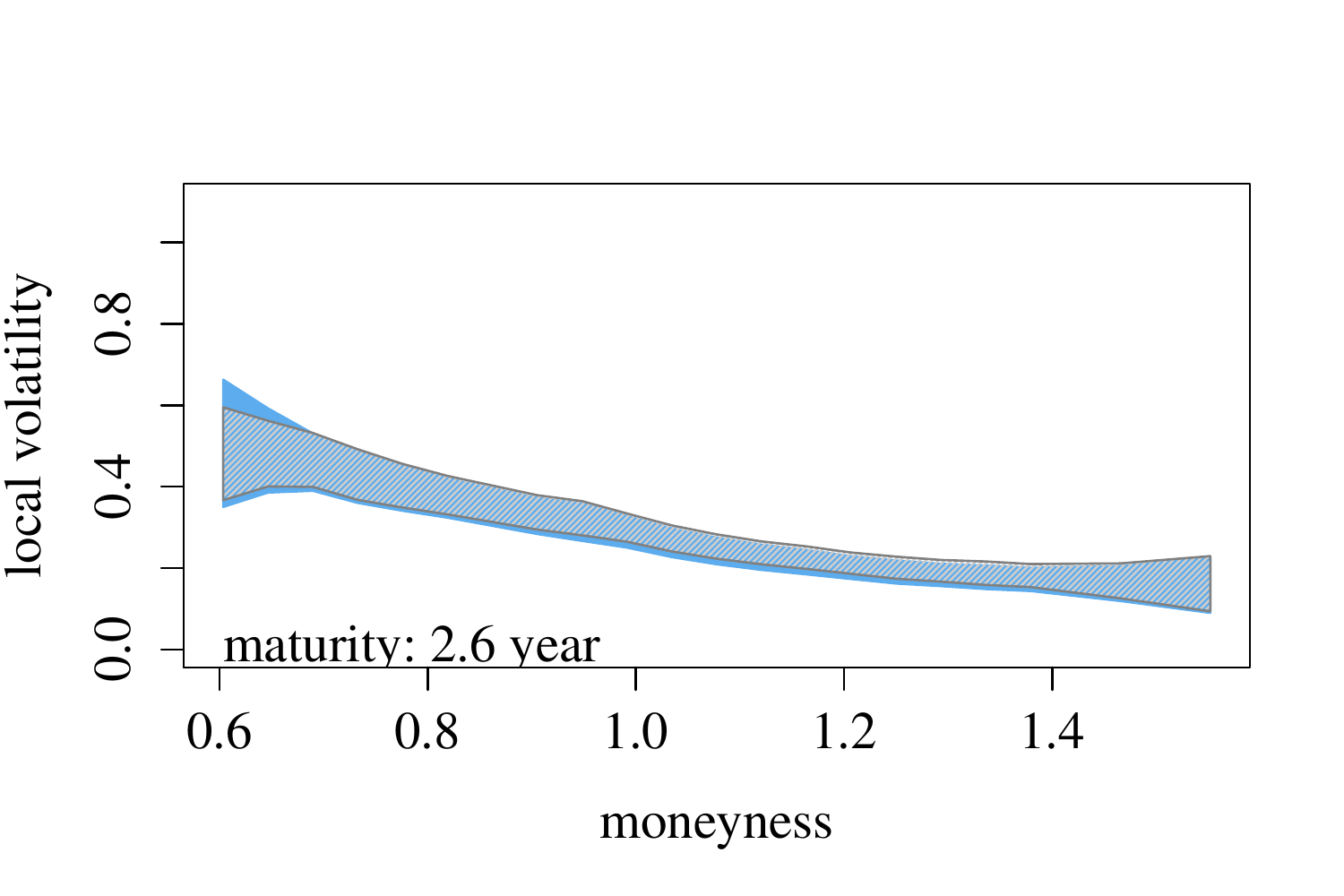} 
\includegraphics[scale=0.5,trim=  0 10 25 50,clip]{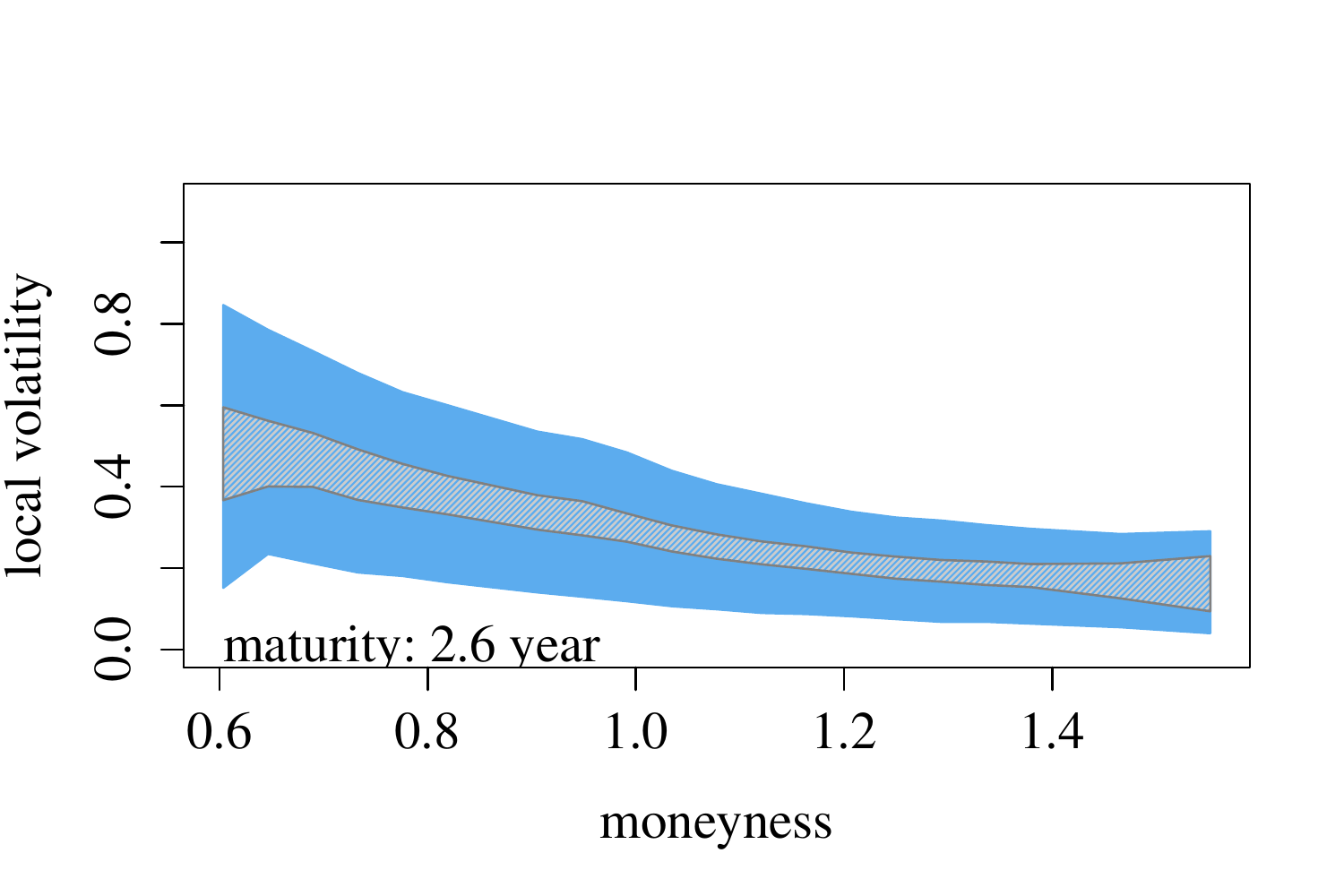} \\
\caption{Credible intervals representing the predictive distribution over local volatility as of 10 May 2010. Realised (``true'') calibration intervals shown in grey. \textbf{Left column:} 1-step prediction  from 3 May 2010. \textbf{Right column:} 3-step prediction from 19 April 2010.  }
\label{fig77}
\end{figure}

As an example, we consider  1-week and 3-weeks predictive distributions over local volatility at $\tau+s=$ 10 May 2010, i.e.  as predicted from $\tau=$ 3 May 2010 and $\tau=$ 19 April 2010, respectively. We treat \eqref{eqPRED} as an equally-weighted Gaussian mixture where the $i^{\text{th}}$  component is the  conditional prior $p(\bfs_{\tau+s}|\bf_{1:\tau},\kappa,m)$ computed with state $(\bf_{1:\tau}^{(i)},\kappa^{(i)},m^{(i)})$ from the  posterior sample of time $\tau$. For  simplicity we  condition on only the most recent functional values $\bf_{\tau}$ in the conditional prior. The result based on  a prediction sample of 1000 surfaces  
is illustrated in Figure \ref{fig77} for four maturities. In the left column, the 1-step predictive intervals ($\pm2$ std. deviations around the sample mean) are fairly similar to the ``true"  intervals as \textit{realised} with the calibration of 10 May 2010. The 3-step predictive intervals  in the  right column are considerably wider with a high degree of uncertainty, due to  the longer prediction horizon. 

In a similar fashion, $s$-step predictions of the call price are obtained by mapping  predicted local volatility surfaces through the call price operator. Only here we do not have access to the future stock price $S_{\tau+s}$. In stead we  predict implied volatilities over different levels of moneyness. These are invariant to the  stock price level as  $C(S_t,T,K=mS_t)$ and $C(S_t=1,T,K=m)$ give equivalent implied volatilities. We thus predict the latter at a range of different levels of moneyness and back out their implied volatilities.

\begin{figure}[!t]
\centering 
\includegraphics[scale=0.5,trim=  0 65 25 50,clip]{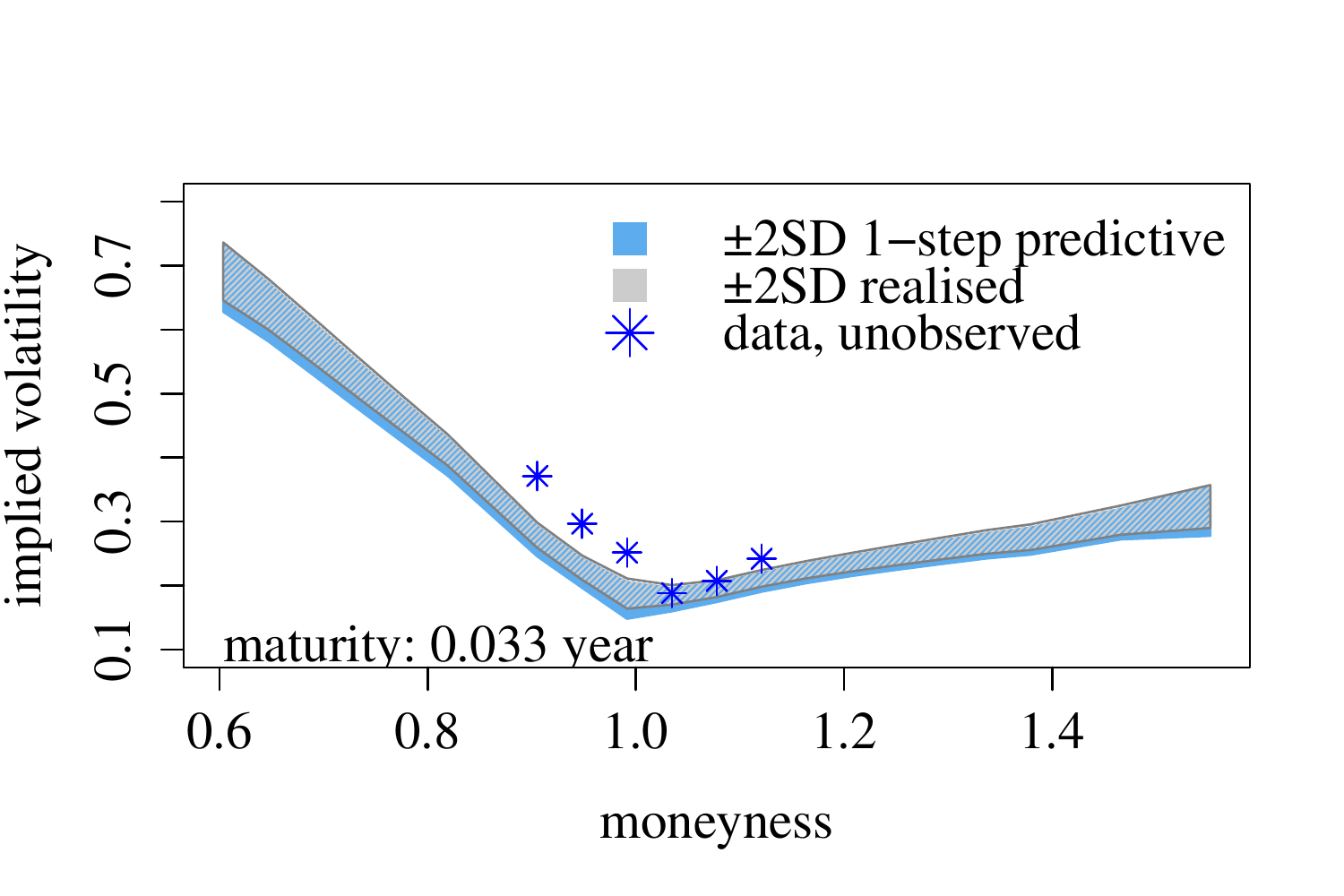} 
\includegraphics[scale=0.5,trim=  0 65 25 50,clip]{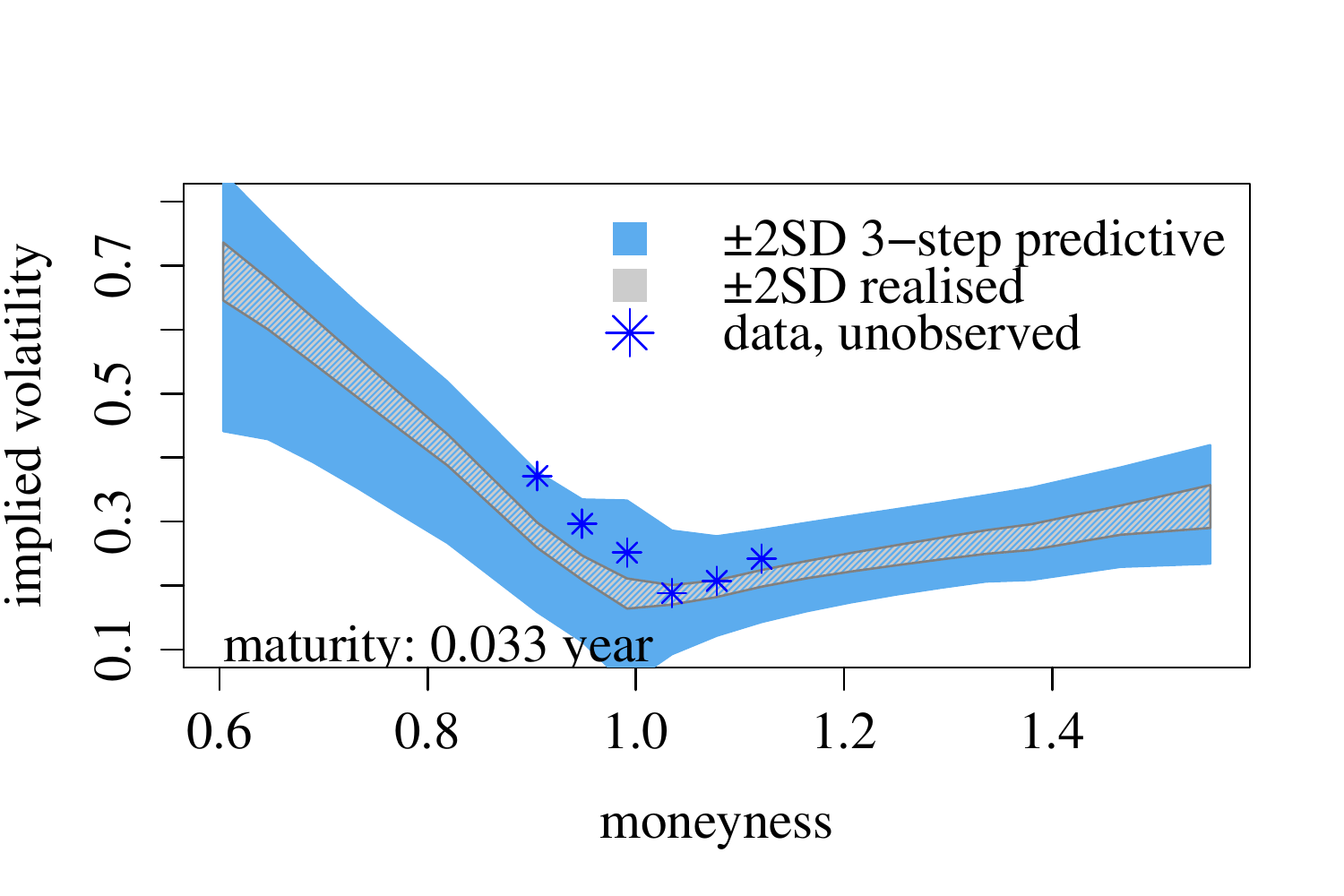} \\
\includegraphics[scale=0.5,trim=  0 65 25 50,clip]{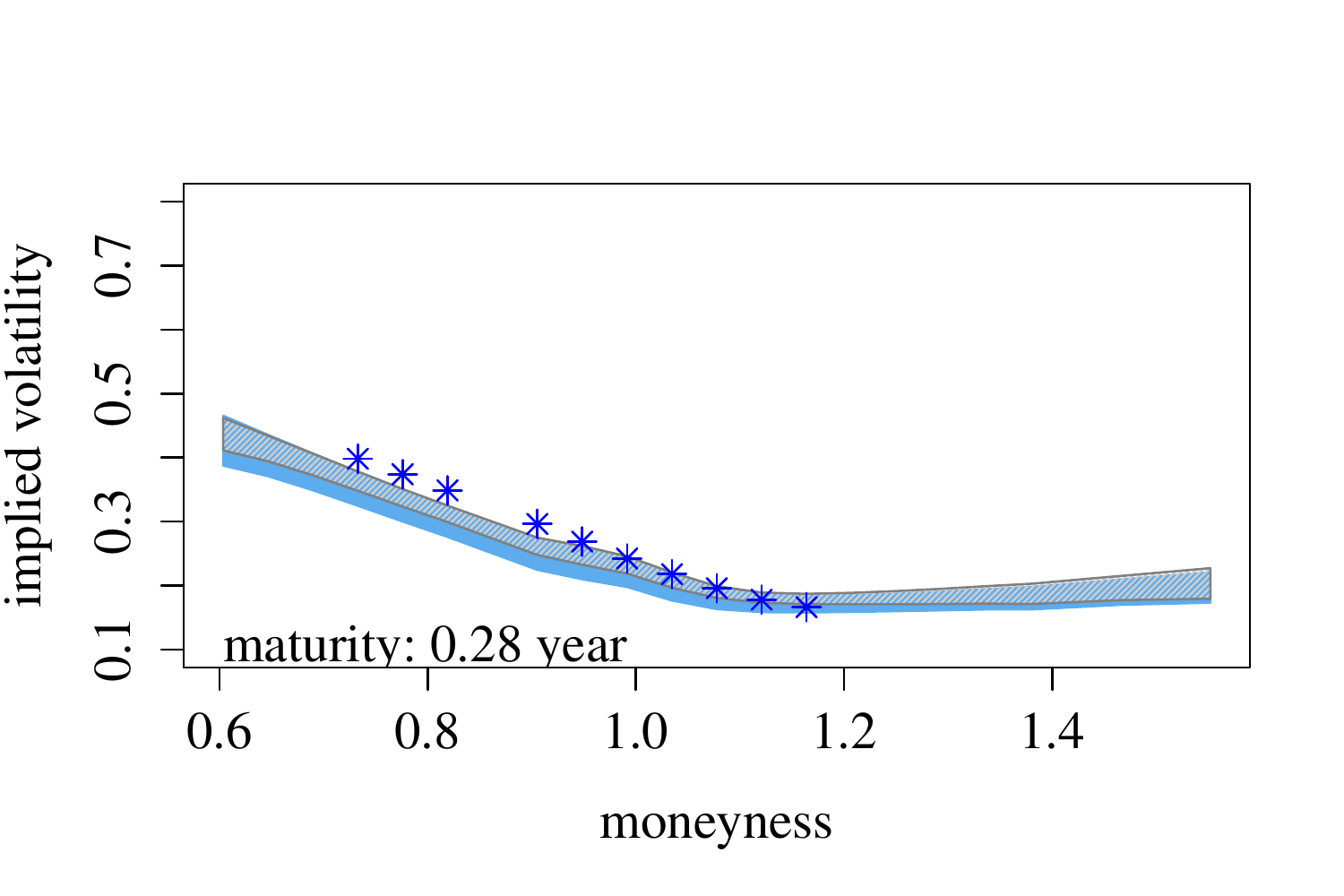} 
\includegraphics[scale=0.5,trim=  0 65 25 50,clip]{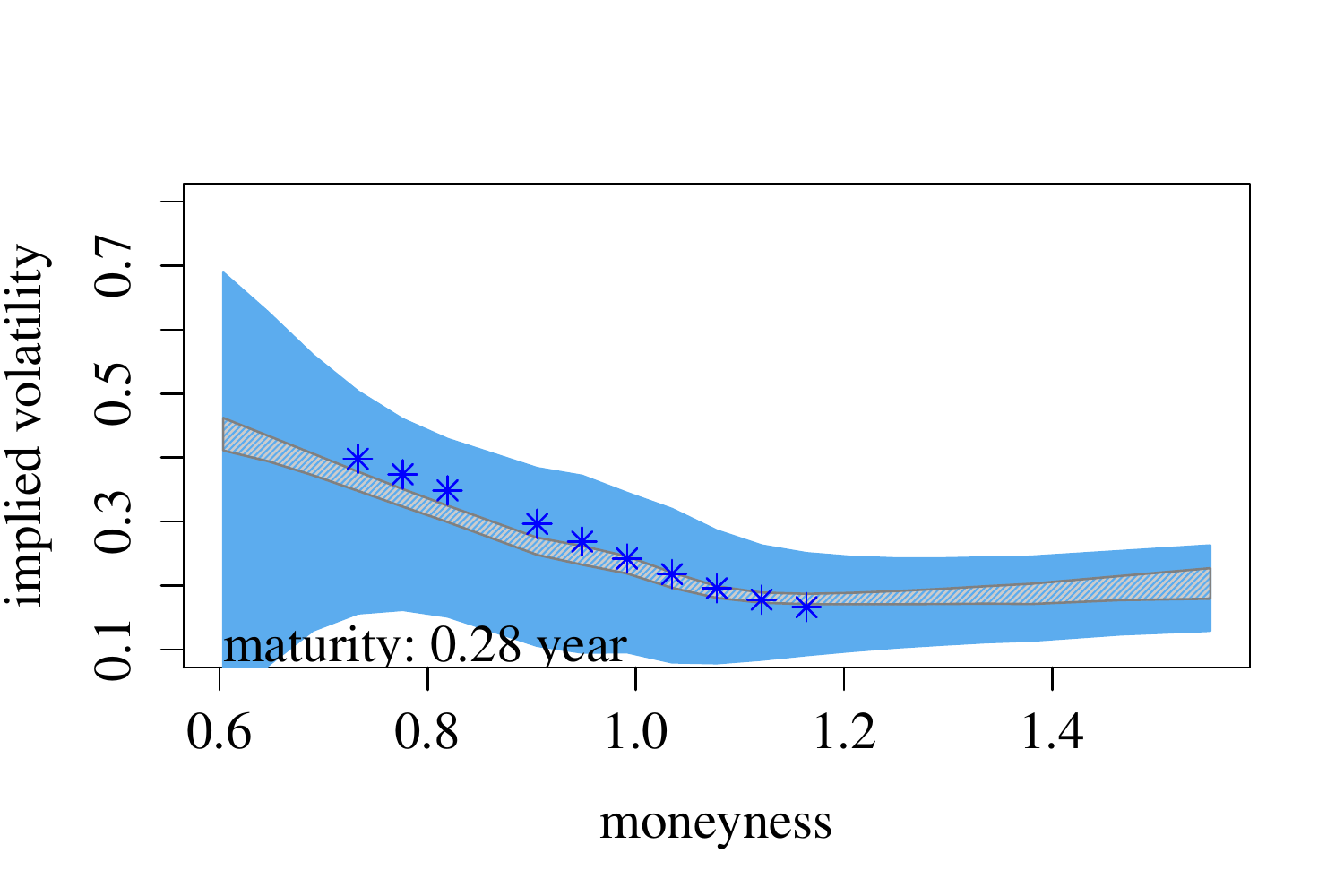} \\
\includegraphics[scale=0.5,trim=  0 65 25 50,clip]{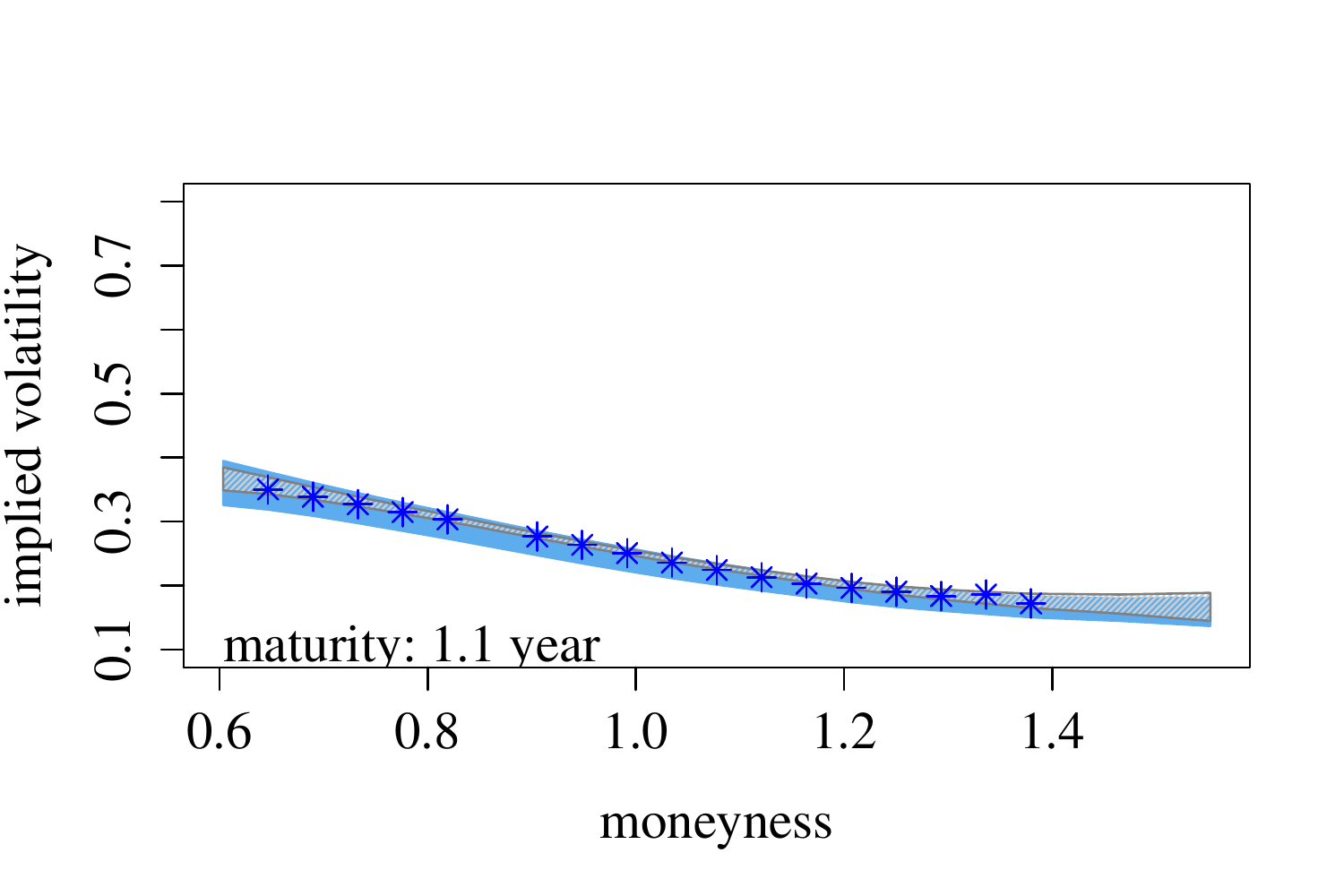} 
\includegraphics[scale=0.5,trim=  0 65 25 50,clip]{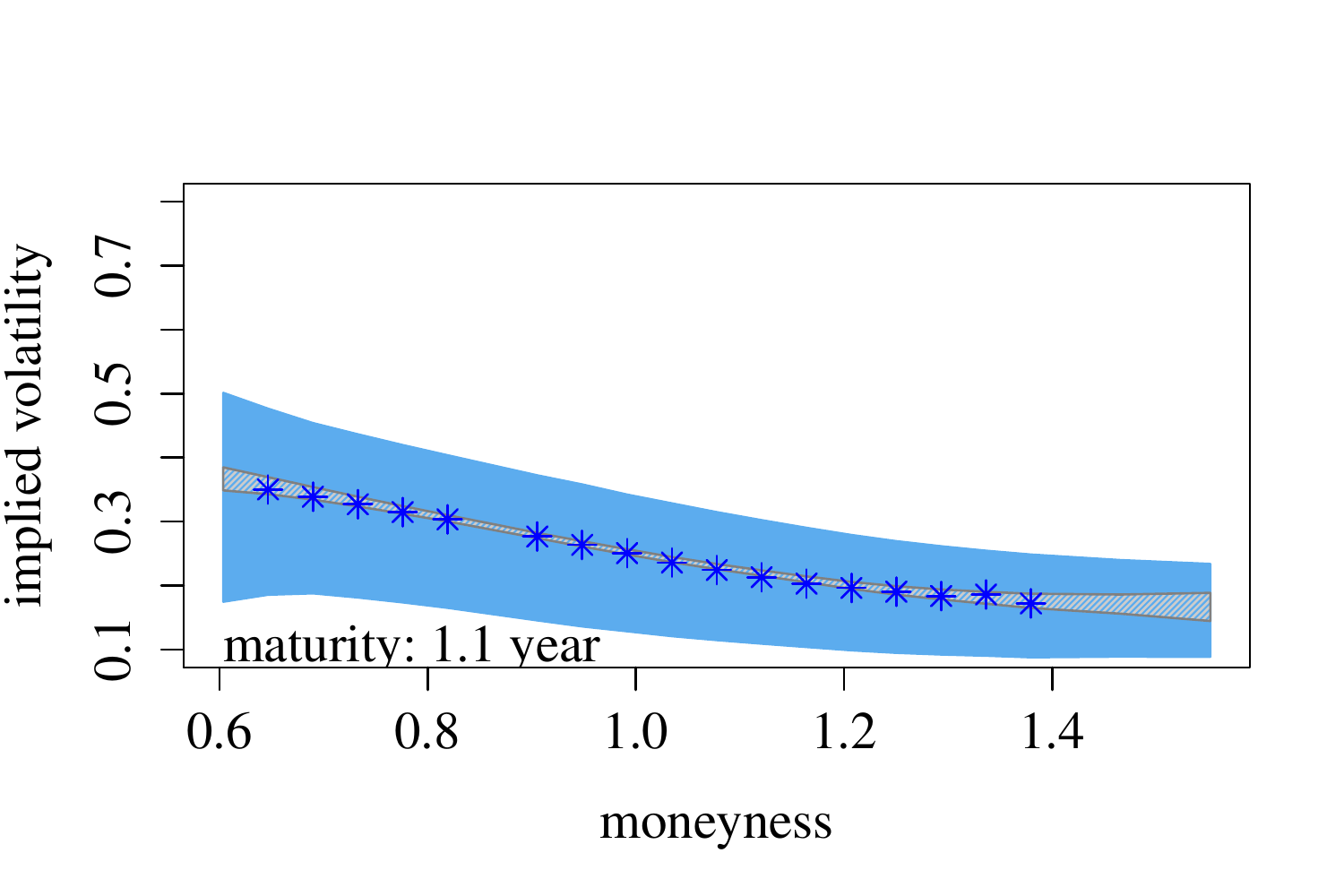} \\
\includegraphics[scale=0.5,trim=  0 10 25 50,clip]{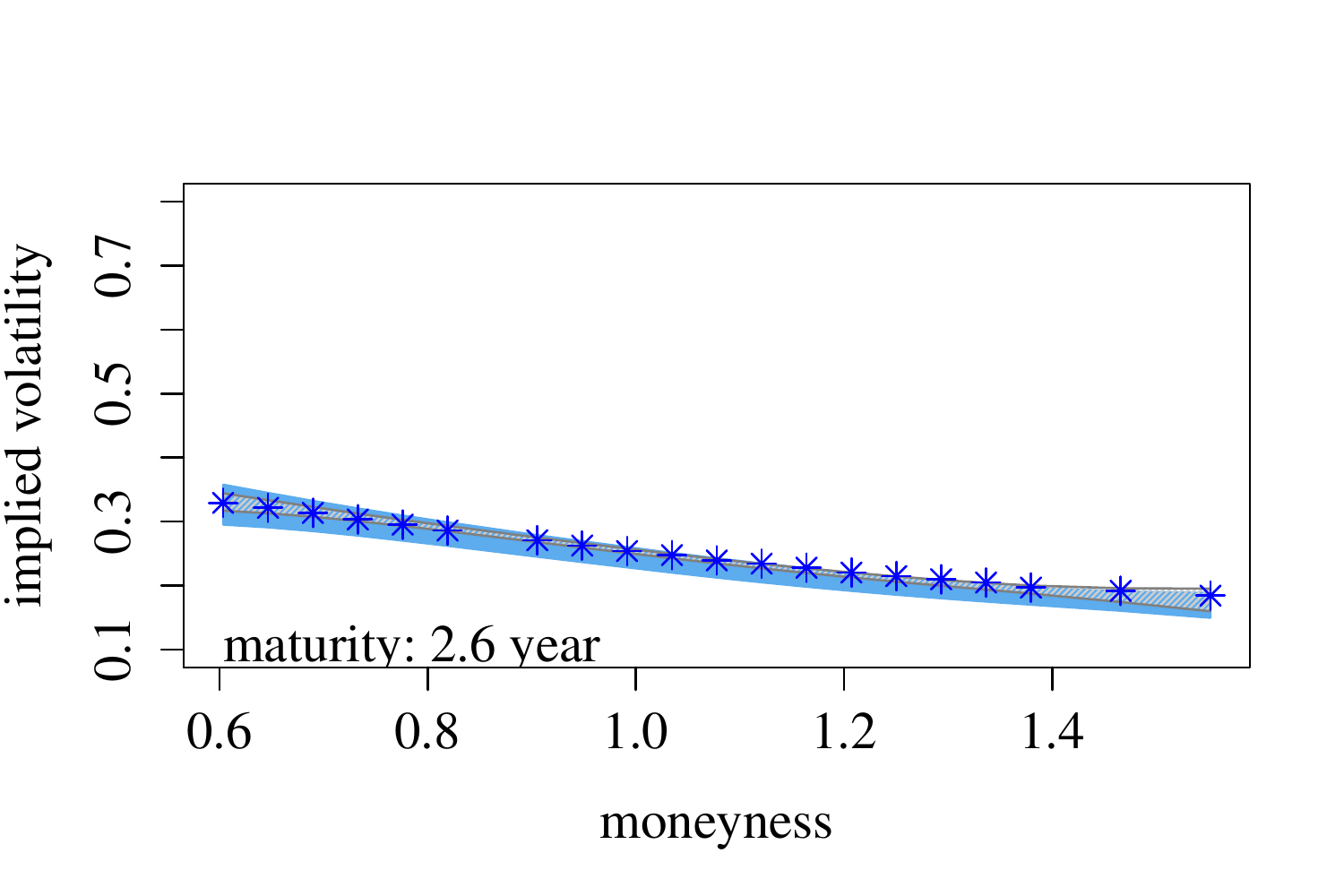} 
\includegraphics[scale=0.5,trim=  0 10 25 50,clip]{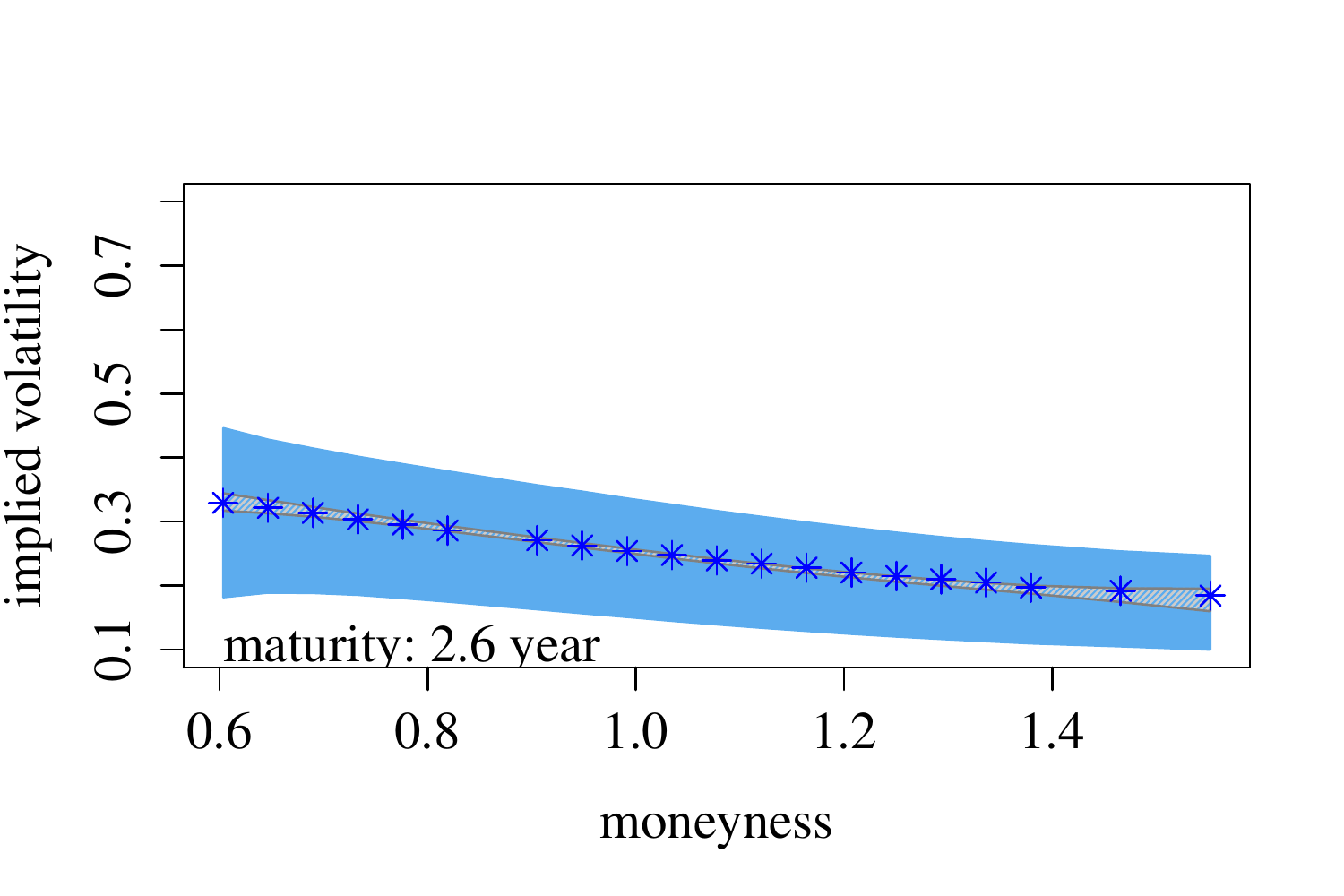} \\
\caption{Credible intervals representing the predictive distribution over implied volatility at 10 May 2010. 1-step prediction  from 3 May  (\textbf{left column}) and 3-step prediction from 19 April (\textbf{right column:}).  }
\label{fig78}
\end{figure}

Figure \ref{fig78} represents the 1-step and 3-step predictive distributions over implied volatility at 10 May 2010, corresponding to the predicted local volatility  in Figure \ref{fig77}. In terms of credible intervals,  1-step predictions are fairly similar to the realised distribution. In contrast, with 3-step predictions, the uncertainty  propagated from the  prediction of local volatility  is profound, especially at later maturities.

\begin{figure}
\makebox[\textwidth][c]{  
\includegraphics[scale=0.35,trim=0 0 10 0,clip]{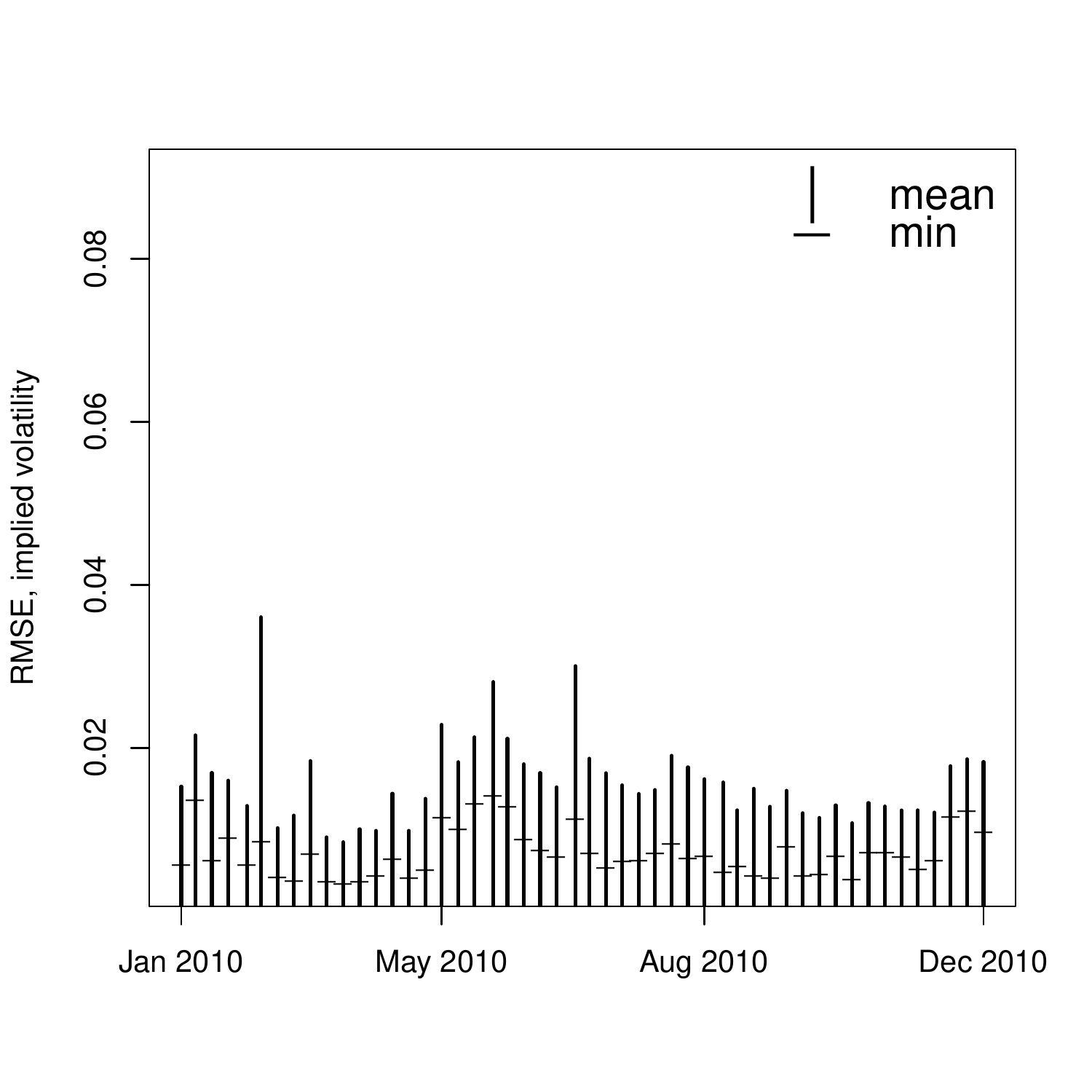} 
\includegraphics[scale=0.35,trim=0 0 10 0,clip]{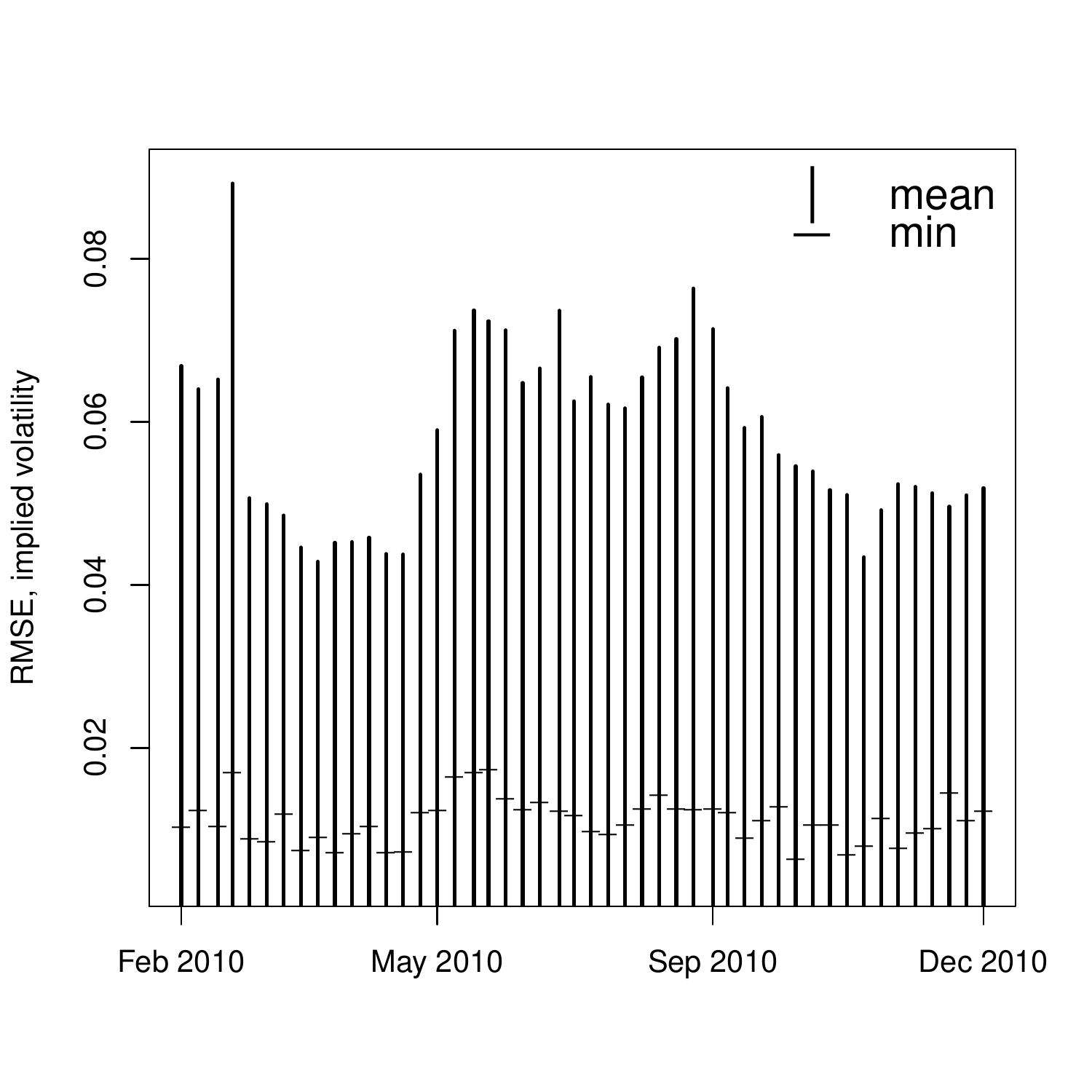} 
\includegraphics[scale=0.35,trim=0 0 10 0,clip]{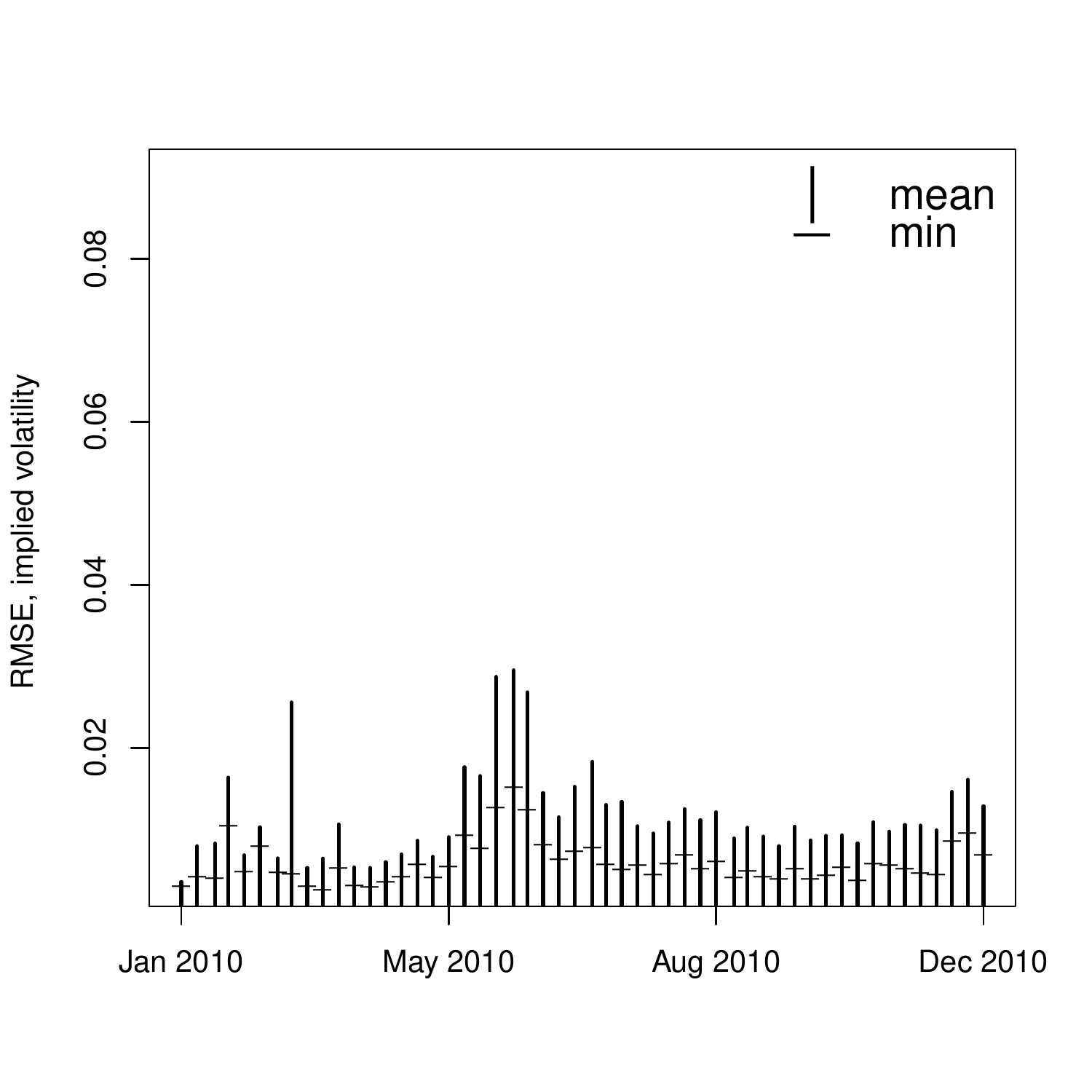} 
}
\caption{Root mean squared errors of implied volatilities at $t=\tau+s$ as predicted at $\tau$. \textbf{Left} with $s=1$ step predictions, \textbf{middle} with $s=3$ step predictions, while \textbf{right} shows $s=0$ for comparison, i.e. the calibration also shown in Figure \ref{fig5bbb}.   
}
\label{fib1}
\end{figure}

For a measure of  predictive accuracy, we use the root mean squared error of implied volatilities \eqref{eqrmse} taken at each predicted surface ${\bfs_{\tau+s}}^{(l)}$, $l=1,\dots,M$. Figure \ref{fib1} (left) shows the 1-step minimum and mean across $l$ for every prediction time $t = \tau+s$ of the data set.
If we contrast these  implied-volatility prediction errors  with their realised correspondents (Figure \ref{fib1}, right and Figure \ref{fig5bbb}, left) we see that errors' sizes are in the same order of magnitude. While  sample-minimum errors of  3-step predictions are slightly higher---see Figure \ref{fib1} (middle)---averages are considerably higher. Again, this is an effect of the longer prediction horizon.

\paragraph{Prediction of the VIX }
For a  more consolidated view on the predictive power of our approach we consider the VIX index.
The Chicago Board of Options Exchange's Volatility Index (VIX) is a benchmark index for stock market volatility. It is based on prices of S\&P 500 options by the general formula
\begin{equation}\label{vix}
VS(T) = 2\sum_i \frac{\Delta K_i}{K_i^2}e^{rT}O(K_i,T) - \left(\frac{F^T}{K_0} - 1 \right)^2
\end{equation}

\begin{figure}
\makebox[\textwidth][c]{  
\includegraphics[scale=0.5,trim=0 0 30 50,clip]{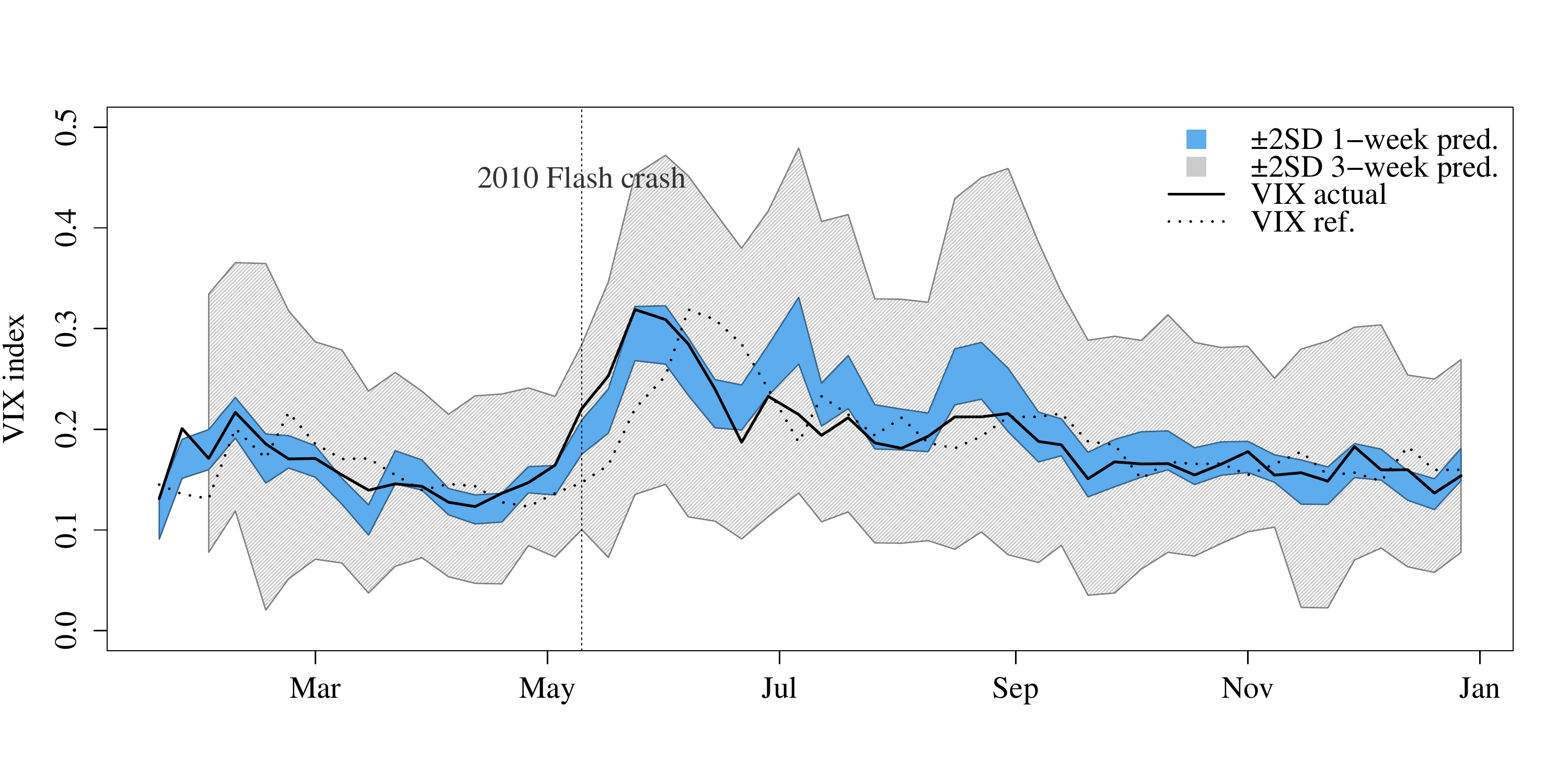} 
}
\caption{1-step and 3-step forward predictions of the VIX index during 2010.
}
\label{fib2}
\end{figure}
at the two shortest maturities $T_1$ and $T_2$ which are greater than 8 days. The index runs through all available options, $i<0$ is associated with out of money puts and $i>0$ with out of money calls; $O(K_i,T)$ denotes their price. For $i=0$, the price $O(K_0,T)$ is the average of the put and call price and the last factor in \eqref{vix} compensates for the call being in the money. $VS(T_1)$ and $VS(T_2)$ are  interpolated for an estimate at 30 days to maturity and then annualised for the final VIX,  see \cite{carr2006tale} for further details. 

 We use \eqref{vix} as a plug-in formula and map the sample of $s$-step prices  for a predictive distribution over VIX. Here we  also use predictions of $C(S_t=1,T,K=m)$ for different levels of moneyness in place of $C(S_t,T,K=mS_t)$ as their VIX calculations \eqref{vix}  are equivalent.  For a benchmark, we construct the ``true'' VIX by mapping corresponding  market prices through \eqref{vix}. The result is shown in Figure \ref{fib2} for both the 1-step and 3-step predictions. While the latter yield very wide intervals, the former show rater encouraging predictions.	For reference, we have included the index calculated with the data as of the prediction date. Compared to this, our approach outperforms.

\section{Conclusion}\label{secC}

Whilst the local volatility model provides  flexibility of fitting any set of traded option prices in theory,  practical estimation  is challenging. Ipso facto, with  discrete and finite sets of data, there is no unique local volatility \`{a} la Dupire.  Thus, along with designing an appropriate model---literature practice is a parametric functional form---comes identifiability issues when calibrating to real market prices. Without sufficient regularisation (also imposing structure on local volatility) the optimisation of common  objectives   is ill-posed. This has beckoned us to attempt probabilistic learning for  local volatility  with a cleaner, nonparametric functional hypothesis model.

In this paper, we proposed to approach the calibration problem of local volatility with Bayesian statistics to infer a conditional distribution over functions given observed data. Notably, common calibration objectives such as squared errors directly correspond to likelihood assumptions. We introduced the Gaussian process framework for local volatility modelling  and defined a prior distribution over functions. The motivation behind  was at least two-fold: First, prior domain knowledge, such as smoothness properties, was easily encoded through the functional prior. Second, while the model is nonparametric in the sense that the number of parameters  adapts with data, its \textit{complexity} is automatically tuned  in the Bayesian inference process. We further leveraged the latter for  approaching the model selection process itself, simply by considering  yet another level of inference. Lastly, we built on the rich literature on Gaussian processes   to propose efficient numerical algorithms for  training the model to data, i.e. for sampling the posterior distribution over local volatility.

We exemplified the utility of our approach with price quotes from S\&P 500 options. This included a practical set-up for the model and for selecting an appropriate set of market data. In light of   calibration results, we discussed  benefits of a probabilistic representation with a certain focus on uncertainty; its origin as well as its propagation to model derivatives. Further, as a machine for prediction almost came  for free with our approach, we  considered their probabilistic representation at maturities much longer than priced by the market, just to see that the uncertainty in such predictions where substantial.

Since local volatility  needs to be \textit{re-calibrated} on a regular basis, we extended our model to account for a time dimension in the input space. We could then infer dependencies of local volatility across time by sequential sampling of the posterior conditional on a series of price surfaces. Galvanized by   our  framework already rolled out for predictions, we immediately tried to predict local and implied volatility forward in time. We showed encouraging results for a 1-week horizon with prediction errors not  far off from corresponding realised calibration errors. Finally, we  demonstrated how straightforward it is to employ the set-up for a predictive distribution over the VIX index.

Several extensions of our work are immediate---we have merely proposed a  framework. The breath of the Gaussian process model through its covariance function  is extensive---\cite{rasmussen2006gaussian} give a basic account while \cite{wilson2013gaussian} and \cite{wilson2016deep} exemplify extensions---and learning the covariance structure can even be included in the inference process, see \cite{duvenaud2013structure} for the case of Gaussian process regression.  Exploring different covariances is one direction for future research. Similarly, we have only considered a simple  data  model with homoscedastic noise.  Still, alternative calibration objectives, such as weighted least squares and absolute pricing error, directly transfer to likelihood models which  can be  adopted in our framework. Here we also stress that more general models with correlated noise (representing believes about the  bid-ask spreads across strike-maturities) would easily fit into our approach such that  noise structure may be inferred.  Finally, for  ease of exposition, we have not looked at model derivatives, such as prices of exotics, neither at hedging quantities nor risk measures.

A more fundamental extension to our approach would be to cast the Brunick-Shreve model  \citep{brunick2013mimicking} in a Bayesian setting. The latent diffusion function is here  extended to depend on a third state, the running maximum of the stock price. Since the payoff of barrier options depends on the  latter, barriers have a similar role for this model as calls have for local volatility. With a pricing equation corresponding to Dupire's equation derived in \cite{hambly2016forward}, we could thus mimic our prevailing approach with an extended input space of the Gaussian process. Albeit,  due to the availability of market data from barriers, this set-up would have its natural limitations. We leave this for future work.

\clearpage

\bibliographystyle{apalike}
\bibliography{biblio}

\end{document}